\documentclass[%
 a4paper, twocolumn,11pt
amsmath,amssymb
aps,floatfix,superscriptaddress,
]
{revtex4-2}
\pdfoutput = 1

\usepackage[utf8]{inputenc}
\usepackage[english]{babel}
\usepackage[T1]{fontenc}
\usepackage{xkeyval}
\usepackage{graphicx}
\usepackage[export]{adjustbox}
\usepackage{soul}
\usepackage{dcolumn}
\usepackage{bm}
\usepackage{epstopdf}
 \usepackage[usenames,dvipsnames]{pstricks}
 \usepackage{epsfig}
 \usepackage{pst-grad} 
 \usepackage{pst-plot} 
 \usepackage[space]{grffile} 
 \usepackage{etoolbox} 
 \makeatletter 
 \patchcmd\Gread@eps{\@inputcheck#1 }{\@inputcheck"#1"\relax}{}{}
 \makeatother
\usepackage{booktabs}
\usepackage{boxhandler}
\usepackage{amsmath}
\usepackage[caption=false]{subfig}
\usepackage{floatrow}
\usepackage{braket}
\usepackage{amsmath}
\usepackage{amsfonts}
\makeatletter
\let\newfloat\newfloat@ltx
\makeatother

\usepackage{hyperref}
\hypersetup{
  colorlinks   = true,    
  urlcolor     = blue,    
  linkcolor    = blue,    
  citecolor    = blue      
}

\usepackage{multirow}
\usepackage{enumitem}

\newtheorem{definition}{Definition}

\renewcommand{\vec}[1]{\bm{#1}}

\usepackage{soul}
\usepackage[normalem]{ulem}

\begin{document}

\title{Leveraging biased noise for more efficient quantum error correction at the circuit-level with two-level qubits}

\author{Josu {Etxezarreta Martinez}}
\email{jetxezarreta@unav.es}
\affiliation{Department of Basic Sciences, Tecnun - University of Navarra, 20018 San Sebastian, Spain.}
\affiliation{Cavendish Laboratory, Department of Physics, University of Cambridge, Cambridge CB3 0HE, UK.}
\author{Paul Schnabl}
\affiliation{Department of Basic Sciences, Tecnun - University of Navarra, 20018 San Sebastian, Spain.}
\affiliation{Institute for Theoretical Physics, University of Innsbruck, A-6020 Innsbruck, Austria.}
\author{Javier {Oliva del Moral}}
\affiliation{Department of Basic Sciences, Tecnun - University of Navarra, 20018 San Sebastian, Spain.}
\affiliation{Donostia International Physics Center, 20018 San Sebastian, Spain.}
\author{Reza Dastbasteh}
\affiliation{Department of Basic Sciences, Tecnun - University of Navarra, 20018 San Sebastian, Spain.}
\author{Pedro M. Crespo}
\affiliation{Department of Basic Sciences, Tecnun - University of Navarra, 20018 San Sebastian, Spain.}
\author{Ruben M. Otxoa}
\affiliation{Hitachi Cambridge Laboratory, J. J. Thomson Avenue, Cambridge, CB3 0HE, United Kingdom.}


\begin{abstract}
Tailoring quantum error correction codes (QECC) to biased noise has demonstrated significant benefits. However, most of the prior research on this topic has focused on code capacity noise models. Furthermore, a no-go theorem prevents the construction of CNOT gates for two-level qubits in a bias preserving manner which may, in principle, imply that noise bias cannot be leveraged in such systems. 
In this work, we show that a residual bias up to $\eta\sim$5 can be maintained in CNOT gates under certain conditions. Moreover, we employ controlled-phase (CZ) gates in syndrome extraction circuits and show how to natively implement these in a bias-preserving manner for a broad class of qubit platforms. This motivates the introduction of what we call a \textit{hybrid biased-depolarizing} (HBD) circuit-level noise model which captures these features.
We numerically study the performance of the XZZX surface code and observe that bias-preserving CZ gates are critical for leveraging biased noise. Accounting for the residual bias present in the CNOT gates, we observe an increase in the code threshold up to a $1.27\%$ physical error rate, representing a $90\%$ improvement. Additionally, we find that the required qubit footprint can be reduced by up to a $75\%$ at relevant physical error rates.

\end{abstract}

\maketitle
\section{Introduction}
The extremely fragile nature of quantum information implies that quantum error correction (QEC) techniques are going to be necessary to obtain the full benefits that quantum computing offers theoretically \cite{fowlerSC,decoders}. In the quest of fault tolerance, many technologies are being explored as candidates to host the error corrected quantum computers of the future. Superconducting junctions, neutral atoms, ion traps or silicon spin quantum dots are some examples of these technologies \cite{qubitTechs}.

Interestingly, some of those qubit technologies present a strong bias towards dephasing errors, i.e. the probability of experiencing Pauli $Z$ (phase-flip) errors is much higher than of experiencing $X$ (bit-flip) and $Y$ (bit-and-phase-flip) errors. This occurs for qubits with much longer relaxation times than dephasing times, $T_1>>T_2$. The bias or degree of asymmetry is usually quantified by the parameter $\eta = p_z/(p_x+p_y)\approx T_1/T_2 - 1/2$ \cite{xzzx,biasQLDPC,approximatingdecoherence}. This is the case, for example, for trapped-ion qubits \cite{preskillBiased,iontrapbias1,iontrapbias2,iontrapbias3}, silicon spin qubits \cite{preskillBiased,woottonSpin,takeda2022quantum,Noiri2022,DiraqSpinQ}, NV center qubits \cite{NVbias1,NVbias2} or certain superconducting qubit architectures \cite{preskillBiased,biasQLDPC,toninid,catExponentialSup,cat100secT1}. Intuitively, errors that are strongly biased towards dephasing should be helpful for QEC since the entropy of the error source is smaller than for the symmetric or depolarizing case \cite{entropyPauliChann,xzzx}. Indeed, bias tailoring of QECCs has resulted in great performance improvements when the noise experienced by qubits presents such feature \cite{xzzx,tailoredXZZX,biasQLDPC,tuckettBiasSC,tuckettxy,biasColorCodes}. The XZZX surface code is a notable example of such bias tailoring, exhibiting a threshold of $\approx 48\%$ at finite and experimentally relevant bias of $\eta = 1000$ \cite{xzzx}. This significantly surpasses the $10.9\%$ threshold of this code (and for the CSS surface code) for the depolarizing case.

However, most of the bias tailoring of QECCs has been done considering code capacity noise, i.e. considering that only the data qubits of the code experience errors and that the syndrome extraction circuits are ideal \cite{decoders}. In reality, all elements involved in the syndrome extraction are faulty and introduce additional errors. This noise model is referred to as \textit{circuit-level noise}. For this model, each noisy operation in the syndrome extraction circuits is modeled as an ideal operation followed by a noise channel, i.e. if $\tilde{\mathcal{U}}$ is a noisy gate, then it is modeled as $\mathcal{N}\circ\mathcal{U}$. While one may naively think that in a system with biased noise all the noisy operations can be modeled as ideal gates followed by biased noise channels, i.e. as to be \textit{bias-preserving}, this assumption does not hold in general. A no-go theorem in \cite{twolevelCNOTnobiaspreserve} demonstrates that a bias-preserving CNOT gate can not be implemented between two qubits encoded in finite-dimensional Hilbert spaces (we will refer to these as \textit{two-level qubits} through the text as in \cite{catXZZX}). This result is a huge impediment for bias tailoring QECCs due to the fundamental role of the CNOT gate in syndrome extraction circuits.

In this context, several alternatives to circumvent the no-go theorem in \cite{twolevelCNOTnobiaspreserve} have been proposed in groundbreaking results for bias-preserving CNOT gates in superconducting cat qubits  \cite{biaspreservingCNOTs} and for certain neutral atom qubit processors \cite{biaspreservingCNOTsNeutral}. As a result, cat qubit technologies have received special attention \cite{Q2BPreskill}. Furthermore, huge qubit overhead reductions have been reported theoretically as a result of the combination of bias tailoring with quantum low-density-parity-check codes (qLDPC) over cat qubits \cite{catoverhead,catoverhead2}. For example, the calculations in \cite{catoverhead2} predict that around $100000$ cat qubits can be enough to run Shor's algorithm to factor $2048$ RSA integers, while surface code-based QEC over a platform with symmetric noise would require almost $20$ million physical qubits \cite{gidneyShor}. That is a $200$ times overhead reduction. In the quest for constructing the first \textit{Megaquop machine}, or a quantum computer able to realize millions of error free operations, such overhead savings may be crucial for certain technologies to surpass others \cite{PreskillMegaquop}. The construction of a Megaquop machine consisting of around $100$ logical qubits is considered to be the next milestone for the field of quantum computing, posing the end of the so-called Noisy-Intemediate Scale Quantum (NISQ) era.

Nevertheless, a question remains open: can we leverage the natural bias towards dephasing noise present in certain platforms based on two-level qubits (or qubits encoded in finite dimensional systems) even if the CNOT gates implemented in those are not bias-preserving? This has lost some of the attention of the community beyond cat qubits, mainly due to the limitation of constructing bias-preserving CNOT gates in such platforms leading to the study of QECCs over depolarizing circuit-level noise models whenever two-level qubit platforms are considered \cite{beliefmatching,catoverhead,catoverhead2,biasedCLNog,catXZZX}. Furthermore, it is also important to state that while possible, the experimental implementation of bias-preserving CNOT gates on cat qubit platforms is very challenging and remains an open question \cite{catCNOTbpdifficult}. In fact, note that the recent experimental demonstration of a repetition code in a hybrid cat-transmon architecture uses CNOT gates with high bias (higher than $25$), but which are not bias-preserving \cite{catRepExperimental}.

In this work, we aim to answer to such a question, focusing on which elements of the syndrome extraction circuit can, in fact, be done in a bias-preserving manner so that dephasing noise can still be predominant enough in the system that bias tailoring can still be successful. 
Specifically, we begin by defining what bias-preserving gates are, to then discuss which of the usual gates used for syndrome extraction are bias-preserving and which are not. We study the residual or remainder bias of CNOT (and CY) gates by means of a Lindblad equation and determine that while most of the bias is destroyed, there is still a certain bias in their noise channels. Also, we observe that Hadamard gates fully depolarize the noise bias. We then focus on the importance of CZ gates in syndrome extraction protocols and how those gates can be constructed in a bias-preserving manner across several quantum technologies. Furthermore, we discuss the importance of implementing the CZ gate in a bias-preserving manner at the native hardware level. As a result, we propose a \textit{hybrid biased-depolarizing circuit level noise model} (HBD) that captures the noise features of the qubit platforms under discussion in a generic way. In this model, we consider that CZ gates are bias-preserving while the rest of the gates strictly introduce fully symmetric, i.e. depolarizing, noise. In order to determine if tailored codes can leverage noise bias under this noise model, we numerically study the performance of the XZZX variant of the rotated surface code. We compare the obtained performance with the case in which the CZ gates are not natively constructed in a bias-preserving manner. We observe almost no improvement of the threshold for the latter case as a function of bias. This starkly contrasts with the bias-preserving case, with notable improvement in performance (around $40\%$). We consider this generic HBD circuit-level noise model to be a valuable tool for determining whether certain bias tailored codes can exhibit performance improvements when implemented on a two-level qubit technology.

Moreover, motivated by our analysis of the remainder bias noise present in CNOT gates, we numerically study the rotated XZZX surface code over a more fine grained HBD circuit-level noise model for which the CNOT gates do exhibit this residual bias. The main rationale to do so resides in the fact that a small bias does result in significant threshold increase whenever biased code capacity models are considered \cite{xzzx}. These results confirm our intuition with a code threshold surpassing $1.2\%$ whenever $\eta\geq 100$, which is a $90\%$ boost relative to the standard depolarizing case. We continue our study by numerically exploring the Mega-, Giga-, and Teraquop footprints for two experimentally relevant physical error rates, $p=0.003$ and $p=0.001$. We observe that as a result of the bias tailoring, footprint reductions ranging from $33\%$ up to a $75\%$ are obtained for all relevant quantum operation regimes. We envision that these results are relevant for practical QEC since huge physical qubit savings can be obtained.

\section{Syndrome extraction circuits and bias-preserving gates}\label{sec:BPgates}
QECCs rely on extraction circuits to obtain the syndrome that the decoder will use to estimate the error that has occurred and therefore propose a recovery operator \cite{decoders}. Those extraction circuits refer to measuring the stabilisers defining the code and are generally conducted by means of two-qubit entangling gates between the check qubit and the data qubits whose parity is captured by such ancilla. In general, a stabiliser measurement is done by means of a Hadamard test in which the controlled unitary refers to the stabiliser being measured by the check qubit in question \cite{approximatingdecoherence}. Since the stabilisers are Pauli strings, the controlled stabiliser operation is split into CNOT ($X$-components), CZ ($Z$-components) and controlled-Y/CY ($Y$-components) two-qubit gates entangling the associated data qubits and the respective check. It is well known that not all gates can be done in a bias-preserving manner (see Appendix \ref{app:biasP} for a formal definition of bias-preserving gates). As said above, the ubiquitous CNOT gate in syndrome extraction circuits cannot be implemented in a bias-preserving manner for two-level qubits \cite{twolevelCNOTnobiaspreserve,biaspreservingCNOTs}. As a result, it is a general conception that the benefits of bias tailoring are effectively lost for two-level qubits with natural biased noise. In the following, we discuss this issue and analyze the scenario for CZ gates.

\subsection{CNOT gates (and CY)}

CNOT gates are essential gates in order to measure $X$ stabilisers and, thus, detect possible phase-flip errors. However, these gates cannot be constructed in a bias-preserving manner for two-level qubits \cite{twolevelCNOTnobiaspreserve}. As explained in \cite{biaspreservingCNOTs}, the continuous evolution required to implement a CNOT gate can convert phase errors into $Y$ errors \cite{catXZZX}, resulting in a reduction of the bias of the noise toward dephasing. Specifically, a CNOT gate can be implemented with an interaction described by the Hamiltonian \cite{biaspreservingCNOTs}
\begin{equation}\label{eq:CNOTH}
    H_{CNOT} = V \left[\frac{I+Z}{2}\otimes I + \frac{I-Z}{2}\otimes X\right],
\end{equation}
where $V$ refers to the strength of the interaction and the CNOT gate is realized at time $VT=\pi/2$, up to a global phase. During the continuous noisy evolution described by this interaction, some of the phase-flips occurring in the target qubit are transformed due to the anticommutation between the target term and the predominant $Z$ errors implying that a CNOT gate cannot be implemented in a bias-preserving manner, i.e. $\tilde{\mathcal{U}}_{CNOT} \neq \Lambda_P(\eta_{sys})\circ\mathcal{U}_{CNOT}$, where $\eta_{sys}$ refers to the bias of the system. While there are alternatives to implement a CNOT gate, no bias-preserving CNOT gate can be constructed for two-level qubits \cite{twolevelCNOTnobiaspreserve}.

However, the fact that some phase-flips may be transformed into bit-flips does not imply that the noise is fully depolarized, i.e. all possible Pauli errors are equiprobable. To elucidate this, we analyze the underlying structure that the noise presents after the application of the noisy CNOT gate implemented by means of the interaction in equation \eqref{eq:CNOTH}.

We present the remainder or residual noise bias towards dephasing errors in Figure \ref{fig:biaslossCNOT} (see Appendix \ref{app:lindblad} for details on the numerical simulations). The results show that the actual bias of the noise towards dephasing for the CNOT gate is reduced very significantly, rapidly saturating at $\eta_{CNOT} \approx 5$ for $\eta_{sys} > 1000$. While the bias is reduced by orders of magnitude, it is noteworthy to recall that, for example, the XZZX code showed almost a $25\%$ increase of the threshold in the code capacity model for a $\eta =5$ bias \cite{xzzx}. We will further discuss how this moderate bias could be very beneficial later on.

\begin{figure}[!h]
\centering
\includegraphics[width=\columnwidth]{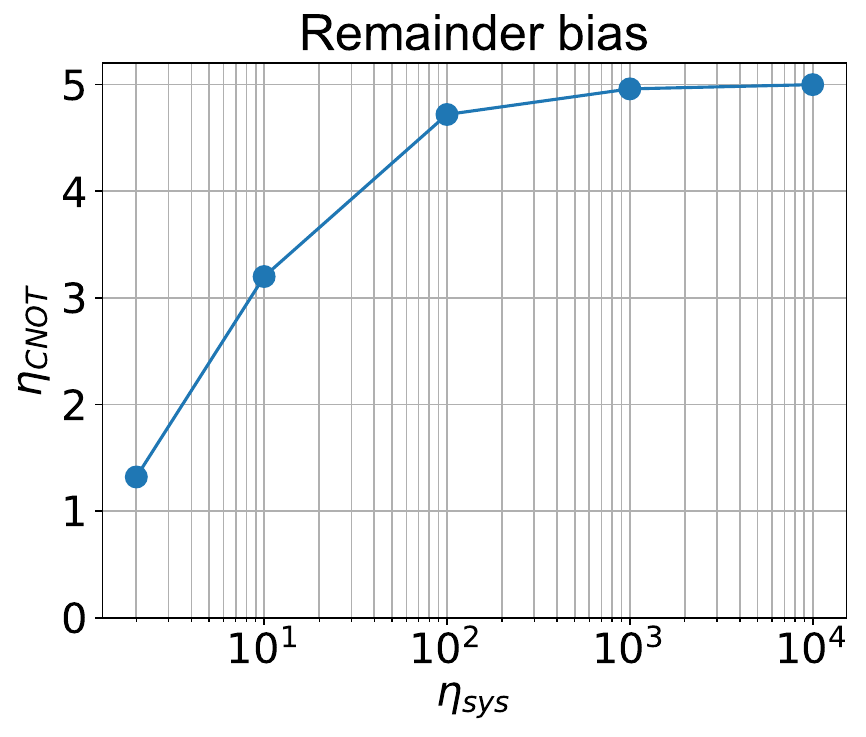}
\caption{Remainder bias towards dephasing errors when implementing a CNOT gate with an interaction in equation \eqref{eq:CNOTH}. We set the coupling rates of the dissipators so that the gate fidelity is around $99.7\%$ and change the system bias $\eta_{sys}$. We name the residual bias after the CNOT gate as $\eta_{CNOT}.$}
\label{fig:biaslossCNOT}
\end{figure}

Some other code proposals tailored to deal with biased noise involve $Y$ stabilisers, which are susceptible to both bit-flip and phase-flip errors \cite{beliefmatching,ZZZY}. In order to measure such stabilisers, controlled-Y or CY gates are employed to entangle the data and check qubits. These gates can be implemented with an interaction similar to equation~\eqref{eq:CNOTH}, but with a $Y$ operator on the target qubit instead of the $X$ operator. Note that for these gates, the predominant dephasing noise in the system will also not commute with such operator of the target qubit, indicating that the bias is also similarly lowered.

\subsection{Hadamard gates}
Hadamard gates also play a pivotal role in syndrome extraction circuits and therefore it becomes crucial to elucidate what happens with the bias when applying these single qubit gates. A Hadamard gate is a $\pi$ rotation over the $(X+Z)/\sqrt{2}$ axis in the Bloch sphere. Hence, it can already be seen that, since the interaction does not commute with the predominant dephasing noise, it will not result in a bias-preserving gate. 

Furthermore, we conducted a similar analysis as for the CNOT and CY gates in order to quantify the residual noise after the execution of a Hadamard gate (see Appendix \ref{app:lindblad} for details on the numerical simulations). For doing so, we perform the rotation by means of the interaction
\begin{equation}\label{eq:HamHad}
    H_{Had} = V \left[\frac{X+Z}{\sqrt{2}}\right],
\end{equation}
which results in the desired gate at time $VT = \pi/2$ up to a global phase. The rest of the setting is similar, i.e. the resulting gate fidelity is around $99.7\%$, with the slight difference that there are three dissipators, $\{X,Y,Z\}$. We observed that for this case the noise bias is completely depolarized, i.e. the structure of the resulting noise channel is almost symmetric independently of the system noise bias \footnote{We observe that the output bias is in the range $\eta_{Had}\in[0.5,0.6]$, i.e almost depolarizing for every case. Recall that for the single-qubit case the bias is quantified as $\eta = p_z/(p_x+p_y)$, with $1/2$ being the depolarizing case.}. Thus, we conclude that the Hadamard gates will introduce depolarizing errors, at least if implemented by the interaction in equation \eqref{eq:HamHad}.

\subsection{CZ gates}
Controlled-Z (CZ) gates are also ubiquitous in syndrome extraction circuits, with the aim of measuring $Z$-stabilisers susceptible to bit-flip errors. The essential difference between this entangling gate and the previously discussed ones is that it is a rotation around the $ZZ$ basis. Hence, this gate commutes with the predominant error channel, i.e. pure dephasing errors, and is a bias-preserving gate in the sense of definition \ref{def:biaspreserving}. Similar to the CY gate, the CZ gate can be realized by means of substituting the $X$ term in the control qubit of the Hamiltonian in \eqref{eq:CNOTH} by a $Z$ term. Furthermore, as explained in \cite{biaspreservingCNOTs}, it can also be achieved by means of an interaction described by Hamiltonian
\begin{equation}
        H_{ZZ} = -V \left[ Z\otimes Z\right],
\end{equation}
and some local Pauli $Z$ rotations, i.e. also commuting with the predominant noise \cite{ZZCZdecomposition}.

To sum up, the main point is that CZ gates can be implemented in a bias-preserving manner, as was pointed out before in \cite{preskillBiased,biaspreservingCNOTs,catXZZX}. As we will later discuss, this fact is critical in order to leverage biased noise when working with two-level qubits.

\subsection{Native implementations of the CZ gate}
In the previous section, we have seen that a CZ gate can be enabled in a bias-preserving manner for two-level qubits. However, this does not imply that using this gate for syndrome extraction circuits will result unavoidably in biased noise. We emphasize here the term \textit{native gate}, which refers to the implementation of a specific quantum operation in a physical hardware without requiring further decomposition into other gates. The interaction between qubits, as well as interactions between them and an external electromagnetic field, depends on the underlying hardware architecture. Achieving a desired quantum operation requires precise control over these interactions. In Table \ref{tab:nativeGates}, we summarize ways of enabling CZ gates \footnote{Note that there can be many other ways of enabling such gates in those platforms. Our intention here is to show that there are ways that preserve bias and ways that do not.} in several technologies exhibiting biased noise as well as if those would be bias-preserving. See Appendix \ref{app:native} for a description of how those native CZ gates are implemented.

\begin{table}[h!]
 \centering
\caption{Certain native interactions to enable CZ gates for several qubit technologies. See Appendix \ref{app:native} for descriptions of these gates.}
\label{tab:nativeGates}
\begin{tabular}{ |c|c|c| }
 \hline
 Platform & Interaction & Bias-preserving \\ 
 \hline\hline
Silicon spin qubits & Exchange (strong limit) & No \\ 
 \hline
  & Exchange (weak limit) & Yes \\ 
 \hline
  Trapped ions & Mølmer–Sørensen (XX) & No \\ 
 \hline
   & Liebfried–Sørensen & Yes \\ 
 \hline
  Neutral Atoms & Rydberg blockade & Yes \\ 
 \hline
  NV centers & Magnetic dipolar interaction & Yes \\ 
 \hline
  & Hyperfine interaction & Yes \\ 
 \hline
 Superconducting & Cross resonance gate & No \\ 
 \hline
  & STC coupler & Yes \\ 
 \hline
\end{tabular}
\end{table}

\section{Quantum error correction for two-level qubits exhibiting biased noise}

A pertinent question is whether the noise bias can still be leveraged for more efficient QEC even if some of the gates do not preserve the bias. In this section, we answer this question by studying the XZZX surface code \cite{xzzx} (see Appendix \ref{app:xzzxcode} for a description of the rotated XZZX code). Specifically, we first study a \textit{hybrid biased-depolarizing circuit-level noise model} as a generic tool for studying QEC tailored to biased noise when dealing with two-level qubit and study the performance of the rotated XZZX code over such model (see Appendix \ref{app:HBD} for our noise model proposal). We then proceed to study a more tailored hybrid circuit-level noise model considering our previous discussion on how some residual bias still exists when implementing CNOT gates.

\subsection{The hybrid biased-depolarizing circuit-level noise model}

\floatsetup[figure]{style=plain,subcapbesideposition=top}
\captionsetup[subfloat]{labelformat=brace}
\begin{figure*}[!ht]
\centering
\sidesubfloat[]{\includegraphics[width=7.9cm]{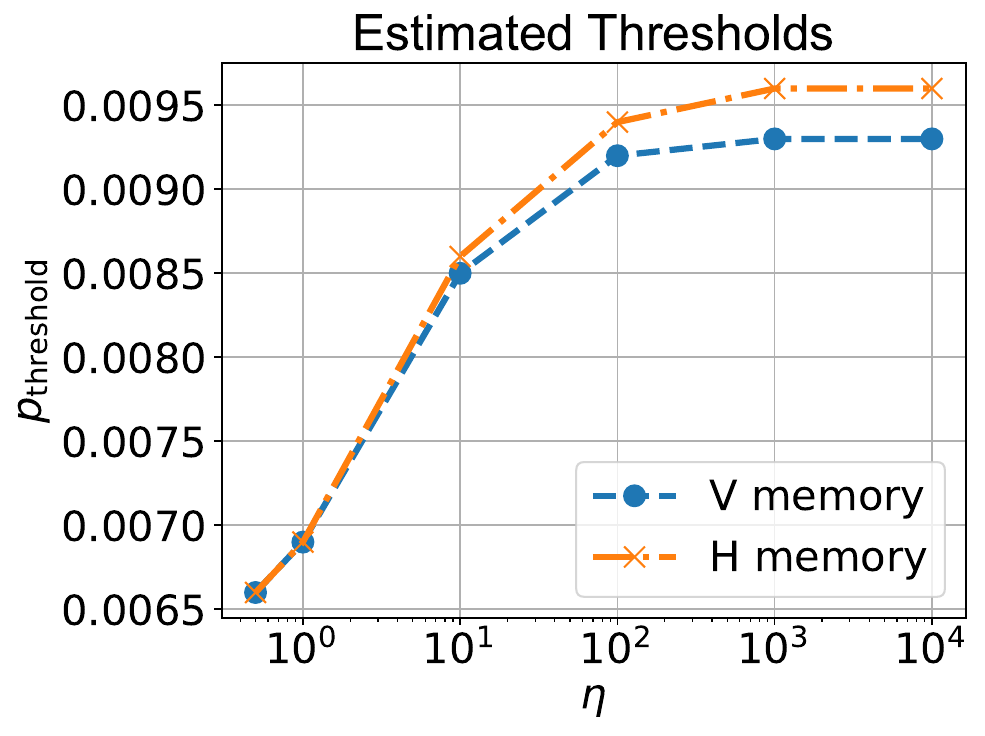}\label{fig:threshsub1}}\quad%
\sidesubfloat[]{\includegraphics[width=7.9cm]{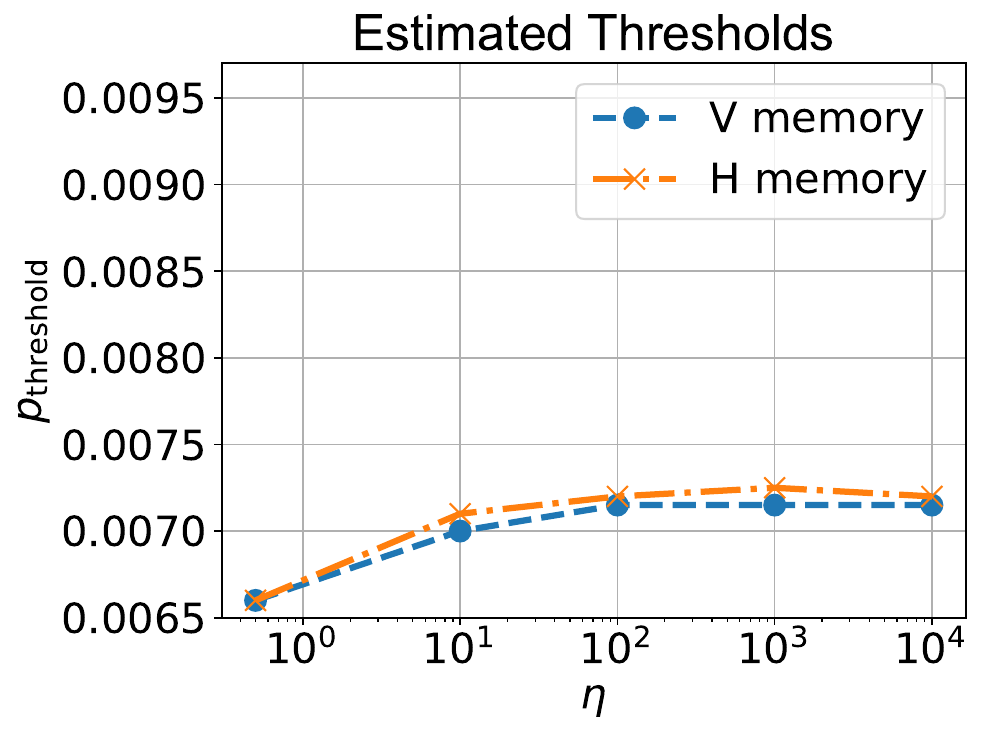}\label{fig:threshsub2}}
\caption{Threshold of the rotated XZZX code as a function of the system bias for the HBD circuit-level noise model with and without bias-preserving CZ gates. \textbf{a)} Bias-preserving CZ gates. The threshold improved from around $0.66\%$ for the SD noise model up to $0.93\%$ when $\eta \geq 100$. This is a $40\%$ improvement. Furthermore, a saturation of the threshold can be observed. \textbf{b)} CZ gates that are not bias-preserving, modeled here as introducing two-qubit depolarizing errors. There is no significant improvement for this case. In fact thresholds saturate early on around the value obtained for $\eta=1$ with bias-preserving CZ gates.}
\label{fig:hbdThresh}
\end{figure*}

Figure \ref{fig:threshsub1} shows the estimated thresholds for the XZZX variant of the rotated surface code when operating over the proposed HBD circuit-level noise model. Note that here we refer to the standard depolarizing circuit-level noise model with an $\eta = 1/2$. Note also that the thresholds for the H memory are slightly higher than the ones for the V memory. However, we observed that their subthreshold, i.e. at physical error rates substantially lower than the threshold, logical error rates are very similar. Specifically, even if the threshold is higher, the logical error rates of the H memory are slightly worse (almost insignificantly) than the ones for the V memory for the deep subthreshold physical error rates. The cause for this effect is probably that for the H memory there are $\lceil d/2 \rceil$ logical strings consisting of $\lceil d/2\rceil$ $Z$ operators and $\lfloor d/2 \rfloor$ $X$ operators, while for the V memory there are $\lfloor d/2 \rfloor$ of those. Since the considered distances are odd and the noise is predominantly biased, this makes the logical error probability slightly higher for the H memory.

\begin{table}[h!]
 \centering
\caption{Improvement of the threshold as a function of the bias for the HBD circuit-level noise model.}
\label{tab:hdbthreshgains}
\begin{tabular}{ |c|c|c| }
 \hline
 $\eta$ & $p_{threshold}$ & Improvement ($\approx$) \\ 
 \hline\hline
$1/2$ (SD) & $0.66\%$ & - \\ 
 \hline
 $1$ & $0.69\%$ & $4.5\%$ \\ 
 \hline
  $10$ & $0.85\%$ & $29\%$ \\ 
 \hline
  $100$ & $0.92\%$ & $40\%$ \\ 
 \hline
  $1000$ & $0.93\%$ & $40\%$ \\ 
 \hline
  $10000$ & $0.93\%$ & $40\%$ \\ 
 \hline
\end{tabular}
\end{table}

Furthermore, Figure \ref{fig:threshsub2} shows the estimated thresholds for the XZZX code if the CZ gates used for syndrome extraction are not done in a bias-preserving manner, i.e. they introduce depolarizing noise. For such a case, it can be seen that the threshold barely increases and rapidly saturates around $0.7\%$. This is very close to the threshold of the $\eta=1$ case when the CZ gates are done so that they preserve bias, i.e. in Figure \ref{fig:threshsub1}, indicating that there is no real performance gain for this case. Hence, it is very important to enable the CZ interactions in such a way (see Section \ref{sec:BPgates}) to leverage the bias present in certain two-level qubits. Otherwise, no performance gain will be obtained.

\subsection{A more tailored hybrid circuit-level noise model with residually biased CNOT gates}\label{sec:tailoredCNOTs}

If we recall section \ref{sec:BPgates}, CNOT gates do show some small residual bias in the structure of the noise. In this section, we will study the importance of such residual bias.

The noise model considered in this section is the HBD circuit-level noise model with the difference that the CNOT gates present some residual biased noise. Specifically, the model considers the tuple $(\eta_{sys},\eta_{CNOT})$ with the values obtained in Section \ref{sec:BPgates}, in which the CNOT gate is enabled via the interaction in equation \eqref{eq:CNOTH}. For clarity, we will use the value $\eta = \eta_{sys}$ for the plots and discussion, but it should be clear that this does not imply that the CNOT gates preserve this bias, i.e. for each $\eta$, the CNOT will have its own residual bias. Finally, we will consider that the CZ gates are constructed by means of native interactions that preserve the system bias.
\subsubsection{Threshold}
\begin{figure}[!h]
\centering
\includegraphics[width=\columnwidth]{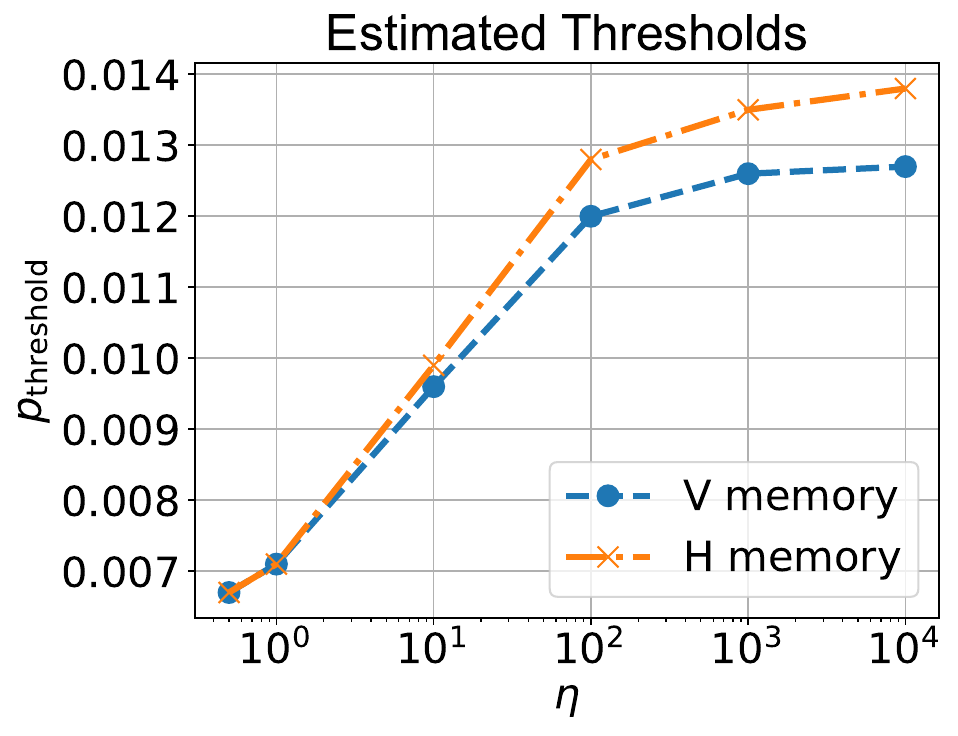}
\caption{Threshold of the rotated XZZX code as a function of the system bias considering residual CNOT bias. We use the residual bias values as studied in section \ref{sec:BPgates}. The threshold improved from around $0.66\%$ for the SD noise model up to $1.27\%$. This is a $93\%$ improvement. Furthermore, a saturation of the threshold can still be observed for high bias values.}
\label{fig:residualCNOTbiasThresh}
\end{figure}

In Figure \ref{fig:residualCNOTbiasThresh}, we present the threshold improvement as a result of the system bias for the studied code. As it can be observed, the actual threshold improvement is significantly higher than for the HBD model without considering CNOT gates with a residual bias. In fact, the threshold improves up to $\approx 1.27\%$ for system biases exceeding a thousand. It is noteworthy to comment that this is a $91\%$ improvement with respect to the standard depolarizing (SD) circuit-level noise case.

Thus, the fact that the CNOT gates present residual bias is very important for leveraging the bias towards dephasing even for two-level qubits. Furthermore, the obtained threshold improvements are significant considering moderate residual values saturating at $\eta_{CNOT} = 5$, as presented in section \ref{sec:BPgates}. To understand this, it is important to recall that this is only possible due to the presence of bias-preserving CZ gates, indicating that any improvement in the residual bias on the CNOTs results in significantly increasing the overall bias in the whole noisy extraction circuits. As seen for code capacity models in \cite{xzzx}, even a slight increase in the bias resulted in huge threshold improvements and, here, we are observing the same effect for the residual bias in the noisy CNOT gates used for extraction.

\begin{table}[h!]
 \centering
\caption{Improvement of the threshold as a function of the bias for the HBD circuit-level noise model and CNOT gates exhibiting residual bias as studied in Section \ref{sec:BPgates}.}
\label{tab:biasCNOTthreshgains}
\begin{tabular}{ |c|c|c| }
 \hline
 $\eta$ & $p_{threshold}$ & Improvement ($\approx$) \\ 
 \hline\hline
$1/2$ (SD) & $0.66\%$ & - \\ 
 \hline
 $1$ & $0.71\%$ & $7.5\%$ \\ 
 \hline
  $10$ & $0.96\%$ & $45.5\%$ \\ 
 \hline
  $100$ & $1.2\%$ & $81.9\%$ \\ 
 \hline
  $1000$ & $1.26\%$ & $91\%$ \\ 
 \hline
  $10000$ & $1.27\%$ & $91\%$ \\ 
 \hline
\end{tabular}
\end{table}

We quantify the overall threshold improvements for different system bias values in Table \ref{tab:biasCNOTthreshgains}. Note how the gain does also saturate for this case as it happened for the HBD model. However, it saturates for higher biases. This makes sense since the residual bias for the CNOT gates does also present such similar saturation tendency as seen in Figure \ref{fig:biaslossCNOT}.

\begin{table*}[t]
 \centering
\caption{Qubit footprints and relative decrease in percentage with respect to the symmetric SD circuit-level noise model for each bias value considered at $p=0.003$. The footprints are computed by selecting the first odd distance value below the target logical error probability in Supplementary Figure S1 and using $d^2+(d^2-1)$ as the number of qubits (data and check qubits) required for such distance. }
\label{tab:foot0_003}
\begin{tabular}{|c|c|c|c|c|c|c| }
 \hline
 $\eta$ & Megaquop footprint & Decrease ($\approx$) & Gigaquop footprint & Decrease ($\approx$) & Teraquop footprint & Decrease ($\approx$) \\ 
 \hline\hline
$1/2$ (SD) & $1681$ & - &  $4417$ & - & $8977$ & -\\ 
 \hline
  $10$ & $721$ & $57\%$ & $1921$ & $57\%$ & $3361$ & $63\%$\\ 
 \hline
  $100$ & $449$ & $74\%$ & $1249$ & $72\%$ & $2449$ & $73\%$\\ 
 \hline
  $1000$ & $449$ & $74\%$ & $1249$ & $72\%$ & $2177$ & $76\%$\\ 
 \hline
  $10000$ & $449$ & $74\%$ & $1249$ &  $72\%$ & $2177$ & $76\%$\\ 
 \hline
\end{tabular}
\end{table*}

\begin{table*}[t]
 \centering
\caption{Qubit footprints and relative decrease in percentage with respect to the symmetric SD circuit-level noise model for each bias value considered at $p=0.001$. The footprints are computed by selecting the first odd distance value below the target logical error probability in Supplementary Figure S2 and using $d^2+(d^2-1)$ as the number of qubits (data and check qubits) required for such distance. }
\label{tab:foot0_001}
\begin{tabular}{|c|c|c|c|c|c|c| }
 \hline
 $\eta$ & Megaquop footprint & Decrease ($\approx$) & Gigaquop footprint & Decrease ($\approx$) & Teraquop footprint & Decrease ($\approx$) \\ 
 \hline\hline
$1/2$ (SD) & $241$ & - &  $721$ & - & $1249$ & -\\ 
 \hline
  $10$ & $161$ & $34\%$ & $449$ & $38\%$ & $881$ & $30\%$\\ 
 \hline
  $100$ & $161$ & $34\%$ & $449$ & $38\%$ & $721$ & $43\%$\\ 
 \hline
  $1000$ & $161$ & $34\%$ & $337$ & $54\%$ & $721$ & $43\%$\\ 
 \hline
  $10000$ & $161$ & $34\%$ & $337$ &  $54\%$ & $721$ & $43\%$\\ 
 \hline
\end{tabular}
\end{table*}

\subsubsection{Footprints}
We now discuss the required qubit footprints to reach a certain regime of error free quantum operations as a function of the bias (see Appendix \ref{app:footprints} for a description of QEC footprints).

We numerically computed the logical error rates per round, $p_L$, as a function of the code distance for $d\in\{5,7,9,11,13,15\}$ and used the values to project further. Simulations were done for the same circuit-level noise models as in the threshold section (considering residual bias in CNOTs). We did so for two experimentally relevant physical error rates: $p=0.003$ representing state-of-the-art CZ gate error rates reported for the Google Quantum AI Willow processor \cite{willow}; and $p=0.001$ representing the usual physical error rate for practical QEC, i.e. the $99.9\%$ gate fidelity \cite{teraHoneycomb}. The results are shown in Supplementary Figures S1 and S2, respectively, which convincingly illustrate that the required distances to reach certain logical error probabilities are significantly reduced when a biased system is considered. Furthermore, it can be observed that the reduction in logical error rate provided by the bias shows a similar saturation effect as for the code threshold for $\eta\geq 100$. Table \ref{tab:foot0_003} quantifies the qubit footprints required to reach the Mega-, Giga-, and Teraquop regimes at a physical error rate of $p=0.003$. The reduction in qubit footprint ranges from around $60\%$ for a moderate bias of $\eta = 10$ to around $75\%$ for biases exceeding the value of a hundred compared to an SD circuit-level noise model. Furthermore, we quantify the same data for the $p=0.001$ physical error rate case in Table \ref{tab:foot0_001}. Inspecting the table, we can observe that for this case, the decrease in qubit footprint ranges from $33\%$ to around $50\%$.

The main conclusion that can be taken out of these results is that noise bias towards dephasing errors can still be leveraged for two-level qubits for which bias-preserving CNOT gates cannot be constructed, resulting in a significant reduction in the required number of qubits to achieve certain fault-tolerant quantum computational regimes of interest. It is noteworthy to state that the footprint reductions discussed here come ``for free'' for those qubit platforms if the CZ gates are enabled natively in a bias-preserving manner.

\section{Discussion}
In this article, we discussed how to leverage the bias towards dephasing noise that certain two-level qubit platforms exhibit. Our results indicate that by leveraging the system bias towards dephasing one can reach practical surface code based QEC with a significantly lower amount of resources than for systems whose noise is symmetric. First, the obtained thresholds for the rotated XZZX surface code with $\eta \geq 100$ stand above $1.2\%$. This implies that technologies exhibiting such amount of bias such as silicon spin qubits or ion traps, for example, could operate well below threshold right now \cite{RiverlaneQEC}. Importantly, it is common for the CNOTs to be compiled by means of a CZ gate sandwiched by Hadamard gates on the target qubit in many technologies such as silicon spin qubits or neutral atoms, among others \cite{Cai2019siliconsurfacecode,biaspreservingCNOTsNeutral}. We present the results of the XZZX code with extraction circuits following such a compilation in the Supplementary Material. As it can be seen there, the conclusions are similar.

The methods presented provide a clear improvement to qubit technologies exhibiting bias towards dephasing. The next milestone towards the goal of fault-tolerant quantum computing that would actually make the field go beyond the NISQ-era stands in the construction of the Megaquop machine \cite{PreskillMegaquop}. Currently, many technologies are competing to reach such milestone as fast as possible and, in an era where each physical qubit counts, reducing the required footprint up to a $75\%$ may give the edge to technologies chasing the predominant superconducting qubits. Also, this is especially relevant since, as we have studied in this article, qubit savings are much more significant for the physical error rates around $0.3\%$ that are common at this point in time.

Furthermore, it is also important to comment not only on the space (footprint) savings but also on the time savings that are obtained. In this sense, the time cost to perform fault-tolerant operations essentially defines the logical clock rate of a fault-tolerant quantum computer, and it is related to the number of syndrome extraction rounds \cite{Litinski2019gameofsurfacecodes}. Thus, our results indicate that a reduction in the required code distance also results in a higher logical clock speed (note that for fully understanding this, stability experiments would be required too \cite{Gidney2022stability}). The overall space-time cost, in terms of (physical qubits)$\cdot$(code cycles), of a computation depends on how the logical operations are enabled, but the leading order has a cubic dependence on the code distance \cite{Litinski2019gameofsurfacecodes}. Therefore, we can determine the space-time cost savings between equivalent protocols (i.e. the logical operations are done similarly, indicating that the constants of the cost are the same) whenever biased noise is leveraged. We summarize these space-time cost savings in Tables \ref{tab:spacetime0_003} and \ref{tab:spacetime0_001}. The overall space-time cost reductions are relevant and undoubtedly convenient to make fault-tolerant quantum machines in a significantly more efficient manner.

One of the key points of the obtained results is that the improvement in the QEC performance is obtained with no additional overhead. We have seen that the key element to leverage bias towards dephasing for two-level qubit technologies for which a bias-preserving CNOT cannot be constructed is to get tailored codes that use CZ gates to extract the syndrome and to natively enable those entangling gates to preserve the bias. Implementing native bias-preserving CZ gates is rather straightforward for many technologies, as discussed in section \ref{sec:BPgates}. Additionally, we have observed how the remainder bias present in the CNOT gates required for syndrome extraction boosts the performance of the tailored code even further. It is noteworthy to state that our analysis considers a generic Hamiltonian and error model to enable the CNOT gate. However, the residual bias of CNOT gates with other pulses and error models may be larger and, thus, the tailored code is expected to perform even better. Hence, even if a bias-preserving CNOT gate cannot be done for two-level qubits, trying to maximize the residual bias is an important future line of research, as it will make the QEC gains even bigger.

Furthermore, note that the considered HBD circuit-level noise models, both with depolarizing and with residual bias CNOT gates, consider that each element of the circuit fails at the same rate. In reality, the errors will be non-identically distributed, i.e. each element will present a different error rate \cite{toninid,willow,teraHoneycomb,rotvsunrot,spinhex}. This means that some operations will be more fault-prone than others. For example, it is very typical that single-qubit gates, e.g. Hadamard gates, are an order of magnitude less prone to experience errors than entangling gates such as CNOT or CZ gates \cite{teraHoneycomb,rotvsunrot,spinhex}. This bit is important for our discussion since the Hadamard gates used in the syndrome extraction circuits are the only gates exhibiting no residual bias, i.e. they introduce depolarizing noise. As for real systems those elements will introduce errors with orders of magnitude less frequency than the entangling gates, it will make overall noise in the system more predominantly biased. Therefore, we expect certain performance increases whenever such more fine-grained noise models with bias are considered. This becomes especially relevant for the compilation just using CZ gates that involves two extra layers of Hadamard gates per round on the data qubits (see Supplementary Material).

\begin{table}[t]
 \centering
\caption{Overall space-time cost decrease as a result of bias tailoring at $p=0.003$. The decreases are computed by selecting the first odd distance value below the target logical error probability in Supplementary Figure S2 and using the leading order space-time cost $d^3$ \cite{Litinski2019gameofsurfacecodes} to compute the decrease when compared to the standard depolarizing case.}
\label{tab:spacetime0_003}
\begin{tabular}{|c|c|c|c| }
 \hline
 $\eta$ & Megaquop &  Gigaquop  & Teraquop  \\ 
 \hline\hline
  $10$ & $72\%$ & $72\%$ & $77\%$\\ 
 \hline
  $100$ & $86\%$ & $85\%$ & $86\%$\\ 
 \hline
  $1000$ & $86\%$  & $85\%$ & $88\%$\\ 
 \hline
  $10000$ & $86\%$ & $85\%$ & $88\%$\\ 
 \hline
\end{tabular}
\end{table}
\begin{table}[t]
 \centering
\caption{Overall space-time cost decrease as a result of bias tailoring at $p=0.001$. The decreases are computed by selecting the first odd distance value below the target logical error probability in Supplementary Figure S2 and using the leading order space-time cost $d^3$ \cite{Litinski2019gameofsurfacecodes} to compute the decrease when compared to the standard depolarizing case.}
\label{tab:spacetime0_001}
\begin{tabular}{|c|c|c|c| }
 \hline
 $\eta$ & Megaquop &  Gigaquop  & Teraquop  \\ 
 \hline\hline
  $10$ & $45\%$ & $51\%$ & $41\%$\\ 
 \hline
  $100$ & $45\%$ & $51\%$ & $56\%$\\ 
 \hline
  $1000$ & $45\%$  & $68\%$ & $56\%$\\ 
 \hline
  $10000$ & $45\%$ & $68\%$ & $56\%$\\ 
 \hline
\end{tabular}
\end{table}

It is also important to discuss the way in which we have modeled state preparation and measurement (SPAM) errors. Note how in our HBD models, state preparation errors imply flipping to the orthogonal state and measurement errors flipping to the contrary measurement, each of them with probability $p$. This contrasts with some other circuit-level noise models including bias present in the literature \cite{beliefmatching,biasedCLNog,catXZZX}. Those models consider that the probability of preparing or measuring a state in the $Z$ basis is $\eta$ times smaller than for doing so in the $X$ basis. This implies that those operations on the $Z$ basis would fail orders of magnitude less frequently than the other elements in the system. This contrasts with the way in which bias is included for the gates in our model, i.e. the fidelity remains the same, but the frequency in which each of the Pauli errors occurs changes. Moreover, the physics behind SPAM operations is different to the ones regarding quantum gates, indicating that adding such bias is not as clear as for the other elements we have discussed in this article \cite{woottonSpin}. Studying how to realistically model bias in SPAM for different qubit technologies is considered as a future work.

Another important consequence of the distance reduction achieved by the tailoring to leverage bias towards dephasing relates with the decoder. Specifically, decoders must be fast enough in order to be able to implement non-Clifford operations in a fault-tolerant manner \cite{decoders}. In the main text we have used the Sparse Blossom implementation of the MWPM decoder for aiming to correct the errors \cite{sparseBlossom}. This implementation can decode distance-$17$ surface code syndromes in less than a microsecond per round at the target physical error rate of $0.1\%$, which is enough to avoid the backlog problem \cite{sparseBlossom}. As can be seen in Table \ref{tab:foot0_001}, the required footprint for all relevant regimes when the bias exceeds $100$ does not involve a distance beyond $17$, indicating that the used decoder should be able to operate below the microsecond per round. The microsecond per round target is not necessary for all qubit technologies since neutral atom or ion trap systesms require much more relaxed latencies \cite{AmbiguityClusteringOSD}. However, technologies with fast gates such as spin qubits will have stringent decoding latencies of around $1 \mu s$. Moreover, more precise decoding algorithms could also be used in order to reduce the footprints even more. For example, using belief propagation as a predecoder has shown to be beneficial for biased noise models \cite{beliefmatching}, so using beliefmatching or the almost-linear time belief propagation plus ordered Tanner forest (BP+OTF) decoder may improve performance and reductions even further \cite{beliefmatching,bpotf_2024}.

Regarding code construction, the present work also provides insights and future paths to explore. First, the present manuscript has been centered around the rotated XZZX variant of the surface code \cite{xzzx,spinhex}, mainly due to it being one of the strongest proposals to leverage bias and its planar nature, making it easier to implement experimentally in many platforms. However, there exist many bias tailored code proposals in the literature considering code capacity noise \cite{xzzx,tailoredXZZX,biasQLDPC,tuckettBiasSC,tuckettxy,beliefmatching,ZZZY,biasColorCodes}. The impossibility of constructing a bias-preserving CNOT gate in two-level qubits has made the community think that bias cannot be leveraged for such platforms that naturally exhibit such a feature. We have refuted such a statement in the present article noting that if the code involves CZ gates, constructed in a bias-preserving manner, substantial improvement in error correction performance can be obtained. This indicates that bias tailored codes based on using CY gates such as the XY or the ZZZY surface codes may not be good candidates to exploit bias in two-level qubits \cite{beliefmatching,ZZZY}. The main reason for that is that those proposals rely on using $Y$ stabilisers instead of $Z $ ones and, as we discussed in section \ref{sec:BPgates}, the CY gates used to perform said measurements cannot be enabled in a bias-preserving manner, similar to CNOT gates for $X$ stabiliser measurements. Hence, the syndrome extraction circuits for those codes will mainly consist of gates not preserving bias and, similar to what happens for CZ gates constructed in a non-bias-preserving way, will not be able to leverage bias at the circuit level. Anyway, there are many other code tailoring proposals involving $X$ and $Z$ stabilisers such as the domain wall color codes \cite{biasColorCodes}, XZZX toric codes or even general bias tailored qLDPC codes \cite{biasQLDPC}. Furthermore, qLDPC codes have shown the ability to reduce the required qubit footprints up to a factor of ten for circuit-level noise models \cite{BBcodes} so tailoring those codes and circuits with our findings may result in further qubit savings for qubit technologies with noise biased towards dephasing, making those closer to fault-tolerance. Note also that this reduction in footprint and required code distance would also simplify the decoding problem for such codes.

To sum up, in this article we have discussed how to leverage bias for QEC at the circuit level even when bias-preserving CNOT gates cannot be constructed. In a time in which qubits are expensive resources, reducing the required physical qubit per logical qubit rates is critical in order to make the first fault-tolerant realizations of quantum algorithms a reality. Here we recover the case that noise bias towards dephasing can make technologies that present such characteristic to have an edge in such a quest. Importantly, this physical qubit footprint reduction comes for free. We hope that bias tailoring for two-level qubits flourishes again as an efficient path towards fault-tolerance as it did during the last five years when QEC theorists mostly considered code capacity models. We hope that following this path Megaquop machines consisting of $100$ logical qubits can be constructed with physical qubit counts in the orders of thousands rather than several tens of thousands, making such first milestone and further ones closer to current times.

\section{Code availability}
The code supporting the findings
of this study can be found on GitHub: \url{https://github.com/jetxezarreta/qec-two-level-qubits-circuit-noise-bias}

\section{Acknowledgements}
We would like to thank Antonio de Marti i Olius for many useful discussions and recommendations. Furthermore, we also thank other members of the Cavendish Laboratory, the Hitachi-Cambridge Laboratory, and the Quantum Information Group at Tecnun for their support and many useful discussions.

This work was supported by the Spanish Ministry of Economy and Competitiveness through the MADDIE project (Grant No. PID2022-137099NBC44), by the Spanish Ministry of Science and Innovation through the project ``Few-qubit quantum hardware, algorithms and codes, on photonic and solidstate systems'' (PLEC2021-008251); by the Diputación Foral de Gipuzkoa through the ``Biased quantum error mitigation and applications of quantum computing to gene regulatory networks'' project (2024-QUAN-000020); and by the Ministry for Digital Transformation and of Civil Service of the Spanish Government through the QUANTUM ENIA project call - QUANTUM SPAIN project, and by the European Union through the Recovery, Transformation and Resilience Plan - NextGenerationEU within the framework of the Digital Spain 2026 Agenda. 

J.E.M. is funded by the Spanish Ministry of Science, Innovation and Universities through a Jos\'e Castillejo mobility grant for his stay at the Cavendish Laboratory of the University of Cambridge.

\appendix
\section{Bias-preserving gates}\label{app:biasP}
For code capacity noise models, the elements in the syndrome extraction circuits, i.e. gates and SPAM; are considered to be ideal and noiseless \cite{decoders}. However, each of those are actually imperfect and introduce errors in reality. This results in the so-called \textit{circuit-level noise} error model, considered to be realistic enough to represent what happens in actual hardware. As discussed before, stabiliser measurements are split into sequences of two-qubit gates that cannot be done simultaneously and, thus, an extraction schedule should be defined \cite{scheduleZS}. Many extraction schedules are valid, but generally a good schedule should aim to parallelize as many stabiliser operations as possible while trying to maximize the \textit{circuit-level distance} \cite{BBcodes}. The circuit-level distance refers to the minimum number of circuit errors that result in an undetectable logical error. This value is upper bounded by the code distance, and depends on the actual extraction schedule proposed.

Biased noise has shown the possibility of considerably improving the performance of QECCs when those are tailored to operate over models with an asymmetric noise structure \cite{xzzx,tailoredXZZX,biasQLDPC,tuckettBiasSC,tuckettxy,biasColorCodes}. This, however, has been mainly done from the perspective of code capacity noise models. The extension of this noise asymmetry to the circuit-level noise picture is far from trivial, since the noisy circuit operations should in principle maintain the noise bias, $\eta = p_z/(p_x+p_y)$, for leveraging such bias tailoring. 

\begin{definition}[Bias-preserving gate]\label{def:biaspreserving}
\textit{A bias-preserving gate is a noisy gate, $\tilde{\mathcal{U}}$, that can be modeled as the perfect gate, $\mathcal{U}$, followed by a noise channel maintaining the noise bias, $\Lambda_P(\eta$), present in the system, i.e. 
\begin{equation}
    \tilde{\mathcal{U}} = \Lambda_P(\eta)\circ\mathcal{U}.
\end{equation}}
\end{definition}

Note that the subscript $P$ in the noise channel refers to the fact that we are considering Pauli noise \footnote{This is sufficient since the noise can be transformed into stochastic Pauli noise by means of Pauli twirling \cite{decoders,approximatingdecoherence}. In fact, Pauli noise is the one generally considered for theoretical QEC analysis \cite{stim}}. It is important to comment that there are other conceptions of bias-preserving gates, related with how Pauli operators propagate through those gates \cite{biaspother1,biaspother2}. While such conception might be of interest for quantum error mitigation related studies, in the context of QEC, a bias-preserving gate refers to the one in Definition \ref{def:biaspreserving}. Importantly, it is known that not all gates can be constructed in a bias-preserving manner, i.e. $\tilde{\mathcal{U}} \neq \Lambda_P(\eta)\circ\mathcal{U}$.

\section{Numerical simulations of error channels}\label{app:lindblad}
In section \ref{sec:BPgates}, we presented a numerical study of the error channels resulting from the application of certain gates in order to determine the remainder bias after applying such gates. For doing so, we performed master equation simulations of the gates in question using the QuTiP software package \cite{qutip}. Specifically, we numerically simulate the gates in question defined by a Hamiltonian, $H$, under Markovian noise. This results in solving the following Lindblad equation for the time at which the gate of interest is realized,
\begin{align}\label{eq:master}
    \dot{\rho} = -\frac{i}{\hbar} [H,\rho] + \sum_i \lambda_i\mathcal{L}_i\{\rho\},
\end{align}
where $\rho$ is the density matrix, the dissipators $\mathcal{L}_i$ are given by the Pauli matrices, $\{X,Y,Z\}$ for single qubit gates and $\{I,X,Y,Z\}^{\otimes 2}/I^{\otimes 2}$ for two-qubit gates; and $\lambda_i$ are their coupling rates. The error channels for each gate are then determined by means of the so-called Pauli transfer matrix, defined as \cite{biaspreservingCNOTs}
\begin{equation}
    R_{ij} = \frac{1}{2^n} \mathrm{Tr}{\left[P_i \Lambda(P_j) \right]},
\end{equation}
where $n$ is the number of qubits, $\Lambda$ represents the noise map and $P_i,P_j\in\{I,X,Y,Z\}^{\otimes n}$. This way, the Pauli transfer matrix describes the fidelity that each of the elements of the density matrix shows after experiencing the noise channel. Thus, in order to extract the noise Pauli transfer matrix, we simulate equation \eqref{eq:master} to get the $R_{noisy}^{gate}$ transfer matrix and its associated ideal evolution ,$R_{ideal}^{gate}$, by means of the von Neumann equation. The noise transfer matrix is then obtained as $R_{noise}=R_{noisy}^{gate}(R_{ideal}^{gate})^{-1}$, i.e. getting just the part referring to the noise in the evolution \cite{biaspreservingCNOTs}. Finally, since the diagonal elements of Pauli transfer matrix are related to fidelities, we use a symplectic Walsh-Hadamard transform to convert those into the error probabilities in an operator-sum representation \cite{flammiaHadTransf,PEC}.

For the simulations using the Hamiltonian in equation \eqref{eq:CNOTH}, we set $V=1$ and the coupling rates of the dissipators of the Lindblad equation so that the combined rates of the $\{ZI,IZ,ZZ\}$ elements are $\eta_{sys}$ times higher than the ones of the rest of the non-trivial Pauli matrices. Furthermore, we set $\lambda_{ZI} + \lambda_{IZ} + \lambda_{ZZ}= 0.002$, that results in an overall gate fidelity of around $99.7\%$, i.e. an experimentally reasonable value.

\section{Native CZ gates for several qubit tehcnologies}\label{app:native}

In this section, we review several native implementations of CZ gates in different physical platforms and analyze whether they preserve noise bias. First, it is known that a CZ gate can be implemented conjugating the control of a CNOT gates with Hadamard gates. Nevertheless, this realization of a CZ gate will not preserve the noise bias since CNOT and Hadamard gates do not commute with the dephasing noise. Hence, CZ gate compilations using native CNOT gates enabled by means of $ZX$ interactions will not preserve bias, as explained in the main text.

Spin qubits are encoded in the spin state of an electron bounded to a quantum dot (QD) generated on a semiconductor and present a high bias towards $Z$ errors \cite{preskillBiased,woottonSpin,takeda2022quantum,Noiri2022,DiraqSpinQ}. In these devices, the electron's spin state ($\vec{s}_i$) interacts with an external magnetic field $\vec{B}(t)$ through a Zeeman Hamiltonian $\text{H}_{Zeeman}=\vec{B}(t)\cdot\vec{s}_i$ and with other electron spin qubit $\vec{s}_j$ through the Heisenberg exchange interaction $\text{H}_{exc}=J_\varepsilon(t) \vec{s}_i\cdot \vec{s}_j$, whose coupling strength depends on the voltage detuning $\varepsilon(t)$ \cite{RevModPhys.95.025003}. The control is done by means of electron spin resonance (ESR) using oscillating magnetic fields \cite{ESR}  or by means of electron dipole spin resonator (EDSR) using micromagnets \cite{EDRS1,EDSR2}. The relationship between the coupling strength, $J_\varepsilon(t)$, and the Zeeman gradient, $\delta B_z = B_{z_1}- B_{z_2}$,  defines two regimes in which CZ gates can be enabled. In the strong exchange regime ($J_\varepsilon(t) \gg \delta B_z$), the exchange interaction generates the mixed singlet and neutral triplet states, i.e. it allows the transition $\ket{\uparrow \downarrow} $ to $ \ket{\downarrow \uparrow} $, which does not commute with the dephasing noise. CZs are usually compiled as sequences of native $\sqrt{\text{SWAP}}$ gates \cite{Cai2019siliconsurfacecode},  which as explained will not preserve bias. In the weak exchange limit ($J_\varepsilon(t) \ll \delta B_z$), however, these transitions in the antiparallel spin subspace can be treated as a perturbation, resulting in a dipole-dipole interaction which is diagonal in the $Z$-basis \cite{Cai2019siliconsurfacecode}. Letting the system evolve for a certain time results in a CZ gate up to single qubit $Z$ rotations \cite{PhysRevB.97.085421,PhysRevB.83.121403,Veldhorst2015, Cai2019siliconsurfacecode,dzurakBPCZ}. Thus, the resulting gate will be bias-preserving.

Trapped-ion qubits are encoded in the hyperfine energy levels of a trapped ion's ground state. Qubits are manipulated using external lasers which couple to the spin and motion states of the ion \cite{PhysRevA.62.042307}. The Mølmer–Sørensen (MS) two-qubit gate evolution operator is defined as $\hat{U}\left(\theta\right)= \text{exp}\left(-i \frac{\theta}{2} \hat{\sigma}_1^{\phi_{S}}\hat{\sigma}_2^{\phi_{S}}\right)$ with $\hat{\sigma}_i^{\phi_{S}}= X_i \cos \phi_S + Y_i\sin \phi_S$, where $\phi_S$ is the effective spin rotation angle. The MS gate does not preserve the bias since the Hamiltoinan rotates the state around the $X$ and $Y$ axis \cite{Manovitz2022}. However, for the Liebfried–Sørensen (LS) gate \cite{Leibfried2003-bf}, a variation of the MS gate, performs a rotation defined by the Hamiltonian $\text{H}_I = \hbar \chi Z_i Z_j$, where $\chi= \eta^2 \Omega^2/(N\delta)$, being $\eta$ the Lamb-Dicke parameter, $\Omega$ is the Rabi frequency of the oscillations between two internal states of the ion, $N$ the number of ions and $\delta$ is the detuning between the laser and ion transitions \cite{Milburn2000}. The evolution described by this operator is a CZ gate up to single qubit $Z$-rotations and, thus, preserves the bias.

In neutral or Rydberg atom devices, qubits are encoded in the hyperfine ground states of the atoms and are manipulated using laser beams. The electrons bound to an atom can be excited to a high-energy state, known as the Rydberg state. When an atom is excited to this state, it induces a shift in the energy levels of nearby atoms through dipole-dipole interactions, preventing multi-atom excitation. This phenomenon, known as Rydberg blockade, is a fundamental mechanism for implementing two-qubit gates \cite{PhysRevLett.85.2208, PhysRevLett.87.037901}. The controlled phase gate can be done by applying three different laser pulses which couple the state $\ket{1}$ with the Rydberg state $\ket{r}$: the first one performs a $\pi$ rotation on the control atom, then,  a $2 \pi$ rotation is applied to the target atom, and finally another $\pi$ rotation is applied to the control atom \cite{RevModPhys.82.2313}. The Rydberg CZ gate is bias-preserving after appropriate conversion of leakage errors to $Z$-type errors \cite{biaspreservingCNOTsNeutral}.

Nitrogen-vacancy (NV) centers in diamond store quantum information in an electron bound to the defect. In its ground state, the NV center hosts an electronic spin system with total spin S = 1, and spin states $( m_S = \{ -1, 0, 1\})$ \cite{Childress2013}. There are two ways to perform a two-qubit gate. The first one is performed by using the magnetic dipolar interaction between two NV centers. The damping operators of the dipolar interaction can be neglected since the dipolar coupling $ \nu_{dip}$ is much weaker than the energy gap between the qubit states, so the interaction Hamiltonian can be expressed as $\text{H}_{dip} \approx \frac12 h \nu_{dip} Z_A Z_B$ \cite{Dolde2013}. The evolution obtained is a CZ gate up to single qubit rotations over the $Z$-axis. The second way is by using the hyperfine interaction between a NV center and a nitrogen or carbonous nucleus. Usually, a $^{14}$N atom is used with the information encoded in the states $\ket{m_n=+1/2} \equiv \ket{0}$ and $\ket{m_n=-1/2} \equiv \ket{1}$. The interaction Hamiltonian is $\text{H}_I= \vec{S} \mathcal{A} \vec{I}$, where $\vec{S}$ and $\vec{I}$ are the electron spin and nuclear spin operators, respectively. $\mathcal{A}$ is the hyperfine interaction coupling tensor, where the perpendicular hyperfine interaction $\mathcal{A}_\bot = (A_x,A_y,0)$ can be neglected \cite{NV-hyper,PhysRevA.89.052317,PhysRevA.87.012301,Liang_2016}. Therefore, the controlled phase gate, which preserves the noise bias, can be implemented natively using both interactions. 

Superconducting qubits generally do not exhibit highly biased noise when used as two-level systems, with bias values around $\eta\approx 10$ \cite{approximatingdecoherence}. However, there are recent proposals based on two-level superconducting qubits that show high bias values \cite{oliver25}. Entangling gates can be enabled by means of the cross resonance (CR) gate \cite{cr1,cr2}. This gate results from a $ZX$ interaction Hamiltonian that results in a native CNOT gate as shown in the main text. As explained before, the CZ gate using this interaction will not be bias-preserving. Furthermore, a native CZ gate can also be enabled for transmon qubits by means of single frequency-tunable transmon couplers (STC) that enable $ZZ$ interactions between qubtis. As seen before, this interactions result in CZ gates up to single qubit $Z$ rotations indicating that they preserve bias \cite{stc1,stc2,oliver25}.

\section{The XZZX surface code}\label{app:xzzxcode}
\begin{figure}[!h]
\centering
\includegraphics[width=\columnwidth]{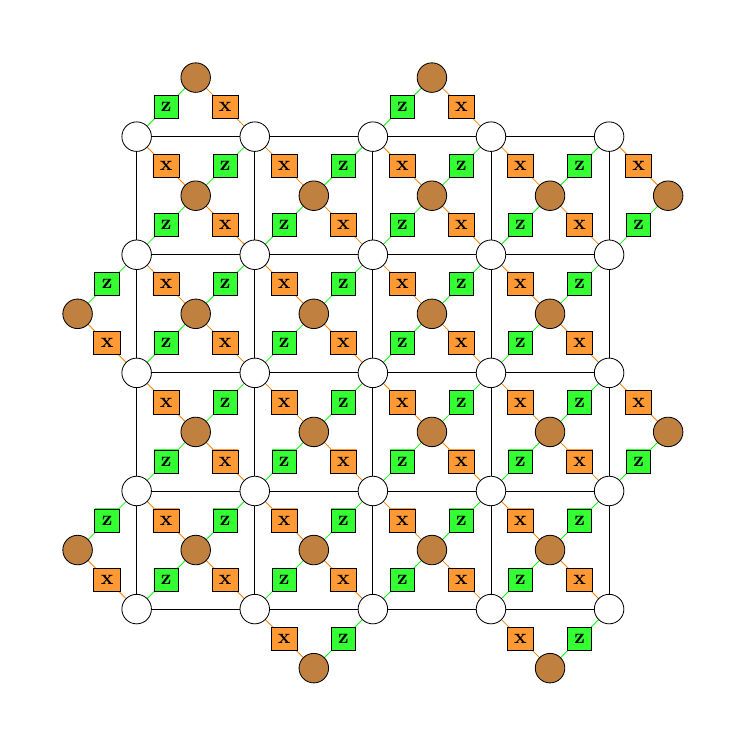}
\caption{Layout of the rotated XZZX surface code. White circles represent data qubits, while brown circles represent check qubits. Green and orange squares represent $Z$ and $X$ stabilisers, respectively. It can be seen that the check qubits measure XZZX patterns.}
\label{fig:xzzxLayout}
\end{figure}

The XZZX variant of the surface code can be obtained by means of a Clifford deformation of the CSS surface code \cite{decoders,cliffordDeformation}. Specifically, a Hadamard deformation is applied to every other qubit of the CSS surface code \cite{xzzx,cliffordDeformation,spinhex}. Up to the boundaries, the stabilisers of the resulting code are all of the $S_i = X_aZ_bZ_cX_d$ type, where the labels refer to the nearest neighbor qubits. Specifically, we will consider the XZZX code on a rotated layout with open boundary conditions, yielding parameters $[[d^2,1,d]]$ \cite{rotvsunrot}, where $d$ is the code distance. In contrast to the standard CSS surface code, every check qubit performs $Z$ and $X$-check measurements and, thus, it is a non-CSS code. The layout for this rotated XZZX code is shown in Figure \ref{fig:xzzxLayout}. While the code is locally equivalent to the CSS surface code (inheriting its properties), it has an extremely good performance when operating over biased noise, reaching a threshold of $50\%$ at the infinite bias limit for code capacity noise \cite{xzzx}. The main reasons for such performance boost comes from the fact that for the XZZX code there is a single logical operator purely consisting of $Z$ errors, significantly reducing the probability for it to happen \cite{decoders}; and because the defects created by $X$ and $Z$ errors have particular directions (and perpendicular among them) implying that weighting a decoder for leveraging the fact that the noise is biased is very beneficial for decoding \cite{catXZZX}. Furthermore, under biased noise, one can refine the weight of each stabilizer operator depending on the amount of noise. 
This refined notion of weight is called the {\em effective weight} of an operator \cite{tailoredXZZX}. For instance, under depolarizing noise, as all the errors $X$, $Y$, and $Z$ are equally likely, the effective weight of an operator is simply the number of non-trivial Pauli operators in that operator. 
 This coincides with the conventional definition of Hamming weight. 
However, in the case of infinite biased noise ($P_Z \gg P_X$), the effective weight of each operator is $\infty$ if it contains at least one $X$ or $Y$ Pauli, and otherwise, its effective weight is the number of $Z$ Pauli matrices in its decomposition. 
Therefore, the set of likely errors is reduced to those consisting of only $Z$ errors, also depending on their effective weights.
Under this scenario, as there exists only one logical operator in the rotated XZZX code consisting solely of $Z$ errors (main diagonal error), the probability of a logical error in such a code is significantly reduced. Under other bias rates, using a similar argument, one can see that the probability of likely logical errors to happen is reduced (in comparison to depolarizing noise), which can boost the performance of such codes.

\begin{figure}[!h]
\centering
\includegraphics[width=1.15\columnwidth]{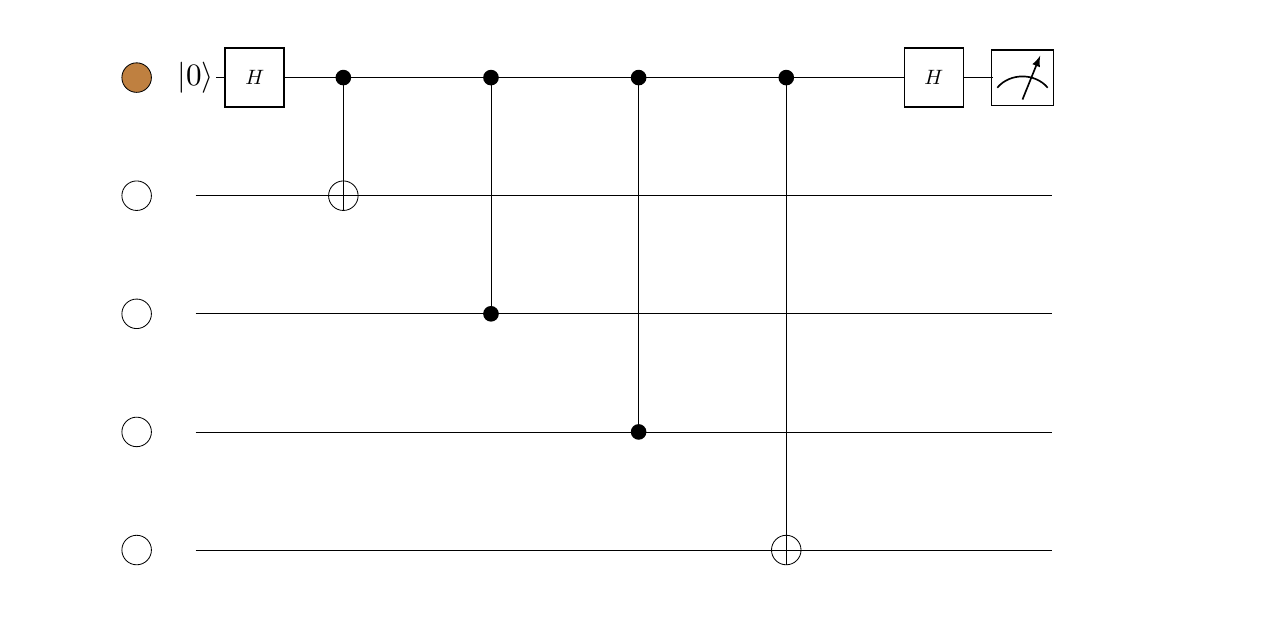}
\caption{An XZZX check measurement. The check qubit is colored in brown and the $4$ data qubits involved in such parity measurement are in white. The check qubit is prepared in the $\ket{0}$ and mapped to the $X$ basis by means of a Hadamard gate. Check measurement involves the same basis transformation to then be measured in the $Z$ basis. The CNOT and CZ gates represent the XZZX pattern.}
\label{fig:xzzxCheck}
\end{figure}

A fundamental step for a QEC protocol is the syndrome extraction circuit that is used in order to measure the stabilisers. The fist important decision is which actual gates will be used to measure the stabilisers. For this rotated XZZX code in consideration, we will use sequences of CNOT and CZ gates to measure the stabilisers by means of the check qubits, which will be prepared and measured in the $Z$-basis and converted to the $X$-basis by means of Hadamard gates. An XZZX stabiliser check measurement is depicted in Figure \ref{fig:xzzxCheck}. Furthermore, the operations involved in the stabiliser measurements of each check should be parallelized as much as possible to reduce the time steps in which the qubits are idling and, thus, subjected to more errors. This should be done in a way that the circuit-level distance is as high as possible, since certain errors at the circuit-level can reduce the distance that the code should show in theory, e.g. hook errors \cite{BBcodes,scheduleZS}.  Thus, it is necessary to define a syndrome extraction schedule. As explained in \cite{scheduleZS}, the order of the physical data qubits in the circuits of a check measurement, as in Figure \ref{fig:xzzxLayout}, should follow \textbf{S} or \textbf{Z} patterns since, otherwise, commutation of nearby stabilisers is lost. Thus, the possible schedules involve XZZX or ZXXZ patterns. For our circuits, we select an XZZX pattern specially since as pointed out by Darmawan et al. in \cite{catXZZX}, an $X$ error after the second entangling gate propagates the next two data qubits and aligns with the logical error strings. In \cite{catXZZX}, the authors discuss that since their CNOT gates are bias-preserving, such hook error is a low rate error and, thus, benign. In our circuits, CNOT gates are not bias-preserving, implying that in the ZXXZ extraction pattern, the hook event will not be rare. Thus, we use the XZZX pattern as the CZ gate can be done in a bias-preserving manner and such hook error is again a low rate event.

Furthermore, in the Supplementary Material we discuss the compilation of the XZZX surface code by means of CZ gates and Hadamard gates. For this case, the CNOT gates used for extracting the $X$ parities are enabled by means of a CZ gate and two Hadamard gates sandwiching the target qubit. For this case, the extra Hadamards are included in the data qubits. Furthermore, the scheduling used to extract the syndrome allows the cancelation of pairs of Hadamard gates in the bulk of the extraction circuits, leading to a sequence of CZ gates, a layer of Hadamard gates over all data qubits, two sequences of CZ gates, another layer of Hadamard gates over all data qubits and a last sequence of CZ gates. The resulting extraction circuit is the same as the one used by Google Quantum AI in \cite{googleSurf,willow}. The specific schedule can be found in the Supplementary Material of \cite{googleSurf}. The superconducting processor used by Google in those works does not present bias towards dephasing noise, but they compiled the XZZX code using only CZ gates because those are native to their technology. Note that the gate pattern is the one we also used for the XZZX surface codes compiled with CNOT gates. Finally, note that in this case all entangling gates will be bias-preserving, indicating that the harmful hook errors commented on before will be very low rate events for high bias values.

\section{The hybrid biased-depolarizing circuit-level noise model}\label{app:HBD}
Generic noise models that capture the most important characteristics of the noise in certain systems are invaluable for quantum error correction theorists to evaluate the methods they propose without having to dig into the specific details of a certain implementation. In this sense, the \textit{standard depolarizing} (SD) circuit-level noise model is the most studied model considering imperfect syndrome extraction circuits and it is extremely useful to determine if a certain QEC protocol can work whenever noise at the circuit-level is considered \cite{decoders,beliefmatching,biasedCLNog,fowlerSC,rotvsunrot,BBcodes,bpgdg,biasedCLNog,gidneyShor,woottonSpin,strikisSC,closedbranch,bpotf_2024,LSD,pymatching,sparseBlossom}. Biased circuit-level noise models have also been considered in the literature, however, those models are specific for qubit platforms for which bias-preserving single- and two-qubit gates can be constructed \cite{beliefmatching,catXZZX,catoverhead,catoverhead2,biasedCLNog}. From our discussion in Section \ref{sec:BPgates}, none of those models capture the specific characteristics of two-level qubits whose noise is biased towards dephasing. Thus, we will propose a noise model generic enough so that theoretical evaluations of biased tailored QEC over those platforms can be done while still capturing the essential properties of the noise in those.

The \textit{hybrid biased-depolarizing circuit-level noise model} is defined by the tuple of parameters physical error rate, $p$, and bias, $\eta$, as:
\begin{itemize}
    \item Hadamard gates are followed by single-qubit depolarizing errors at rate $p$, i.e. each $X,Y$ or $Z$ error occurs with probability $p/3$ \footnote{We only discuss Hadamard gates for the noise model since those are generally the ones appearing in QEC syndrome extraction circuits. In a more generic case, single qubit gates related to rotations involving $X$ or $Y$ basis will be followed by depolarizing noise, while rotations strictly over the $Z$ axis can be done in a bias-preserving manner.}.
    \item CNOT (or CY) gates are followed by two-qubit depolarizing errors at rate $p$, i.e. each error from the set $\{I,X,Y,Z\}^{\otimes 2}/I^{\otimes 2}$ occurs with probability $p/15$.
    \item CZ gates are followed by two-qubit biased Pauli errors with bias $\eta$ at rate $p$. Specifically, each pure dephasing error $\{ZI,IZ,ZZ\}$ occurs with probability $\frac{\eta p}{3(1+\eta)}$, while the rest occur with probability $\frac{p}{12(1+\eta)}$. By this encoding, the overall error rate of the gate is $p$ independent of the bias, and the bias just changes how each of the errors contributes. The bias quantifies how much more probable pure dephasing errors are compared to the other ones.
    \item Preparing a state in the $\ket{0}$ ($\ket{+}$) state instead results in preparing the orthogonal state $\ket{1}$ ($\ket{-}$) with probability $p$.
    \item Measurement results in the $X$ or $Z$ basis experience a flip reporting an incorrect outcome with probability $p$.
    \item Idling steps experience single-qubit biased Pauli errors with bias $\eta$ at rate $p$. Specifically, $Z$ errors occur with probability $\frac{\eta p}{1+\eta}$, while $X$ and $Y$ errors each occur with probability $\frac{p}{2(1+\eta)}$.
\end{itemize}

A slight difference of this model with previous biased circuit-level noise models is the way in which we define the probabilities of the biased and unbiased errors. In \cite{beliefmatching,catXZZX,catoverhead,catoverhead2,biasedCLNog}, the authors used a noise model in which the biased errors occur with probability $\frac{p}{4^{n}-1}$, while the rest occur at $\frac{p}{(4^{n}-1)\eta}$, with $n$ the number of qubits involved in the gate. Although this may be convenient for some cases, the overall gate error probability changes as a function of the bias,  $\eta$. In fact, a higher bias implies a lower gate error rate. On the contrary, in our model the overall gate error rate is the same for all bias values, with the change being in how the probabilities of each individual event are distributed \footnote{Note that the standard depolarizing case does not occur at $\eta=1/2$ since at such value, the two-qubit gate errors will occur with probability $p/9$ for the biased errors and with $p/18$ for the rest. Thus, the standard depolarizing case has to be defined on its own.}. We consider this to be a fairer comparison between depolarizing and biased cases since the probability that any element fails will be the same, independently of the bias.

Note that for this model we have ignored the possibility that some of the gates may exhibit some residual bias, case that is analyzed in section \ref{sec:tailoredCNOTs}. We do so because such residual bias will probably depend on the interaction used to form those gates, and the goal is to have some generic circuit-level noise model considering bias in two-level qubits. Since the depolarizing case is the most demanding one (any residual bias will only improve the tailoring), then this is useful to study the performance of QECCs agnostic to the way in which gates that do not preserve bias are constructed. The main feature is that the CZ gates are actually done in a bias-preserving manner, i.e. by means of a native interaction that exhibits such a feature.

\section{Quantum error correction footprints}\label{app:footprints}
While the threshold is an interesting metric to study since it basically provides the maximum physical error rate for which the QEC method is effective, the required physical qubit footprints to be able to implement those are actually what matters in practice. Footprints refer to the space overhead in terms of physical qubits per logical qubit in order to execute quantum circuits of certain size with a probability circa $1/e$ \cite{teraHoneycomb,teraBurst}. In this context, usually the \textit{Megaquop} (a million quantum operations), the \textit{Gigaquop} (a billion quantum operations) and the \textit{Teraquop} (a trillion quantum operations) regimes are considered as the interesting regimes for fault-tolerant quantum computing. Those relate to logical error rates per round of the order of $10^{-6}$, $10^{-9}$, and $10^{-12}$, respectively. In order to make a fair comparison in terms of numbers, we will consider that the required distance to reach a certain regime is the first one exactly below the target error rates presented before. However, this should be taken with a grain of salt since, as we comment in the main text, there are cases for which a smaller distance is very close to the actual error rate. 

\section{Numerical simulations}\label{app:numerical}
Extensive Monte Carlo simulations of the XZZX code have been performed in order to estimate the performances of the codes considering the circuit-level noise model described. We implemented the noisy extraction circuits in order to perform the sampling of the errors by using Stim \cite{stim}. Stim considers the check measurements upon a set of syndrome extractions altogether with a final measurement of the data qubits. In order to conduct the memory experiments of the XZZX rotated surface code, two different memory experiments are conducted depending on which initial state is aimed to be preserved: The horizontal (H) memory experiment in which data qubits are initialized in an interchanging pattern of $\ket{+}$ and $\ket{0}$ states starting from $\ket{+}$ for the top left data qubit and finally measured in their corresponding bases; and the vertical (V) memory experiment in which data qubits are initialized in an interchanging pattern of $\ket{0}$ and $\ket{+}$ states starting from $\ket{0}$ for the top left data qubit and finally measured in the corresponding basis. The horizontal and vertical memory names come from the direction of the strings that form logical operators in each of the cases. Note that the results of memory experiments are also applicable to understand logical operations performed by means of lattice surgery \cite{surgery,rotvsunrot}. For fully understanding these protocols, stability experiments would also be required \cite{Gidney2022stability}.

The operational figure of merit we use to evaluate the performance is the logical error probability, $p_L$, per extraction round. We ran $3d$ rounds of syndrome extraction to reduce time-boundary edge effects coming from the fact that the first and last extraction rounds are less noisy than the ones in the bulk \cite{rotvsunrot}. Once the circuits are run to collect the samples, we use the \textit{pymatching} implementation the minimum-weight perfect-matching (MWPM) decoder in order to aim recovery and determine if a logical error has occurred or not \cite{decoders,pymatching,sparseBlossom}. In order to have good enough statistical accuracy from the Monte Carlo simulation estimations, we began collecting samples with a ceiling of twenty million circuit shots and a hundred thousand logical errors. For the lowest logical error probability points, i.e. $d=13,15$ at $p=10^{-3}$, we increased the shot ceiling up to ten billion shots and five hundred thousand logical errors. We highlight regions showing $p_L$ values for which the conditional probabilities $P(p_L|k)$ are within a factor of $1000$ of the maximum likelihood estimation, $p_L = k/n$, assuming a binomial distribution, only for the numerically estimated points, some are too small to be observed \cite{rotvsunrot}.

\bibliography{bibliography}

\begin{thebibliography}{98}%
\makeatletter
\providecommand \@ifxundefined [1]{%
 \@ifx{#1\undefined}
}%
\providecommand \@ifnum [1]{%
 \ifnum #1\expandafter \@firstoftwo
 \else \expandafter \@secondoftwo
 \fi
}%
\providecommand \@ifx [1]{%
 \ifx #1\expandafter \@firstoftwo
 \else \expandafter \@secondoftwo
 \fi
}%
\providecommand \natexlab [1]{#1}%
\providecommand \enquote  [1]{``#1''}%
\providecommand \bibnamefont  [1]{#1}%
\providecommand \bibfnamefont [1]{#1}%
\providecommand \citenamefont [1]{#1}%
\providecommand \href@noop [0]{\@secondoftwo}%
\providecommand \href [0]{\begingroup \@sanitize@url \@href}%
\providecommand \@href[1]{\@@startlink{#1}\@@href}%
\providecommand \@@href[1]{\endgroup#1\@@endlink}%
\providecommand \@sanitize@url [0]{\catcode `\\12\catcode `\$12\catcode `\&12\catcode `\#12\catcode `\^12\catcode `\_12\catcode `\%12\relax}%
\providecommand \@@startlink[1]{}%
\providecommand \@@endlink[0]{}%
\providecommand \url  [0]{\begingroup\@sanitize@url \@url }%
\providecommand \@url [1]{\endgroup\@href {#1}{\urlprefix }}%
\providecommand \urlprefix  [0]{URL }%
\providecommand \Eprint [0]{\href }%
\providecommand \doibase [0]{https://doi.org/}%
\providecommand \selectlanguage [0]{\@gobble}%
\providecommand \bibinfo  [0]{\@secondoftwo}%
\providecommand \bibfield  [0]{\@secondoftwo}%
\providecommand \translation [1]{[#1]}%
\providecommand \BibitemOpen [0]{}%
\providecommand \bibitemStop [0]{}%
\providecommand \bibitemNoStop [0]{.\EOS\space}%
\providecommand \EOS [0]{\spacefactor3000\relax}%
\providecommand \BibitemShut  [1]{\csname bibitem#1\endcsname}%
\let\auto@bib@innerbib\@empty
\bibitem [{\citenamefont {Fowler}\ \emph {et~al.}(2012)\citenamefont {Fowler}, \citenamefont {Mariantoni}, \citenamefont {Martinis},\ and\ \citenamefont {Cleland}}]{fowlerSC}%
  \BibitemOpen
  \bibfield  {author} {\bibinfo {author} {\bibfnamefont {A.~G.}\ \bibnamefont {Fowler}}, \bibinfo {author} {\bibfnamefont {M.}~\bibnamefont {Mariantoni}}, \bibinfo {author} {\bibfnamefont {J.~M.}\ \bibnamefont {Martinis}},\ and\ \bibinfo {author} {\bibfnamefont {A.~N.}\ \bibnamefont {Cleland}},\ }\bibfield  {title} {\bibinfo {title} {Surface codes: Towards practical large-scale quantum computation},\ }\href {https://doi.org/10.1103/PhysRevA.86.032324} {\bibfield  {journal} {\bibinfo  {journal} {Phys. Rev. A}\ }\textbf {\bibinfo {volume} {86}},\ \bibinfo {pages} {032324} (\bibinfo {year} {2012})}\BibitemShut {NoStop}%
\bibitem [{\citenamefont {deMarti iOlius}\ \emph {et~al.}(2024)\citenamefont {deMarti iOlius}, \citenamefont {Fuentes}, \citenamefont {Or{\'{u}}s}, \citenamefont {Crespo},\ and\ \citenamefont {Etxezarreta~Martinez}}]{decoders}%
  \BibitemOpen
  \bibfield  {author} {\bibinfo {author} {\bibfnamefont {A.}~\bibnamefont {deMarti iOlius}}, \bibinfo {author} {\bibfnamefont {P.}~\bibnamefont {Fuentes}}, \bibinfo {author} {\bibfnamefont {R.}~\bibnamefont {Or{\'{u}}s}}, \bibinfo {author} {\bibfnamefont {P.~M.}\ \bibnamefont {Crespo}},\ and\ \bibinfo {author} {\bibfnamefont {J.}~\bibnamefont {Etxezarreta~Martinez}},\ }\bibfield  {title} {\bibinfo {title} {Decoding algorithms for surface codes},\ }\href {https://doi.org/10.22331/q-2024-10-10-1498} {\bibfield  {journal} {\bibinfo  {journal} {{Quantum}}\ }\textbf {\bibinfo {volume} {8}},\ \bibinfo {pages} {1498} (\bibinfo {year} {2024})}\BibitemShut {NoStop}%
\bibitem [{\citenamefont {Gill}\ \emph {et~al.}(2022)\citenamefont {Gill}, \citenamefont {Kumar}, \citenamefont {Singh}, \citenamefont {Singh}, \citenamefont {Kaur}, \citenamefont {Usman},\ and\ \citenamefont {Buyya}}]{qubitTechs}%
  \BibitemOpen
  \bibfield  {author} {\bibinfo {author} {\bibfnamefont {S.~S.}\ \bibnamefont {Gill}}, \bibinfo {author} {\bibfnamefont {A.}~\bibnamefont {Kumar}}, \bibinfo {author} {\bibfnamefont {H.}~\bibnamefont {Singh}}, \bibinfo {author} {\bibfnamefont {M.}~\bibnamefont {Singh}}, \bibinfo {author} {\bibfnamefont {K.}~\bibnamefont {Kaur}}, \bibinfo {author} {\bibfnamefont {M.}~\bibnamefont {Usman}},\ and\ \bibinfo {author} {\bibfnamefont {R.}~\bibnamefont {Buyya}},\ }\bibfield  {title} {\bibinfo {title} {Quantum computing: A taxonomy, systematic review and future directions},\ }\href {https://doi.org/https://doi.org/10.1002/spe.3039} {\bibfield  {journal} {\bibinfo  {journal} {Software: Practice and Experience}\ }\textbf {\bibinfo {volume} {52}},\ \bibinfo {pages} {66} (\bibinfo {year} {2022})},\ \Eprint {https://arxiv.org/abs/https://onlinelibrary.wiley.com/doi/pdf/10.1002/spe.3039} {https://onlinelibrary.wiley.com/doi/pdf/10.1002/spe.3039} \BibitemShut {NoStop}%
\bibitem [{\citenamefont {Bonilla~Ataides}\ \emph {et~al.}(2021)\citenamefont {Bonilla~Ataides}, \citenamefont {Tuckett}, \citenamefont {Bartlett}, \citenamefont {Flammia},\ and\ \citenamefont {Brown}}]{xzzx}%
  \BibitemOpen
  \bibfield  {author} {\bibinfo {author} {\bibfnamefont {J.~P.}\ \bibnamefont {Bonilla~Ataides}}, \bibinfo {author} {\bibfnamefont {D.~K.}\ \bibnamefont {Tuckett}}, \bibinfo {author} {\bibfnamefont {S.~D.}\ \bibnamefont {Bartlett}}, \bibinfo {author} {\bibfnamefont {S.~T.}\ \bibnamefont {Flammia}},\ and\ \bibinfo {author} {\bibfnamefont {B.~J.}\ \bibnamefont {Brown}},\ }\bibfield  {title} {\bibinfo {title} {The xzzx surface code},\ }\href {https://doi.org/10.1038/s41467-021-22274-1} {\bibfield  {journal} {\bibinfo  {journal} {Nature Communications}\ }\textbf {\bibinfo {volume} {12}},\ \bibinfo {pages} {2172} (\bibinfo {year} {2021})}\BibitemShut {NoStop}%
\bibitem [{\citenamefont {Roffe}\ \emph {et~al.}(2023)\citenamefont {Roffe}, \citenamefont {Cohen}, \citenamefont {Quintavalle}, \citenamefont {Chandra},\ and\ \citenamefont {Campbell}}]{biasQLDPC}%
  \BibitemOpen
  \bibfield  {author} {\bibinfo {author} {\bibfnamefont {J.}~\bibnamefont {Roffe}}, \bibinfo {author} {\bibfnamefont {L.~Z.}\ \bibnamefont {Cohen}}, \bibinfo {author} {\bibfnamefont {A.~O.}\ \bibnamefont {Quintavalle}}, \bibinfo {author} {\bibfnamefont {D.}~\bibnamefont {Chandra}},\ and\ \bibinfo {author} {\bibfnamefont {E.~T.}\ \bibnamefont {Campbell}},\ }\bibfield  {title} {\bibinfo {title} {Bias-tailored quantum {LDPC} codes},\ }\href {https://doi.org/10.22331/q-2023-05-15-1005} {\bibfield  {journal} {\bibinfo  {journal} {{Quantum}}\ }\textbf {\bibinfo {volume} {7}},\ \bibinfo {pages} {1005} (\bibinfo {year} {2023})}\BibitemShut {NoStop}%
\bibitem [{\citenamefont {{Etxezarreta Martinez}}\ \emph {et~al.}(2020)\citenamefont {{Etxezarreta Martinez}}, \citenamefont {Fuentes}, \citenamefont {Crespo},\ and\ \citenamefont {Garcia-Frias}}]{approximatingdecoherence}%
  \BibitemOpen
  \bibfield  {author} {\bibinfo {author} {\bibfnamefont {J.}~\bibnamefont {{Etxezarreta Martinez}}}, \bibinfo {author} {\bibfnamefont {P.}~\bibnamefont {Fuentes}}, \bibinfo {author} {\bibfnamefont {P.~M.}\ \bibnamefont {Crespo}},\ and\ \bibinfo {author} {\bibfnamefont {J.}~\bibnamefont {Garcia-Frias}},\ }\bibfield  {title} {\bibinfo {title} {Approximating decoherence processes for the design and simulation of quantum error correction codes on classical computers},\ }\href {https://doi.org/10.1109/ACCESS.2020.3025619} {\bibfield  {journal} {\bibinfo  {journal} {IEEE Access}\ }\textbf {\bibinfo {volume} {8}},\ \bibinfo {pages} {172623} (\bibinfo {year} {2020})}\BibitemShut {NoStop}%
\bibitem [{\citenamefont {Aliferis}\ and\ \citenamefont {Preskill}(2008)}]{preskillBiased}%
  \BibitemOpen
  \bibfield  {author} {\bibinfo {author} {\bibfnamefont {P.}~\bibnamefont {Aliferis}}\ and\ \bibinfo {author} {\bibfnamefont {J.}~\bibnamefont {Preskill}},\ }\bibfield  {title} {\bibinfo {title} {Fault-tolerant quantum computation against biased noise},\ }\href {https://doi.org/10.1103/PhysRevA.78.052331} {\bibfield  {journal} {\bibinfo  {journal} {Phys. Rev. A}\ }\textbf {\bibinfo {volume} {78}},\ \bibinfo {pages} {052331} (\bibinfo {year} {2008})}\BibitemShut {NoStop}%
\bibitem [{\citenamefont {Seis}\ \emph {et~al.}(2023)\citenamefont {Seis}, \citenamefont {Brown}, \citenamefont {S\o{}rensen},\ and\ \citenamefont {Goodwin}}]{iontrapbias1}%
  \BibitemOpen
  \bibfield  {author} {\bibinfo {author} {\bibfnamefont {Y.}~\bibnamefont {Seis}}, \bibinfo {author} {\bibfnamefont {B.~J.}\ \bibnamefont {Brown}}, \bibinfo {author} {\bibfnamefont {A.~S.}\ \bibnamefont {S\o{}rensen}},\ and\ \bibinfo {author} {\bibfnamefont {J.~F.}\ \bibnamefont {Goodwin}},\ }\bibfield  {title} {\bibinfo {title} {Improving trapped-ion-qubit memories via code-mediated error-channel balancing},\ }\href {https://doi.org/10.1103/PhysRevA.107.052417} {\bibfield  {journal} {\bibinfo  {journal} {Phys. Rev. A}\ }\textbf {\bibinfo {volume} {107}},\ \bibinfo {pages} {052417} (\bibinfo {year} {2023})}\BibitemShut {NoStop}%
\bibitem [{\citenamefont {Sepiol}\ \emph {et~al.}(2019)\citenamefont {Sepiol}, \citenamefont {Hughes}, \citenamefont {Tarlton}, \citenamefont {Nadlinger}, \citenamefont {Ballance}, \citenamefont {Ballance}, \citenamefont {Harty}, \citenamefont {Steane}, \citenamefont {Goodwin},\ and\ \citenamefont {Lucas}}]{iontrapbias2}%
  \BibitemOpen
  \bibfield  {author} {\bibinfo {author} {\bibfnamefont {M.~A.}\ \bibnamefont {Sepiol}}, \bibinfo {author} {\bibfnamefont {A.~C.}\ \bibnamefont {Hughes}}, \bibinfo {author} {\bibfnamefont {J.~E.}\ \bibnamefont {Tarlton}}, \bibinfo {author} {\bibfnamefont {D.~P.}\ \bibnamefont {Nadlinger}}, \bibinfo {author} {\bibfnamefont {T.~G.}\ \bibnamefont {Ballance}}, \bibinfo {author} {\bibfnamefont {C.~J.}\ \bibnamefont {Ballance}}, \bibinfo {author} {\bibfnamefont {T.~P.}\ \bibnamefont {Harty}}, \bibinfo {author} {\bibfnamefont {A.~M.}\ \bibnamefont {Steane}}, \bibinfo {author} {\bibfnamefont {J.~F.}\ \bibnamefont {Goodwin}},\ and\ \bibinfo {author} {\bibfnamefont {D.~M.}\ \bibnamefont {Lucas}},\ }\bibfield  {title} {\bibinfo {title} {Probing qubit memory errors at the part-per-million level},\ }\href {https://doi.org/10.1103/PhysRevLett.123.110503} {\bibfield  {journal} {\bibinfo  {journal} {Phys. Rev. Lett.}\ }\textbf {\bibinfo {volume} {123}},\ \bibinfo {pages} {110503} (\bibinfo {year} {2019})}\BibitemShut
  {NoStop}%
\bibitem [{\citenamefont {Tan}\ \emph {et~al.}(2015)\citenamefont {Tan}, \citenamefont {Gaebler}, \citenamefont {Lin}, \citenamefont {Wan}, \citenamefont {Bowler}, \citenamefont {Leibfried},\ and\ \citenamefont {Wineland}}]{iontrapbias3}%
  \BibitemOpen
  \bibfield  {author} {\bibinfo {author} {\bibfnamefont {T.~R.}\ \bibnamefont {Tan}}, \bibinfo {author} {\bibfnamefont {J.~P.}\ \bibnamefont {Gaebler}}, \bibinfo {author} {\bibfnamefont {Y.}~\bibnamefont {Lin}}, \bibinfo {author} {\bibfnamefont {Y.}~\bibnamefont {Wan}}, \bibinfo {author} {\bibfnamefont {R.}~\bibnamefont {Bowler}}, \bibinfo {author} {\bibfnamefont {D.}~\bibnamefont {Leibfried}},\ and\ \bibinfo {author} {\bibfnamefont {D.~J.}\ \bibnamefont {Wineland}},\ }\bibfield  {title} {\bibinfo {title} {Multi-element logic gates for trapped-ion qubits},\ }\href {https://doi.org/10.1038/nature16186} {\bibfield  {journal} {\bibinfo  {journal} {Nature}\ }\textbf {\bibinfo {volume} {528}},\ \bibinfo {pages} {380} (\bibinfo {year} {2015})}\BibitemShut {NoStop}%
\bibitem [{\citenamefont {Het\'enyi}\ and\ \citenamefont {Wootton}(2024)}]{woottonSpin}%
  \BibitemOpen
  \bibfield  {author} {\bibinfo {author} {\bibfnamefont {B.}~\bibnamefont {Het\'enyi}}\ and\ \bibinfo {author} {\bibfnamefont {J.~R.}\ \bibnamefont {Wootton}},\ }\bibfield  {title} {\bibinfo {title} {Tailoring quantum error correction to spin qubits},\ }\href {https://doi.org/10.1103/PhysRevA.109.032433} {\bibfield  {journal} {\bibinfo  {journal} {Phys. Rev. A}\ }\textbf {\bibinfo {volume} {109}},\ \bibinfo {pages} {032433} (\bibinfo {year} {2024})}\BibitemShut {NoStop}%
\bibitem [{\citenamefont {Takeda}\ \emph {et~al.}(2022)\citenamefont {Takeda}, \citenamefont {Noiri}, \citenamefont {Nakajima}, \citenamefont {Kobayashi},\ and\ \citenamefont {Tarucha}}]{takeda2022quantum}%
  \BibitemOpen
  \bibfield  {author} {\bibinfo {author} {\bibfnamefont {K.}~\bibnamefont {Takeda}}, \bibinfo {author} {\bibfnamefont {A.}~\bibnamefont {Noiri}}, \bibinfo {author} {\bibfnamefont {T.}~\bibnamefont {Nakajima}}, \bibinfo {author} {\bibfnamefont {T.}~\bibnamefont {Kobayashi}},\ and\ \bibinfo {author} {\bibfnamefont {S.}~\bibnamefont {Tarucha}},\ }\bibfield  {title} {\bibinfo {title} {Quantum error correction with silicon spin qubits},\ }\href@noop {} {\bibfield  {journal} {\bibinfo  {journal} {Nature}\ }\textbf {\bibinfo {volume} {608}},\ \bibinfo {pages} {682} (\bibinfo {year} {2022})}\BibitemShut {NoStop}%
\bibitem [{\citenamefont {Noiri}\ \emph {et~al.}(2022)\citenamefont {Noiri}, \citenamefont {Takeda}, \citenamefont {Nakajima}, \citenamefont {Kobayashi}, \citenamefont {Sammak}, \citenamefont {Scappucci},\ and\ \citenamefont {Tarucha}}]{Noiri2022}%
  \BibitemOpen
  \bibfield  {author} {\bibinfo {author} {\bibfnamefont {A.}~\bibnamefont {Noiri}}, \bibinfo {author} {\bibfnamefont {K.}~\bibnamefont {Takeda}}, \bibinfo {author} {\bibfnamefont {T.}~\bibnamefont {Nakajima}}, \bibinfo {author} {\bibfnamefont {T.}~\bibnamefont {Kobayashi}}, \bibinfo {author} {\bibfnamefont {A.}~\bibnamefont {Sammak}}, \bibinfo {author} {\bibfnamefont {G.}~\bibnamefont {Scappucci}},\ and\ \bibinfo {author} {\bibfnamefont {S.}~\bibnamefont {Tarucha}},\ }\bibfield  {title} {\bibinfo {title} {Fast universal quantum gate above the fault-tolerance threshold in silicon},\ }\href {https://doi.org/10.1038/s41586-021-04182-y} {\bibfield  {journal} {\bibinfo  {journal} {Nature}\ }\textbf {\bibinfo {volume} {601}},\ \bibinfo {pages} {338} (\bibinfo {year} {2022})}\BibitemShut {NoStop}%
\bibitem [{\citenamefont {Steinacker}\ \emph {et~al.}(2024)\citenamefont {Steinacker}, \citenamefont {Stuyck}, \citenamefont {Lim}, \citenamefont {Tanttu}, \citenamefont {Feng}, \citenamefont {Nick}, \citenamefont {Serrano}, \citenamefont {Candido}, \citenamefont {Cifuentes}, \citenamefont {Hudson}, \citenamefont {Chan}, \citenamefont {Kubicek}, \citenamefont {Jussot}, \citenamefont {Canvel}, \citenamefont {Beyne}, \citenamefont {Shimura}, \citenamefont {Loo}, \citenamefont {Godfrin}, \citenamefont {Raes}, \citenamefont {Baudot}, \citenamefont {Wan}, \citenamefont {Laucht}, \citenamefont {Yang}, \citenamefont {Saraiva}, \citenamefont {Escott}, \citenamefont {Greve},\ and\ \citenamefont {Dzurak}}]{DiraqSpinQ}%
  \BibitemOpen
  \bibfield  {author} {\bibinfo {author} {\bibfnamefont {P.}~\bibnamefont {Steinacker}}, \bibinfo {author} {\bibfnamefont {N.~I.~D.}\ \bibnamefont {Stuyck}}, \bibinfo {author} {\bibfnamefont {W.~H.}\ \bibnamefont {Lim}}, \bibinfo {author} {\bibfnamefont {T.}~\bibnamefont {Tanttu}}, \bibinfo {author} {\bibfnamefont {M.}~\bibnamefont {Feng}}, \bibinfo {author} {\bibfnamefont {A.}~\bibnamefont {Nick}}, \bibinfo {author} {\bibfnamefont {S.}~\bibnamefont {Serrano}}, \bibinfo {author} {\bibfnamefont {M.}~\bibnamefont {Candido}}, \bibinfo {author} {\bibfnamefont {J.~D.}\ \bibnamefont {Cifuentes}}, \bibinfo {author} {\bibfnamefont {F.~E.}\ \bibnamefont {Hudson}}, \bibinfo {author} {\bibfnamefont {K.~W.}\ \bibnamefont {Chan}}, \bibinfo {author} {\bibfnamefont {S.}~\bibnamefont {Kubicek}}, \bibinfo {author} {\bibfnamefont {J.}~\bibnamefont {Jussot}}, \bibinfo {author} {\bibfnamefont {Y.}~\bibnamefont {Canvel}}, \bibinfo {author} {\bibfnamefont {S.}~\bibnamefont {Beyne}}, \bibinfo {author} {\bibfnamefont
  {Y.}~\bibnamefont {Shimura}}, \bibinfo {author} {\bibfnamefont {R.}~\bibnamefont {Loo}}, \bibinfo {author} {\bibfnamefont {C.}~\bibnamefont {Godfrin}}, \bibinfo {author} {\bibfnamefont {B.}~\bibnamefont {Raes}}, \bibinfo {author} {\bibfnamefont {S.}~\bibnamefont {Baudot}}, \bibinfo {author} {\bibfnamefont {D.}~\bibnamefont {Wan}}, \bibinfo {author} {\bibfnamefont {A.}~\bibnamefont {Laucht}}, \bibinfo {author} {\bibfnamefont {C.~H.}\ \bibnamefont {Yang}}, \bibinfo {author} {\bibfnamefont {A.}~\bibnamefont {Saraiva}}, \bibinfo {author} {\bibfnamefont {C.~C.}\ \bibnamefont {Escott}}, \bibinfo {author} {\bibfnamefont {K.~D.}\ \bibnamefont {Greve}},\ and\ \bibinfo {author} {\bibfnamefont {A.~S.}\ \bibnamefont {Dzurak}},\ }\bibfield  {title} {\bibinfo {title} {A 300 mm foundry silicon spin qubit unit cell exceeding 99\% fidelity in all operations},\ }\bibfield  {journal} {\bibinfo  {journal} {arXiv}\ }\href {https://doi.org/10.48550/arXiv.2410.15590} {10.48550/arXiv.2410.15590} (\bibinfo {year} {2024}),\ \bibinfo
  {note} {\url{https://doi.org/10.48550/arXiv.2410.15590}},\ \Eprint {https://arxiv.org/abs/2410.15590} {arXiv:2410.15590 [quant-ph]} \BibitemShut {NoStop}%
\bibitem [{\citenamefont {Wang}\ \emph {et~al.}(2022)\citenamefont {Wang}, \citenamefont {Liu}, \citenamefont {Fan}, \citenamefont {Feng}, \citenamefont {Leong}, \citenamefont {Finkler}, \citenamefont {Denisenko}, \citenamefont {Wrachtrup}, \citenamefont {Li},\ and\ \citenamefont {Liu}}]{NVbias1}%
  \BibitemOpen
  \bibfield  {author} {\bibinfo {author} {\bibfnamefont {N.}~\bibnamefont {Wang}}, \bibinfo {author} {\bibfnamefont {C.-F.}\ \bibnamefont {Liu}}, \bibinfo {author} {\bibfnamefont {J.-W.}\ \bibnamefont {Fan}}, \bibinfo {author} {\bibfnamefont {X.}~\bibnamefont {Feng}}, \bibinfo {author} {\bibfnamefont {W.-H.}\ \bibnamefont {Leong}}, \bibinfo {author} {\bibfnamefont {A.}~\bibnamefont {Finkler}}, \bibinfo {author} {\bibfnamefont {A.}~\bibnamefont {Denisenko}}, \bibinfo {author} {\bibfnamefont {J.}~\bibnamefont {Wrachtrup}}, \bibinfo {author} {\bibfnamefont {Q.}~\bibnamefont {Li}},\ and\ \bibinfo {author} {\bibfnamefont {R.-B.}\ \bibnamefont {Liu}},\ }\bibfield  {title} {\bibinfo {title} {Zero-field magnetometry using hyperfine-biased nitrogen-vacancy centers near diamond surfaces},\ }\href {https://doi.org/10.1103/PhysRevResearch.4.013098} {\bibfield  {journal} {\bibinfo  {journal} {Phys. Rev. Res.}\ }\textbf {\bibinfo {volume} {4}},\ \bibinfo {pages} {013098} (\bibinfo {year} {2022})}\BibitemShut {NoStop}%
\bibitem [{\citenamefont {March}\ \emph {et~al.}(2023)\citenamefont {March}, \citenamefont {Wood}, \citenamefont {Stephen}, \citenamefont {Fervenza}, \citenamefont {Breeze}, \citenamefont {Mandal}, \citenamefont {Edmonds}, \citenamefont {Twitchen}, \citenamefont {Markham}, \citenamefont {Williams},\ and\ \citenamefont {Morley}}]{NVbias2}%
  \BibitemOpen
  \bibfield  {author} {\bibinfo {author} {\bibfnamefont {J.~E.}\ \bibnamefont {March}}, \bibinfo {author} {\bibfnamefont {B.~D.}\ \bibnamefont {Wood}}, \bibinfo {author} {\bibfnamefont {C.~J.}\ \bibnamefont {Stephen}}, \bibinfo {author} {\bibfnamefont {L.~D.}\ \bibnamefont {Fervenza}}, \bibinfo {author} {\bibfnamefont {B.~G.}\ \bibnamefont {Breeze}}, \bibinfo {author} {\bibfnamefont {S.}~\bibnamefont {Mandal}}, \bibinfo {author} {\bibfnamefont {A.~M.}\ \bibnamefont {Edmonds}}, \bibinfo {author} {\bibfnamefont {D.~J.}\ \bibnamefont {Twitchen}}, \bibinfo {author} {\bibfnamefont {M.~L.}\ \bibnamefont {Markham}}, \bibinfo {author} {\bibfnamefont {O.~A.}\ \bibnamefont {Williams}},\ and\ \bibinfo {author} {\bibfnamefont {G.~W.}\ \bibnamefont {Morley}},\ }\bibfield  {title} {\bibinfo {title} {Long spin coherence and relaxation times in nanodiamonds milled from polycrystalline ${}^{12}$$\mathrm{C}$ diamond},\ }\href {https://doi.org/10.1103/PhysRevApplied.20.044045} {\bibfield  {journal} {\bibinfo  {journal} {Phys.
  Rev. Appl.}\ }\textbf {\bibinfo {volume} {20}},\ \bibinfo {pages} {044045} (\bibinfo {year} {2023})}\BibitemShut {NoStop}%
\bibitem [{\citenamefont {{deMarti iOlius}}\ \emph {et~al.}(2022)\citenamefont {{deMarti iOlius}}, \citenamefont {{Etxezarreta Martinez}}, \citenamefont {Fuentes}, \citenamefont {Crespo},\ and\ \citenamefont {Garcia-Frias}}]{toninid}%
  \BibitemOpen
  \bibfield  {author} {\bibinfo {author} {\bibfnamefont {A.}~\bibnamefont {{deMarti iOlius}}}, \bibinfo {author} {\bibfnamefont {J.}~\bibnamefont {{Etxezarreta Martinez}}}, \bibinfo {author} {\bibfnamefont {P.}~\bibnamefont {Fuentes}}, \bibinfo {author} {\bibfnamefont {P.~M.}\ \bibnamefont {Crespo}},\ and\ \bibinfo {author} {\bibfnamefont {J.}~\bibnamefont {Garcia-Frias}},\ }\bibfield  {title} {\bibinfo {title} {Performance of surface codes in realistic quantum hardware},\ }\href {https://doi.org/10.1103/PhysRevA.106.062428} {\bibfield  {journal} {\bibinfo  {journal} {Phys. Rev. A}\ }\textbf {\bibinfo {volume} {106}},\ \bibinfo {pages} {062428} (\bibinfo {year} {2022})}\BibitemShut {NoStop}%
\bibitem [{\citenamefont {Lescanne}\ \emph {et~al.}(2020)\citenamefont {Lescanne}, \citenamefont {Villiers}, \citenamefont {Peronnin}, \citenamefont {Sarlette}, \citenamefont {Delbecq}, \citenamefont {Huard}, \citenamefont {Kontos}, \citenamefont {Mirrahimi},\ and\ \citenamefont {Leghtas}}]{catExponentialSup}%
  \BibitemOpen
  \bibfield  {author} {\bibinfo {author} {\bibfnamefont {R.}~\bibnamefont {Lescanne}}, \bibinfo {author} {\bibfnamefont {M.}~\bibnamefont {Villiers}}, \bibinfo {author} {\bibfnamefont {T.}~\bibnamefont {Peronnin}}, \bibinfo {author} {\bibfnamefont {A.}~\bibnamefont {Sarlette}}, \bibinfo {author} {\bibfnamefont {M.}~\bibnamefont {Delbecq}}, \bibinfo {author} {\bibfnamefont {B.}~\bibnamefont {Huard}}, \bibinfo {author} {\bibfnamefont {T.}~\bibnamefont {Kontos}}, \bibinfo {author} {\bibfnamefont {M.}~\bibnamefont {Mirrahimi}},\ and\ \bibinfo {author} {\bibfnamefont {Z.}~\bibnamefont {Leghtas}},\ }\bibfield  {title} {\bibinfo {title} {Exponential suppression of bit-flips in a qubit encoded in an oscillator},\ }\href {https://doi.org/10.1038/s41567-020-0824-x} {\bibfield  {journal} {\bibinfo  {journal} {Nature Physics}\ }\textbf {\bibinfo {volume} {16}},\ \bibinfo {pages} {509} (\bibinfo {year} {2020})}\BibitemShut {NoStop}%
\bibitem [{\citenamefont {Berdou}\ \emph {et~al.}(2023)\citenamefont {Berdou}, \citenamefont {Murani}, \citenamefont {R\'eglade}, \citenamefont {Smith}, \citenamefont {Villiers}, \citenamefont {Palomo}, \citenamefont {Rosticher}, \citenamefont {Denis}, \citenamefont {Morfin}, \citenamefont {Delbecq}, \citenamefont {Kontos}, \citenamefont {Pankratova}, \citenamefont {Rautschke}, \citenamefont {Peronnin}, \citenamefont {Sellem}, \citenamefont {Rouchon}, \citenamefont {Sarlette}, \citenamefont {Mirrahimi}, \citenamefont {Campagne-Ibarcq}, \citenamefont {Jezouin}, \citenamefont {Lescanne},\ and\ \citenamefont {Leghtas}}]{cat100secT1}%
  \BibitemOpen
  \bibfield  {author} {\bibinfo {author} {\bibfnamefont {C.}~\bibnamefont {Berdou}}, \bibinfo {author} {\bibfnamefont {A.}~\bibnamefont {Murani}}, \bibinfo {author} {\bibfnamefont {U.}~\bibnamefont {R\'eglade}}, \bibinfo {author} {\bibfnamefont {W.}~\bibnamefont {Smith}}, \bibinfo {author} {\bibfnamefont {M.}~\bibnamefont {Villiers}}, \bibinfo {author} {\bibfnamefont {J.}~\bibnamefont {Palomo}}, \bibinfo {author} {\bibfnamefont {M.}~\bibnamefont {Rosticher}}, \bibinfo {author} {\bibfnamefont {A.}~\bibnamefont {Denis}}, \bibinfo {author} {\bibfnamefont {P.}~\bibnamefont {Morfin}}, \bibinfo {author} {\bibfnamefont {M.}~\bibnamefont {Delbecq}}, \bibinfo {author} {\bibfnamefont {T.}~\bibnamefont {Kontos}}, \bibinfo {author} {\bibfnamefont {N.}~\bibnamefont {Pankratova}}, \bibinfo {author} {\bibfnamefont {F.}~\bibnamefont {Rautschke}}, \bibinfo {author} {\bibfnamefont {T.}~\bibnamefont {Peronnin}}, \bibinfo {author} {\bibfnamefont {L.-A.}\ \bibnamefont {Sellem}}, \bibinfo {author} {\bibfnamefont {P.}~\bibnamefont
  {Rouchon}}, \bibinfo {author} {\bibfnamefont {A.}~\bibnamefont {Sarlette}}, \bibinfo {author} {\bibfnamefont {M.}~\bibnamefont {Mirrahimi}}, \bibinfo {author} {\bibfnamefont {P.}~\bibnamefont {Campagne-Ibarcq}}, \bibinfo {author} {\bibfnamefont {S.}~\bibnamefont {Jezouin}}, \bibinfo {author} {\bibfnamefont {R.}~\bibnamefont {Lescanne}},\ and\ \bibinfo {author} {\bibfnamefont {Z.}~\bibnamefont {Leghtas}},\ }\bibfield  {title} {\bibinfo {title} {One hundred second bit-flip time in a two-photon dissipative oscillator},\ }\href {https://doi.org/10.1103/PRXQuantum.4.020350} {\bibfield  {journal} {\bibinfo  {journal} {PRX Quantum}\ }\textbf {\bibinfo {volume} {4}},\ \bibinfo {pages} {020350} (\bibinfo {year} {2023})}\BibitemShut {NoStop}%
\bibitem [{\citenamefont {Fern}\ and\ \citenamefont {Whaley}(2008)}]{entropyPauliChann}%
  \BibitemOpen
  \bibfield  {author} {\bibinfo {author} {\bibfnamefont {J.}~\bibnamefont {Fern}}\ and\ \bibinfo {author} {\bibfnamefont {K.~B.}\ \bibnamefont {Whaley}},\ }\bibfield  {title} {\bibinfo {title} {Lower bounds on the nonzero capacity of pauli channels},\ }\href {https://doi.org/10.1103/PhysRevA.78.062335} {\bibfield  {journal} {\bibinfo  {journal} {Phys. Rev. A}\ }\textbf {\bibinfo {volume} {78}},\ \bibinfo {pages} {062335} (\bibinfo {year} {2008})}\BibitemShut {NoStop}%
\bibitem [{\citenamefont {Xu}\ \emph {et~al.}(2023)\citenamefont {Xu}, \citenamefont {Mannucci}, \citenamefont {Seif}, \citenamefont {Kubica}, \citenamefont {Flammia},\ and\ \citenamefont {Jiang}}]{tailoredXZZX}%
  \BibitemOpen
  \bibfield  {author} {\bibinfo {author} {\bibfnamefont {Q.}~\bibnamefont {Xu}}, \bibinfo {author} {\bibfnamefont {N.}~\bibnamefont {Mannucci}}, \bibinfo {author} {\bibfnamefont {A.}~\bibnamefont {Seif}}, \bibinfo {author} {\bibfnamefont {A.}~\bibnamefont {Kubica}}, \bibinfo {author} {\bibfnamefont {S.~T.}\ \bibnamefont {Flammia}},\ and\ \bibinfo {author} {\bibfnamefont {L.}~\bibnamefont {Jiang}},\ }\bibfield  {title} {\bibinfo {title} {Tailored xzzx codes for biased noise},\ }\href {https://doi.org/10.1103/PhysRevResearch.5.013035} {\bibfield  {journal} {\bibinfo  {journal} {Phys. Rev. Res.}\ }\textbf {\bibinfo {volume} {5}},\ \bibinfo {pages} {013035} (\bibinfo {year} {2023})}\BibitemShut {NoStop}%
\bibitem [{\citenamefont {Tuckett}\ \emph {et~al.}(2019)\citenamefont {Tuckett}, \citenamefont {Darmawan}, \citenamefont {Chubb}, \citenamefont {Bravyi}, \citenamefont {Bartlett},\ and\ \citenamefont {Flammia}}]{tuckettBiasSC}%
  \BibitemOpen
  \bibfield  {author} {\bibinfo {author} {\bibfnamefont {D.~K.}\ \bibnamefont {Tuckett}}, \bibinfo {author} {\bibfnamefont {A.~S.}\ \bibnamefont {Darmawan}}, \bibinfo {author} {\bibfnamefont {C.~T.}\ \bibnamefont {Chubb}}, \bibinfo {author} {\bibfnamefont {S.}~\bibnamefont {Bravyi}}, \bibinfo {author} {\bibfnamefont {S.~D.}\ \bibnamefont {Bartlett}},\ and\ \bibinfo {author} {\bibfnamefont {S.~T.}\ \bibnamefont {Flammia}},\ }\bibfield  {title} {\bibinfo {title} {Tailoring surface codes for highly biased noise},\ }\href {https://doi.org/10.1103/PhysRevX.9.041031} {\bibfield  {journal} {\bibinfo  {journal} {Phys. Rev. X}\ }\textbf {\bibinfo {volume} {9}},\ \bibinfo {pages} {041031} (\bibinfo {year} {2019})}\BibitemShut {NoStop}%
\bibitem [{\citenamefont {Tuckett}\ \emph {et~al.}(2018)\citenamefont {Tuckett}, \citenamefont {Bartlett},\ and\ \citenamefont {Flammia}}]{tuckettxy}%
  \BibitemOpen
  \bibfield  {author} {\bibinfo {author} {\bibfnamefont {D.~K.}\ \bibnamefont {Tuckett}}, \bibinfo {author} {\bibfnamefont {S.~D.}\ \bibnamefont {Bartlett}},\ and\ \bibinfo {author} {\bibfnamefont {S.~T.}\ \bibnamefont {Flammia}},\ }\bibfield  {title} {\bibinfo {title} {Ultrahigh error threshold for surface codes with biased noise},\ }\href {https://doi.org/10.1103/PhysRevLett.120.050505} {\bibfield  {journal} {\bibinfo  {journal} {Phys. Rev. Lett.}\ }\textbf {\bibinfo {volume} {120}},\ \bibinfo {pages} {050505} (\bibinfo {year} {2018})}\BibitemShut {NoStop}%
\bibitem [{\citenamefont {Tiurev}\ \emph {et~al.}(2024)\citenamefont {Tiurev}, \citenamefont {Pesah}, \citenamefont {Derks}, \citenamefont {Roffe}, \citenamefont {Eisert}, \citenamefont {Kesselring},\ and\ \citenamefont {Reiner}}]{biasColorCodes}%
  \BibitemOpen
  \bibfield  {author} {\bibinfo {author} {\bibfnamefont {K.}~\bibnamefont {Tiurev}}, \bibinfo {author} {\bibfnamefont {A.}~\bibnamefont {Pesah}}, \bibinfo {author} {\bibfnamefont {P.-J. H.~S.}\ \bibnamefont {Derks}}, \bibinfo {author} {\bibfnamefont {J.}~\bibnamefont {Roffe}}, \bibinfo {author} {\bibfnamefont {J.}~\bibnamefont {Eisert}}, \bibinfo {author} {\bibfnamefont {M.~S.}\ \bibnamefont {Kesselring}},\ and\ \bibinfo {author} {\bibfnamefont {J.-M.}\ \bibnamefont {Reiner}},\ }\bibfield  {title} {\bibinfo {title} {Domain wall color code},\ }\href {https://doi.org/10.1103/PhysRevLett.133.110601} {\bibfield  {journal} {\bibinfo  {journal} {Phys. Rev. Lett.}\ }\textbf {\bibinfo {volume} {133}},\ \bibinfo {pages} {110601} (\bibinfo {year} {2024})}\BibitemShut {NoStop}%
\bibitem [{\citenamefont {Guillaud}\ and\ \citenamefont {Mirrahimi}(2019)}]{twolevelCNOTnobiaspreserve}%
  \BibitemOpen
  \bibfield  {author} {\bibinfo {author} {\bibfnamefont {J.}~\bibnamefont {Guillaud}}\ and\ \bibinfo {author} {\bibfnamefont {M.}~\bibnamefont {Mirrahimi}},\ }\bibfield  {title} {\bibinfo {title} {Repetition cat qubits for fault-tolerant quantum computation},\ }\href {https://doi.org/10.1103/PhysRevX.9.041053} {\bibfield  {journal} {\bibinfo  {journal} {Phys. Rev. X}\ }\textbf {\bibinfo {volume} {9}},\ \bibinfo {pages} {041053} (\bibinfo {year} {2019})}\BibitemShut {NoStop}%
\bibitem [{\citenamefont {Darmawan}\ \emph {et~al.}(2021)\citenamefont {Darmawan}, \citenamefont {Brown}, \citenamefont {Grimsmo}, \citenamefont {Tuckett},\ and\ \citenamefont {Puri}}]{catXZZX}%
  \BibitemOpen
  \bibfield  {author} {\bibinfo {author} {\bibfnamefont {A.~S.}\ \bibnamefont {Darmawan}}, \bibinfo {author} {\bibfnamefont {B.~J.}\ \bibnamefont {Brown}}, \bibinfo {author} {\bibfnamefont {A.~L.}\ \bibnamefont {Grimsmo}}, \bibinfo {author} {\bibfnamefont {D.~K.}\ \bibnamefont {Tuckett}},\ and\ \bibinfo {author} {\bibfnamefont {S.}~\bibnamefont {Puri}},\ }\bibfield  {title} {\bibinfo {title} {Practical quantum error correction with the xzzx code and kerr-cat qubits},\ }\href {https://doi.org/10.1103/PRXQuantum.2.030345} {\bibfield  {journal} {\bibinfo  {journal} {PRX Quantum}\ }\textbf {\bibinfo {volume} {2}},\ \bibinfo {pages} {030345} (\bibinfo {year} {2021})}\BibitemShut {NoStop}%
\bibitem [{\citenamefont {Puri}\ \emph {et~al.}(2020)\citenamefont {Puri}, \citenamefont {St-Jean}, \citenamefont {Gross}, \citenamefont {Grimm}, \citenamefont {Frattini}, \citenamefont {Iyer}, \citenamefont {Krishna}, \citenamefont {Touzard}, \citenamefont {Jiang}, \citenamefont {Blais}, \citenamefont {Flammia},\ and\ \citenamefont {Girvin}}]{biaspreservingCNOTs}%
  \BibitemOpen
  \bibfield  {author} {\bibinfo {author} {\bibfnamefont {S.}~\bibnamefont {Puri}}, \bibinfo {author} {\bibfnamefont {L.}~\bibnamefont {St-Jean}}, \bibinfo {author} {\bibfnamefont {J.~A.}\ \bibnamefont {Gross}}, \bibinfo {author} {\bibfnamefont {A.}~\bibnamefont {Grimm}}, \bibinfo {author} {\bibfnamefont {N.~E.}\ \bibnamefont {Frattini}}, \bibinfo {author} {\bibfnamefont {P.~S.}\ \bibnamefont {Iyer}}, \bibinfo {author} {\bibfnamefont {A.}~\bibnamefont {Krishna}}, \bibinfo {author} {\bibfnamefont {S.}~\bibnamefont {Touzard}}, \bibinfo {author} {\bibfnamefont {L.}~\bibnamefont {Jiang}}, \bibinfo {author} {\bibfnamefont {A.}~\bibnamefont {Blais}}, \bibinfo {author} {\bibfnamefont {S.~T.}\ \bibnamefont {Flammia}},\ and\ \bibinfo {author} {\bibfnamefont {S.~M.}\ \bibnamefont {Girvin}},\ }\bibfield  {title} {\bibinfo {title} {Bias-preserving gates with stabilized cat qubits},\ }\href {https://doi.org/10.1126/sciadv.aay5901} {\bibfield  {journal} {\bibinfo  {journal} {Science Advances}\ }\textbf {\bibinfo {volume}
  {6}},\ \bibinfo {pages} {eaay5901} (\bibinfo {year} {2020})},\ \Eprint {https://arxiv.org/abs/https://www.science.org/doi/pdf/10.1126/sciadv.aay5901} {https://www.science.org/doi/pdf/10.1126/sciadv.aay5901} \BibitemShut {NoStop}%
\bibitem [{\citenamefont {Cong}\ \emph {et~al.}(2022)\citenamefont {Cong}, \citenamefont {Levine}, \citenamefont {Keesling}, \citenamefont {Bluvstein}, \citenamefont {Wang},\ and\ \citenamefont {Lukin}}]{biaspreservingCNOTsNeutral}%
  \BibitemOpen
  \bibfield  {author} {\bibinfo {author} {\bibfnamefont {I.}~\bibnamefont {Cong}}, \bibinfo {author} {\bibfnamefont {H.}~\bibnamefont {Levine}}, \bibinfo {author} {\bibfnamefont {A.}~\bibnamefont {Keesling}}, \bibinfo {author} {\bibfnamefont {D.}~\bibnamefont {Bluvstein}}, \bibinfo {author} {\bibfnamefont {S.-T.}\ \bibnamefont {Wang}},\ and\ \bibinfo {author} {\bibfnamefont {M.~D.}\ \bibnamefont {Lukin}},\ }\bibfield  {title} {\bibinfo {title} {Hardware-efficient, fault-tolerant quantum computation with rydberg atoms},\ }\href {https://doi.org/10.1103/PhysRevX.12.021049} {\bibfield  {journal} {\bibinfo  {journal} {Phys. Rev. X}\ }\textbf {\bibinfo {volume} {12}},\ \bibinfo {pages} {021049} (\bibinfo {year} {2022})}\BibitemShut {NoStop}%
\bibitem [{\citenamefont {Preskill}(2023)}]{Q2BPreskill}%
  \BibitemOpen
  \bibfield  {author} {\bibinfo {author} {\bibfnamefont {J.}~\bibnamefont {Preskill}},\ }\href@noop {} {\bibinfo {title} {Crossing the quantum chasm: From nisq to fault tolerance}},\ \bibinfo {howpublished} {\url{http://theory.caltech.edu/~preskill/talks/Preskill-Q2B-2023}} (\bibinfo {year} {2023}),\ \bibinfo {note} {accessed: 2024-10-26}\BibitemShut {NoStop}%
\bibitem [{\citenamefont {Gouzien}\ \emph {et~al.}(2023)\citenamefont {Gouzien}, \citenamefont {Ruiz}, \citenamefont {Le~R\'egent}, \citenamefont {Guillaud},\ and\ \citenamefont {Sangouard}}]{catoverhead}%
  \BibitemOpen
  \bibfield  {author} {\bibinfo {author} {\bibfnamefont {E.}~\bibnamefont {Gouzien}}, \bibinfo {author} {\bibfnamefont {D.}~\bibnamefont {Ruiz}}, \bibinfo {author} {\bibfnamefont {F.-M.}\ \bibnamefont {Le~R\'egent}}, \bibinfo {author} {\bibfnamefont {J.}~\bibnamefont {Guillaud}},\ and\ \bibinfo {author} {\bibfnamefont {N.}~\bibnamefont {Sangouard}},\ }\bibfield  {title} {\bibinfo {title} {Performance analysis of a repetition cat code architecture: Computing 256-bit elliptic curve logarithm in 9 hours with 126 133 cat qubits},\ }\href {https://doi.org/10.1103/PhysRevLett.131.040602} {\bibfield  {journal} {\bibinfo  {journal} {Phys. Rev. Lett.}\ }\textbf {\bibinfo {volume} {131}},\ \bibinfo {pages} {040602} (\bibinfo {year} {2023})}\BibitemShut {NoStop}%
\bibitem [{\citenamefont {Ruiz}\ \emph {et~al.}(2025)\citenamefont {Ruiz}, \citenamefont {Guillaud}, \citenamefont {Leverrier}, \citenamefont {Mirrahimi},\ and\ \citenamefont {Vuillot}}]{catoverhead2}%
  \BibitemOpen
  \bibfield  {author} {\bibinfo {author} {\bibfnamefont {D.}~\bibnamefont {Ruiz}}, \bibinfo {author} {\bibfnamefont {J.}~\bibnamefont {Guillaud}}, \bibinfo {author} {\bibfnamefont {A.}~\bibnamefont {Leverrier}}, \bibinfo {author} {\bibfnamefont {M.}~\bibnamefont {Mirrahimi}},\ and\ \bibinfo {author} {\bibfnamefont {C.}~\bibnamefont {Vuillot}},\ }\bibfield  {title} {\bibinfo {title} {Ldpc-cat codes for low-overhead quantum computing in 2d},\ }\href {https://doi.org/10.1038/s41467-025-56298-8} {\bibfield  {journal} {\bibinfo  {journal} {Nature Communications}\ }\textbf {\bibinfo {volume} {16}},\ \bibinfo {pages} {1040} (\bibinfo {year} {2025})}\BibitemShut {NoStop}%
\bibitem [{\citenamefont {Gidney}\ and\ \citenamefont {Eker{\aa{}}}(2021)}]{gidneyShor}%
  \BibitemOpen
  \bibfield  {author} {\bibinfo {author} {\bibfnamefont {C.}~\bibnamefont {Gidney}}\ and\ \bibinfo {author} {\bibfnamefont {M.}~\bibnamefont {Eker{\aa{}}}},\ }\bibfield  {title} {\bibinfo {title} {How to factor 2048 bit {RSA} integers in 8 hours using 20 million noisy qubits},\ }\href {https://doi.org/10.22331/q-2021-04-15-433} {\bibfield  {journal} {\bibinfo  {journal} {{Quantum}}\ }\textbf {\bibinfo {volume} {5}},\ \bibinfo {pages} {433} (\bibinfo {year} {2021})}\BibitemShut {NoStop}%
\bibitem [{\citenamefont {Preskill}(2025)}]{PreskillMegaquop}%
  \BibitemOpen
  \bibfield  {author} {\bibinfo {author} {\bibfnamefont {J.}~\bibnamefont {Preskill}},\ }\bibfield  {title} {\bibinfo {title} {{Beyond NISQ: The Megaquop Machine}}\ }(\bibinfo {year} {2025})\ \Eprint {https://arxiv.org/abs/2502.17368} {arXiv:2502.17368 [quant-ph]} \BibitemShut {NoStop}%
\bibitem [{\citenamefont {Higgott}\ \emph {et~al.}(2023)\citenamefont {Higgott}, \citenamefont {Bohdanowicz}, \citenamefont {Kubica}, \citenamefont {Flammia},\ and\ \citenamefont {Campbell}}]{beliefmatching}%
  \BibitemOpen
  \bibfield  {author} {\bibinfo {author} {\bibfnamefont {O.}~\bibnamefont {Higgott}}, \bibinfo {author} {\bibfnamefont {T.~C.}\ \bibnamefont {Bohdanowicz}}, \bibinfo {author} {\bibfnamefont {A.}~\bibnamefont {Kubica}}, \bibinfo {author} {\bibfnamefont {S.~T.}\ \bibnamefont {Flammia}},\ and\ \bibinfo {author} {\bibfnamefont {E.~T.}\ \bibnamefont {Campbell}},\ }\bibfield  {title} {\bibinfo {title} {Improved decoding of circuit noise and fragile boundaries of tailored surface codes},\ }\href {https://doi.org/10.1103/PhysRevX.13.031007} {\bibfield  {journal} {\bibinfo  {journal} {Phys. Rev. X}\ }\textbf {\bibinfo {volume} {13}},\ \bibinfo {pages} {031007} (\bibinfo {year} {2023})}\BibitemShut {NoStop}%
\bibitem [{\citenamefont {Chamberland}\ and\ \citenamefont {Campbell}(2022)}]{biasedCLNog}%
  \BibitemOpen
  \bibfield  {author} {\bibinfo {author} {\bibfnamefont {C.}~\bibnamefont {Chamberland}}\ and\ \bibinfo {author} {\bibfnamefont {E.~T.}\ \bibnamefont {Campbell}},\ }\bibfield  {title} {\bibinfo {title} {Universal quantum computing with twist-free and temporally encoded lattice surgery},\ }\href {https://doi.org/10.1103/PRXQuantum.3.010331} {\bibfield  {journal} {\bibinfo  {journal} {PRX Quantum}\ }\textbf {\bibinfo {volume} {3}},\ \bibinfo {pages} {010331} (\bibinfo {year} {2022})}\BibitemShut {NoStop}%
\bibitem [{\citenamefont {{Hann}}\ \emph {et~al.}(2024)\citenamefont {{Hann}}, \citenamefont {{Noh}}, \citenamefont {{Putterman}}, \citenamefont {{Matheny}}, \citenamefont {{Iverson}}, \citenamefont {{Fang}}, \citenamefont {{Chamberland}}, \citenamefont {{Painter}},\ and\ \citenamefont {{Brand{\~a}o}}}]{catCNOTbpdifficult}%
  \BibitemOpen
  \bibfield  {author} {\bibinfo {author} {\bibfnamefont {C.~T.}\ \bibnamefont {{Hann}}}, \bibinfo {author} {\bibfnamefont {K.}~\bibnamefont {{Noh}}}, \bibinfo {author} {\bibfnamefont {H.}~\bibnamefont {{Putterman}}}, \bibinfo {author} {\bibfnamefont {M.~H.}\ \bibnamefont {{Matheny}}}, \bibinfo {author} {\bibfnamefont {J.~K.}\ \bibnamefont {{Iverson}}}, \bibinfo {author} {\bibfnamefont {M.~T.}\ \bibnamefont {{Fang}}}, \bibinfo {author} {\bibfnamefont {C.}~\bibnamefont {{Chamberland}}}, \bibinfo {author} {\bibfnamefont {O.}~\bibnamefont {{Painter}}},\ and\ \bibinfo {author} {\bibfnamefont {F.~G.~S.~L.}\ \bibnamefont {{Brand{\~a}o}}},\ }\bibfield  {title} {\bibinfo {title} {{Hybrid cat-transmon architecture for scalable, hardware-efficient quantum error correction}},\ }\href@noop {} {\bibfield  {journal} {\bibinfo  {journal} {arXiv e-prints}\ ,\ \bibinfo {eid} {arXiv:2410.23363}} (\bibinfo {year} {2024})},\ \Eprint {https://arxiv.org/abs/2410.23363} {arXiv:2410.23363 [quant-ph]} \BibitemShut {NoStop}%
\bibitem [{\citenamefont {Putterman}\ \emph {et~al.}(2025)\citenamefont {Putterman}, \citenamefont {Noh}, \citenamefont {Hann}, \citenamefont {MacCabe}, \citenamefont {Aghaeimeibodi}, \citenamefont {Patel}, \citenamefont {Lee}, \citenamefont {Jones}, \citenamefont {Moradinejad}, \citenamefont {Rodriguez}, \citenamefont {Mahuli}, \citenamefont {Rose}, \citenamefont {Owens}, \citenamefont {Levine}, \citenamefont {Rosenfeld}, \citenamefont {Reinhold}, \citenamefont {Moncelsi}, \citenamefont {Alcid}, \citenamefont {Alidoust}, \citenamefont {Arrangoiz-Arriola}, \citenamefont {Barnett}, \citenamefont {Bienias}, \citenamefont {Carson}, \citenamefont {Chen}, \citenamefont {Chen}, \citenamefont {Chinkezian}, \citenamefont {Chisholm}, \citenamefont {Chou}, \citenamefont {Clerk}, \citenamefont {Clifford}, \citenamefont {Cosmic}, \citenamefont {Curiel}, \citenamefont {Davis}, \citenamefont {DeLorenzo}, \citenamefont {D'Ewart}, \citenamefont {Diky}, \citenamefont {D'Souza}, \citenamefont {Dumitrescu}, \citenamefont
  {Eisenmann}, \citenamefont {Elkhouly}, \citenamefont {Evenbly}, \citenamefont {Fang}, \citenamefont {Fang}, \citenamefont {Fling}, \citenamefont {Fon}, \citenamefont {Garcia}, \citenamefont {Gorshkov}, \citenamefont {Grant}, \citenamefont {Gray}, \citenamefont {Grimberg}, \citenamefont {Grimsmo}, \citenamefont {Haim}, \citenamefont {Hand}, \citenamefont {He}, \citenamefont {Hernandez}, \citenamefont {Hover}, \citenamefont {Hung}, \citenamefont {Hunt}, \citenamefont {Iverson}, \citenamefont {Jarrige}, \citenamefont {Jaskula}, \citenamefont {Jiang}, \citenamefont {Kalaee}, \citenamefont {Karabalin}, \citenamefont {Karalekas}, \citenamefont {Keller}, \citenamefont {Khalajhedayati}, \citenamefont {Kubica}, \citenamefont {Lee}, \citenamefont {Leroux}, \citenamefont {Lieu}, \citenamefont {Ly}, \citenamefont {Madrigal}, \citenamefont {Marcaud}, \citenamefont {McCabe}, \citenamefont {Miles}, \citenamefont {Milsted}, \citenamefont {Minguzzi}, \citenamefont {Mishra}, \citenamefont {Mukherjee}, \citenamefont
  {Naghiloo}, \citenamefont {Oblepias}, \citenamefont {Ortuno}, \citenamefont {Pagdilao}, \citenamefont {Pancotti}, \citenamefont {Panduro}, \citenamefont {Paquette}, \citenamefont {Park}, \citenamefont {Peairs}, \citenamefont {Perello}, \citenamefont {Peterson}, \citenamefont {Ponte}, \citenamefont {Preskill}, \citenamefont {Qiao}, \citenamefont {Refael}, \citenamefont {Resnick}, \citenamefont {Retzker}, \citenamefont {Reyna}, \citenamefont {Runyan}, \citenamefont {Ryan}, \citenamefont {Sahmoud}, \citenamefont {Sanchez}, \citenamefont {Sanil}, \citenamefont {Sankar}, \citenamefont {Sato}, \citenamefont {Scaffidi}, \citenamefont {Siavoshi}, \citenamefont {Sivarajah}, \citenamefont {Skogland}, \citenamefont {Su}, \citenamefont {Swenson}, \citenamefont {Teo}, \citenamefont {Tomada}, \citenamefont {Torlai}, \citenamefont {Wollack}, \citenamefont {Ye}, \citenamefont {Zerrudo}, \citenamefont {Zhang}, \citenamefont {Brand{\~a}o}, \citenamefont {Matheny},\ and\ \citenamefont {Painter}}]{catRepExperimental}%
  \BibitemOpen
  \bibfield  {author} {\bibinfo {author} {\bibfnamefont {H.}~\bibnamefont {Putterman}}, \bibinfo {author} {\bibfnamefont {K.}~\bibnamefont {Noh}}, \bibinfo {author} {\bibfnamefont {C.~T.}\ \bibnamefont {Hann}}, \bibinfo {author} {\bibfnamefont {G.~S.}\ \bibnamefont {MacCabe}}, \bibinfo {author} {\bibfnamefont {S.}~\bibnamefont {Aghaeimeibodi}}, \bibinfo {author} {\bibfnamefont {R.~N.}\ \bibnamefont {Patel}}, \bibinfo {author} {\bibfnamefont {M.}~\bibnamefont {Lee}}, \bibinfo {author} {\bibfnamefont {W.~M.}\ \bibnamefont {Jones}}, \bibinfo {author} {\bibfnamefont {H.}~\bibnamefont {Moradinejad}}, \bibinfo {author} {\bibfnamefont {R.}~\bibnamefont {Rodriguez}}, \bibinfo {author} {\bibfnamefont {N.}~\bibnamefont {Mahuli}}, \bibinfo {author} {\bibfnamefont {J.}~\bibnamefont {Rose}}, \bibinfo {author} {\bibfnamefont {J.~C.}\ \bibnamefont {Owens}}, \bibinfo {author} {\bibfnamefont {H.}~\bibnamefont {Levine}}, \bibinfo {author} {\bibfnamefont {E.}~\bibnamefont {Rosenfeld}}, \bibinfo {author} {\bibfnamefont
  {P.}~\bibnamefont {Reinhold}}, \bibinfo {author} {\bibfnamefont {L.}~\bibnamefont {Moncelsi}}, \bibinfo {author} {\bibfnamefont {J.~A.}\ \bibnamefont {Alcid}}, \bibinfo {author} {\bibfnamefont {N.}~\bibnamefont {Alidoust}}, \bibinfo {author} {\bibfnamefont {P.}~\bibnamefont {Arrangoiz-Arriola}}, \bibinfo {author} {\bibfnamefont {J.}~\bibnamefont {Barnett}}, \bibinfo {author} {\bibfnamefont {P.}~\bibnamefont {Bienias}}, \bibinfo {author} {\bibfnamefont {H.~A.}\ \bibnamefont {Carson}}, \bibinfo {author} {\bibfnamefont {C.}~\bibnamefont {Chen}}, \bibinfo {author} {\bibfnamefont {L.}~\bibnamefont {Chen}}, \bibinfo {author} {\bibfnamefont {H.}~\bibnamefont {Chinkezian}}, \bibinfo {author} {\bibfnamefont {E.~M.}\ \bibnamefont {Chisholm}}, \bibinfo {author} {\bibfnamefont {M.-H.}\ \bibnamefont {Chou}}, \bibinfo {author} {\bibfnamefont {A.}~\bibnamefont {Clerk}}, \bibinfo {author} {\bibfnamefont {A.}~\bibnamefont {Clifford}}, \bibinfo {author} {\bibfnamefont {R.}~\bibnamefont {Cosmic}}, \bibinfo {author}
  {\bibfnamefont {A.~V.}\ \bibnamefont {Curiel}}, \bibinfo {author} {\bibfnamefont {E.}~\bibnamefont {Davis}}, \bibinfo {author} {\bibfnamefont {L.}~\bibnamefont {DeLorenzo}}, \bibinfo {author} {\bibfnamefont {J.~M.}\ \bibnamefont {D'Ewart}}, \bibinfo {author} {\bibfnamefont {A.}~\bibnamefont {Diky}}, \bibinfo {author} {\bibfnamefont {N.}~\bibnamefont {D'Souza}}, \bibinfo {author} {\bibfnamefont {P.~T.}\ \bibnamefont {Dumitrescu}}, \bibinfo {author} {\bibfnamefont {S.}~\bibnamefont {Eisenmann}}, \bibinfo {author} {\bibfnamefont {E.}~\bibnamefont {Elkhouly}}, \bibinfo {author} {\bibfnamefont {G.}~\bibnamefont {Evenbly}}, \bibinfo {author} {\bibfnamefont {M.~T.}\ \bibnamefont {Fang}}, \bibinfo {author} {\bibfnamefont {Y.}~\bibnamefont {Fang}}, \bibinfo {author} {\bibfnamefont {M.~J.}\ \bibnamefont {Fling}}, \bibinfo {author} {\bibfnamefont {W.}~\bibnamefont {Fon}}, \bibinfo {author} {\bibfnamefont {G.}~\bibnamefont {Garcia}}, \bibinfo {author} {\bibfnamefont {A.~V.}\ \bibnamefont {Gorshkov}}, \bibinfo {author}
  {\bibfnamefont {J.~A.}\ \bibnamefont {Grant}}, \bibinfo {author} {\bibfnamefont {M.~J.}\ \bibnamefont {Gray}}, \bibinfo {author} {\bibfnamefont {S.}~\bibnamefont {Grimberg}}, \bibinfo {author} {\bibfnamefont {A.~L.}\ \bibnamefont {Grimsmo}}, \bibinfo {author} {\bibfnamefont {A.}~\bibnamefont {Haim}}, \bibinfo {author} {\bibfnamefont {J.}~\bibnamefont {Hand}}, \bibinfo {author} {\bibfnamefont {Y.}~\bibnamefont {He}}, \bibinfo {author} {\bibfnamefont {M.}~\bibnamefont {Hernandez}}, \bibinfo {author} {\bibfnamefont {D.}~\bibnamefont {Hover}}, \bibinfo {author} {\bibfnamefont {J.~S.~C.}\ \bibnamefont {Hung}}, \bibinfo {author} {\bibfnamefont {M.}~\bibnamefont {Hunt}}, \bibinfo {author} {\bibfnamefont {J.}~\bibnamefont {Iverson}}, \bibinfo {author} {\bibfnamefont {I.}~\bibnamefont {Jarrige}}, \bibinfo {author} {\bibfnamefont {J.-C.}\ \bibnamefont {Jaskula}}, \bibinfo {author} {\bibfnamefont {L.}~\bibnamefont {Jiang}}, \bibinfo {author} {\bibfnamefont {M.}~\bibnamefont {Kalaee}}, \bibinfo {author} {\bibfnamefont
  {R.}~\bibnamefont {Karabalin}}, \bibinfo {author} {\bibfnamefont {P.~J.}\ \bibnamefont {Karalekas}}, \bibinfo {author} {\bibfnamefont {A.~J.}\ \bibnamefont {Keller}}, \bibinfo {author} {\bibfnamefont {A.}~\bibnamefont {Khalajhedayati}}, \bibinfo {author} {\bibfnamefont {A.}~\bibnamefont {Kubica}}, \bibinfo {author} {\bibfnamefont {H.}~\bibnamefont {Lee}}, \bibinfo {author} {\bibfnamefont {C.}~\bibnamefont {Leroux}}, \bibinfo {author} {\bibfnamefont {S.}~\bibnamefont {Lieu}}, \bibinfo {author} {\bibfnamefont {V.}~\bibnamefont {Ly}}, \bibinfo {author} {\bibfnamefont {K.~V.}\ \bibnamefont {Madrigal}}, \bibinfo {author} {\bibfnamefont {G.}~\bibnamefont {Marcaud}}, \bibinfo {author} {\bibfnamefont {G.}~\bibnamefont {McCabe}}, \bibinfo {author} {\bibfnamefont {C.}~\bibnamefont {Miles}}, \bibinfo {author} {\bibfnamefont {A.}~\bibnamefont {Milsted}}, \bibinfo {author} {\bibfnamefont {J.}~\bibnamefont {Minguzzi}}, \bibinfo {author} {\bibfnamefont {A.}~\bibnamefont {Mishra}}, \bibinfo {author} {\bibfnamefont
  {B.}~\bibnamefont {Mukherjee}}, \bibinfo {author} {\bibfnamefont {M.}~\bibnamefont {Naghiloo}}, \bibinfo {author} {\bibfnamefont {E.}~\bibnamefont {Oblepias}}, \bibinfo {author} {\bibfnamefont {G.}~\bibnamefont {Ortuno}}, \bibinfo {author} {\bibfnamefont {J.}~\bibnamefont {Pagdilao}}, \bibinfo {author} {\bibfnamefont {N.}~\bibnamefont {Pancotti}}, \bibinfo {author} {\bibfnamefont {A.}~\bibnamefont {Panduro}}, \bibinfo {author} {\bibfnamefont {J.~P.}\ \bibnamefont {Paquette}}, \bibinfo {author} {\bibfnamefont {M.}~\bibnamefont {Park}}, \bibinfo {author} {\bibfnamefont {G.~A.}\ \bibnamefont {Peairs}}, \bibinfo {author} {\bibfnamefont {D.}~\bibnamefont {Perello}}, \bibinfo {author} {\bibfnamefont {E.~C.}\ \bibnamefont {Peterson}}, \bibinfo {author} {\bibfnamefont {S.}~\bibnamefont {Ponte}}, \bibinfo {author} {\bibfnamefont {J.}~\bibnamefont {Preskill}}, \bibinfo {author} {\bibfnamefont {J.}~\bibnamefont {Qiao}}, \bibinfo {author} {\bibfnamefont {G.}~\bibnamefont {Refael}}, \bibinfo {author} {\bibfnamefont
  {R.}~\bibnamefont {Resnick}}, \bibinfo {author} {\bibfnamefont {A.}~\bibnamefont {Retzker}}, \bibinfo {author} {\bibfnamefont {O.~A.}\ \bibnamefont {Reyna}}, \bibinfo {author} {\bibfnamefont {M.}~\bibnamefont {Runyan}}, \bibinfo {author} {\bibfnamefont {C.~A.}\ \bibnamefont {Ryan}}, \bibinfo {author} {\bibfnamefont {A.}~\bibnamefont {Sahmoud}}, \bibinfo {author} {\bibfnamefont {E.}~\bibnamefont {Sanchez}}, \bibinfo {author} {\bibfnamefont {R.}~\bibnamefont {Sanil}}, \bibinfo {author} {\bibfnamefont {K.}~\bibnamefont {Sankar}}, \bibinfo {author} {\bibfnamefont {Y.}~\bibnamefont {Sato}}, \bibinfo {author} {\bibfnamefont {T.}~\bibnamefont {Scaffidi}}, \bibinfo {author} {\bibfnamefont {S.}~\bibnamefont {Siavoshi}}, \bibinfo {author} {\bibfnamefont {P.}~\bibnamefont {Sivarajah}}, \bibinfo {author} {\bibfnamefont {T.}~\bibnamefont {Skogland}}, \bibinfo {author} {\bibfnamefont {C.-J.}\ \bibnamefont {Su}}, \bibinfo {author} {\bibfnamefont {L.~J.}\ \bibnamefont {Swenson}}, \bibinfo {author} {\bibfnamefont {S.~M.}\
  \bibnamefont {Teo}}, \bibinfo {author} {\bibfnamefont {A.}~\bibnamefont {Tomada}}, \bibinfo {author} {\bibfnamefont {G.}~\bibnamefont {Torlai}}, \bibinfo {author} {\bibfnamefont {E.~A.}\ \bibnamefont {Wollack}}, \bibinfo {author} {\bibfnamefont {Y.}~\bibnamefont {Ye}}, \bibinfo {author} {\bibfnamefont {J.~A.}\ \bibnamefont {Zerrudo}}, \bibinfo {author} {\bibfnamefont {K.}~\bibnamefont {Zhang}}, \bibinfo {author} {\bibfnamefont {F.~G. S.~L.}\ \bibnamefont {Brand{\~a}o}}, \bibinfo {author} {\bibfnamefont {M.~H.}\ \bibnamefont {Matheny}},\ and\ \bibinfo {author} {\bibfnamefont {O.}~\bibnamefont {Painter}},\ }\bibfield  {title} {\bibinfo {title} {Hardware-efficient quantum error correction via concatenated bosonic qubits},\ }\href {https://doi.org/10.1038/s41586-025-08642-7} {\bibfield  {journal} {\bibinfo  {journal} {Nature}\ }\textbf {\bibinfo {volume} {638}},\ \bibinfo {pages} {927} (\bibinfo {year} {2025})}\BibitemShut {NoStop}%
\bibitem [{\citenamefont {Forlivesi}\ \emph {et~al.}(2024)\citenamefont {Forlivesi}, \citenamefont {Valentini},\ and\ \citenamefont {Chiani}}]{ZZZY}%
  \BibitemOpen
  \bibfield  {author} {\bibinfo {author} {\bibfnamefont {D.}~\bibnamefont {Forlivesi}}, \bibinfo {author} {\bibfnamefont {L.}~\bibnamefont {Valentini}},\ and\ \bibinfo {author} {\bibfnamefont {M.}~\bibnamefont {Chiani}},\ }\bibfield  {title} {\bibinfo {title} {Quantum codes for asymmetric channels: Zzzy surface codes},\ }\href {https://doi.org/10.1109/LCOMM.2024.3454804} {\bibfield  {journal} {\bibinfo  {journal} {IEEE Communications Letters}\ }\textbf {\bibinfo {volume} {28}},\ \bibinfo {pages} {2233} (\bibinfo {year} {2024})}\BibitemShut {NoStop}%
\bibitem [{Note1()}]{Note1}%
  \BibitemOpen
  \bibinfo {note} {We observe that the output bias is in the range $\eta _{Had}\in [0.5,0.6]$, i.e almost depolarizing for every case. Recall that for the single-qubit case the bias is quantified as $\eta = p_z/(p_x+p_y)$, with $1/2$ being the depolarizing case.}\BibitemShut {Stop}%
\bibitem [{\citenamefont {{Mueller}}\ \emph {et~al.}(2023)\citenamefont {{Mueller}}, \citenamefont {{Stollenwerk}}, \citenamefont {{Headley}}, \citenamefont {{Epping}},\ and\ \citenamefont {{Wilhelm}}}]{ZZCZdecomposition}%
  \BibitemOpen
  \bibfield  {author} {\bibinfo {author} {\bibfnamefont {T.}~\bibnamefont {{Mueller}}}, \bibinfo {author} {\bibfnamefont {T.}~\bibnamefont {{Stollenwerk}}}, \bibinfo {author} {\bibfnamefont {D.}~\bibnamefont {{Headley}}}, \bibinfo {author} {\bibfnamefont {M.}~\bibnamefont {{Epping}}},\ and\ \bibinfo {author} {\bibfnamefont {F.~K.}\ \bibnamefont {{Wilhelm}}},\ }\bibfield  {title} {\bibinfo {title} {{Coherent and non-unitary errors in ZZ-generated gates}},\ }\href {https://doi.org/10.48550/arXiv.2304.14212} {\bibfield  {journal} {\bibinfo  {journal} {arXiv e-prints}\ ,\ \bibinfo {eid} {arXiv:2304.14212}} (\bibinfo {year} {2023})},\ \Eprint {https://arxiv.org/abs/2304.14212} {arXiv:2304.14212 [quant-ph]} \BibitemShut {NoStop}%
\bibitem [{Note2()}]{Note2}%
  \BibitemOpen
  \bibinfo {note} {Note that there can be many other ways of enabling such gates in those platforms. Our intention here is to show that there are ways that preserve bias and ways that do not.}\BibitemShut {Stop}%
\bibitem [{\citenamefont {{Google Quantum AI}}\ and\ \citenamefont {{Collaborators}}(2024)}]{willow}%
  \BibitemOpen
  \bibfield  {author} {\bibinfo {author} {\bibnamefont {{Google Quantum AI}}}\ and\ \bibinfo {author} {\bibnamefont {{Collaborators}}},\ }\bibfield  {title} {\bibinfo {title} {Quantum error correction below the surface code threshold},\ }\bibfield  {journal} {\bibinfo  {journal} {Nature}\ }\href {https://doi.org/10.1038/s41586-024-08449-y} {10.1038/s41586-024-08449-y} (\bibinfo {year} {2024})\BibitemShut {NoStop}%
\bibitem [{\citenamefont {Gidney}\ \emph {et~al.}(2021)\citenamefont {Gidney}, \citenamefont {Newman}, \citenamefont {Fowler},\ and\ \citenamefont {Broughton}}]{teraHoneycomb}%
  \BibitemOpen
  \bibfield  {author} {\bibinfo {author} {\bibfnamefont {C.}~\bibnamefont {Gidney}}, \bibinfo {author} {\bibfnamefont {M.}~\bibnamefont {Newman}}, \bibinfo {author} {\bibfnamefont {A.}~\bibnamefont {Fowler}},\ and\ \bibinfo {author} {\bibfnamefont {M.}~\bibnamefont {Broughton}},\ }\bibfield  {title} {\bibinfo {title} {A {F}ault-{T}olerant {H}oneycomb {M}emory},\ }\href {https://doi.org/10.22331/q-2021-12-20-605} {\bibfield  {journal} {\bibinfo  {journal} {{Quantum}}\ }\textbf {\bibinfo {volume} {5}},\ \bibinfo {pages} {605} (\bibinfo {year} {2021})}\BibitemShut {NoStop}%
\bibitem [{\citenamefont {Riverlane}(2024)}]{RiverlaneQEC}%
  \BibitemOpen
  \bibfield  {author} {\bibinfo {author} {\bibnamefont {Riverlane}},\ }\href@noop {} {\bibinfo {title} {The quantum error correction report}},\ \bibinfo {howpublished} {\url{https://www.riverlane.com/quantum-error-correction-report-2024}} (\bibinfo {year} {2024}),\ \bibinfo {note} {accessed: 2025-01-05}\BibitemShut {NoStop}%
\bibitem [{\citenamefont {Cai}\ \emph {et~al.}(2019)\citenamefont {Cai}, \citenamefont {Fogarty}, \citenamefont {Schaal}, \citenamefont {Patom{\"{a}}ki}, \citenamefont {Benjamin},\ and\ \citenamefont {Morton}}]{Cai2019siliconsurfacecode}%
  \BibitemOpen
  \bibfield  {author} {\bibinfo {author} {\bibfnamefont {Z.}~\bibnamefont {Cai}}, \bibinfo {author} {\bibfnamefont {M.~A.}\ \bibnamefont {Fogarty}}, \bibinfo {author} {\bibfnamefont {S.}~\bibnamefont {Schaal}}, \bibinfo {author} {\bibfnamefont {S.}~\bibnamefont {Patom{\"{a}}ki}}, \bibinfo {author} {\bibfnamefont {S.~C.}\ \bibnamefont {Benjamin}},\ and\ \bibinfo {author} {\bibfnamefont {J.~J.~L.}\ \bibnamefont {Morton}},\ }\bibfield  {title} {\bibinfo {title} {A {S}ilicon {S}urface {C}ode {A}rchitecture {R}esilient {A}gainst {L}eakage {E}rrors},\ }\href {https://doi.org/10.22331/q-2019-12-09-212} {\bibfield  {journal} {\bibinfo  {journal} {{Quantum}}\ }\textbf {\bibinfo {volume} {3}},\ \bibinfo {pages} {212} (\bibinfo {year} {2019})}\BibitemShut {NoStop}%
\bibitem [{\citenamefont {Litinski}(2019)}]{Litinski2019gameofsurfacecodes}%
  \BibitemOpen
  \bibfield  {author} {\bibinfo {author} {\bibfnamefont {D.}~\bibnamefont {Litinski}},\ }\bibfield  {title} {\bibinfo {title} {A {G}ame of {S}urface {C}odes: {L}arge-{S}cale {Q}uantum {C}omputing with {L}attice {S}urgery},\ }\href {https://doi.org/10.22331/q-2019-03-05-128} {\bibfield  {journal} {\bibinfo  {journal} {{Quantum}}\ }\textbf {\bibinfo {volume} {3}},\ \bibinfo {pages} {128} (\bibinfo {year} {2019})}\BibitemShut {NoStop}%
\bibitem [{\citenamefont {Gidney}(2022)}]{Gidney2022stability}%
  \BibitemOpen
  \bibfield  {author} {\bibinfo {author} {\bibfnamefont {C.}~\bibnamefont {Gidney}},\ }\bibfield  {title} {\bibinfo {title} {Stability {E}xperiments: {T}he {O}verlooked {D}ual of {M}emory {E}xperiments},\ }\href {https://doi.org/10.22331/q-2022-08-24-786} {\bibfield  {journal} {\bibinfo  {journal} {{Quantum}}\ }\textbf {\bibinfo {volume} {6}},\ \bibinfo {pages} {786} (\bibinfo {year} {2022})}\BibitemShut {NoStop}%
\bibitem [{\citenamefont {{O'Rourke}}\ and\ \citenamefont {{Devitt}}(2024)}]{rotvsunrot}%
  \BibitemOpen
  \bibfield  {author} {\bibinfo {author} {\bibfnamefont {A.~R.}\ \bibnamefont {{O'Rourke}}}\ and\ \bibinfo {author} {\bibfnamefont {S.}~\bibnamefont {{Devitt}}},\ }\bibfield  {title} {\bibinfo {title} {{Compare the Pair: Rotated vs. Unrotated Surface Codes at Equal Logical Error Rates}},\ }\href {https://doi.org/10.48550/arXiv.2409.14765} {\bibfield  {journal} {\bibinfo  {journal} {arXiv e-prints}\ ,\ \bibinfo {eid} {arXiv:2409.14765}} (\bibinfo {year} {2024})},\ \Eprint {https://arxiv.org/abs/2409.14765} {arXiv:2409.14765 [quant-ph]} \BibitemShut {NoStop}%
\bibitem [{\citenamefont {{Otxoa}}\ \emph {et~al.}(2025)\citenamefont {{Otxoa}}, \citenamefont {{Etxezarreta Martinez}}, \citenamefont {{Schnabl}}, \citenamefont {{Mertig}}, \citenamefont {{Smith}},\ and\ \citenamefont {{Martins}}}]{spinhex}%
  \BibitemOpen
  \bibfield  {author} {\bibinfo {author} {\bibfnamefont {R.~M.}\ \bibnamefont {{Otxoa}}}, \bibinfo {author} {\bibfnamefont {J.}~\bibnamefont {{Etxezarreta Martinez}}}, \bibinfo {author} {\bibfnamefont {P.}~\bibnamefont {{Schnabl}}}, \bibinfo {author} {\bibfnamefont {N.}~\bibnamefont {{Mertig}}}, \bibinfo {author} {\bibfnamefont {C.}~\bibnamefont {{Smith}}},\ and\ \bibinfo {author} {\bibfnamefont {F.}~\bibnamefont {{Martins}}},\ }\bibfield  {title} {\bibinfo {title} {{SpinHex: A low-crosstalk, spin-qubit architecture based on multi-electron couplers}},\ }\href {https://arxiv.org/abs/2504.03149} {\bibfield  {journal} {\bibinfo  {journal} {arXiv}\ ,\ \bibinfo {pages} {arXiv:2504.03149}} (\bibinfo {year} {2025})}\BibitemShut {NoStop}%
\bibitem [{\citenamefont {Higgott}\ and\ \citenamefont {Gidney}(2025)}]{sparseBlossom}%
  \BibitemOpen
  \bibfield  {author} {\bibinfo {author} {\bibfnamefont {O.}~\bibnamefont {Higgott}}\ and\ \bibinfo {author} {\bibfnamefont {C.}~\bibnamefont {Gidney}},\ }\bibfield  {title} {\bibinfo {title} {Sparse {B}lossom: correcting a million errors per core second with minimum-weight matching},\ }\href {https://doi.org/10.22331/q-2025-01-20-1600} {\bibfield  {journal} {\bibinfo  {journal} {{Quantum}}\ }\textbf {\bibinfo {volume} {9}},\ \bibinfo {pages} {1600} (\bibinfo {year} {2025})}\BibitemShut {NoStop}%
\bibitem [{\citenamefont {{Wolanski}}\ and\ \citenamefont {{Barber}}(2024)}]{AmbiguityClusteringOSD}%
  \BibitemOpen
  \bibfield  {author} {\bibinfo {author} {\bibfnamefont {S.}~\bibnamefont {{Wolanski}}}\ and\ \bibinfo {author} {\bibfnamefont {B.}~\bibnamefont {{Barber}}},\ }\bibfield  {title} {\bibinfo {title} {{Ambiguity Clustering: an accurate and efficient decoder for qLDPC codes}},\ }\href {https://doi.org/10.48550/arXiv.2406.14527} {\bibfield  {journal} {\bibinfo  {journal} {arXiv e-prints}\ ,\ \bibinfo {eid} {arXiv:2406.14527}} (\bibinfo {year} {2024})},\ \Eprint {https://arxiv.org/abs/2406.14527} {arXiv:2406.14527 [quant-ph]} \BibitemShut {NoStop}%
\bibitem [{\citenamefont {{deMarti iOlius}}\ \emph {et~al.}(2024)\citenamefont {{deMarti iOlius}}, \citenamefont {{Etxezarreta Martinez}}, \citenamefont {Roffe},\ and\ \citenamefont {{Etxezarreta Martinez}}}]{bpotf_2024}%
  \BibitemOpen
  \bibfield  {author} {\bibinfo {author} {\bibfnamefont {A.}~\bibnamefont {{deMarti iOlius}}}, \bibinfo {author} {\bibfnamefont {I.}~\bibnamefont {{Etxezarreta Martinez}}}, \bibinfo {author} {\bibfnamefont {J.}~\bibnamefont {Roffe}},\ and\ \bibinfo {author} {\bibfnamefont {J.}~\bibnamefont {{Etxezarreta Martinez}}},\ }\bibfield  {title} {\bibinfo {title} {{An almost-linear time decoding algorithm for quantum LDPC codes under circuit-level noise}},\ }\href {https://arxiv.org/abs/2409.01440} {\bibfield  {journal} {\bibinfo  {journal} {arXiv}\ ,\ \bibinfo {pages} {2409.01440}} (\bibinfo {year} {2024})}\BibitemShut {NoStop}%
\bibitem [{\citenamefont {Bravyi}\ \emph {et~al.}(2024)\citenamefont {Bravyi}, \citenamefont {Cross}, \citenamefont {Gambetta}, \citenamefont {Maslov}, \citenamefont {Rall},\ and\ \citenamefont {Yoder}}]{BBcodes}%
  \BibitemOpen
  \bibfield  {author} {\bibinfo {author} {\bibfnamefont {S.}~\bibnamefont {Bravyi}}, \bibinfo {author} {\bibfnamefont {A.~W.}\ \bibnamefont {Cross}}, \bibinfo {author} {\bibfnamefont {J.~M.}\ \bibnamefont {Gambetta}}, \bibinfo {author} {\bibfnamefont {D.}~\bibnamefont {Maslov}}, \bibinfo {author} {\bibfnamefont {P.}~\bibnamefont {Rall}},\ and\ \bibinfo {author} {\bibfnamefont {T.~J.}\ \bibnamefont {Yoder}},\ }\bibfield  {title} {\bibinfo {title} {High-threshold and low-overhead fault-tolerant quantum memory},\ }\href {https://doi.org/10.1038/s41586-024-07107-7} {\bibfield  {journal} {\bibinfo  {journal} {Nature}\ }\textbf {\bibinfo {volume} {627}},\ \bibinfo {pages} {778} (\bibinfo {year} {2024})}\BibitemShut {NoStop}%
\bibitem [{\citenamefont {Tomita}\ and\ \citenamefont {Svore}(2014)}]{scheduleZS}%
  \BibitemOpen
  \bibfield  {author} {\bibinfo {author} {\bibfnamefont {Y.}~\bibnamefont {Tomita}}\ and\ \bibinfo {author} {\bibfnamefont {K.~M.}\ \bibnamefont {Svore}},\ }\bibfield  {title} {\bibinfo {title} {Low-distance surface codes under realistic quantum noise},\ }\href {https://doi.org/10.1103/PhysRevA.90.062320} {\bibfield  {journal} {\bibinfo  {journal} {Phys. Rev. A}\ }\textbf {\bibinfo {volume} {90}},\ \bibinfo {pages} {062320} (\bibinfo {year} {2014})}\BibitemShut {NoStop}%
\bibitem [{Note3()}]{Note3}%
  \BibitemOpen
  \bibinfo {note} {This is sufficient since the noise can be transformed into stochastic Pauli noise by means of Pauli twirling \cite {decoders,approximatingdecoherence}. In fact, Pauli noise is the one generally considered for theoretical QEC analysis \cite {stim}}\BibitemShut {NoStop}%
\bibitem [{\citenamefont {{Fellous-Asiani}}\ \emph {et~al.}(2023)\citenamefont {{Fellous-Asiani}}, \citenamefont {{Naseri}}, \citenamefont {{Datta}}, \citenamefont {{Streltsov}},\ and\ \citenamefont {{Oszmaniec}}}]{biaspother1}%
  \BibitemOpen
  \bibfield  {author} {\bibinfo {author} {\bibfnamefont {M.}~\bibnamefont {{Fellous-Asiani}}}, \bibinfo {author} {\bibfnamefont {M.}~\bibnamefont {{Naseri}}}, \bibinfo {author} {\bibfnamefont {C.}~\bibnamefont {{Datta}}}, \bibinfo {author} {\bibfnamefont {A.}~\bibnamefont {{Streltsov}}},\ and\ \bibinfo {author} {\bibfnamefont {M.}~\bibnamefont {{Oszmaniec}}},\ }\bibfield  {title} {\bibinfo {title} {{Scalable noisy quantum circuits for biased-noise qubits}},\ }\href {https://doi.org/10.48550/arXiv.2305.02045} {\bibfield  {journal} {\bibinfo  {journal} {arXiv e-prints}\ ,\ \bibinfo {eid} {arXiv:2305.02045}} (\bibinfo {year} {2023})},\ \Eprint {https://arxiv.org/abs/2305.02045} {arXiv:2305.02045 [quant-ph]} \BibitemShut {NoStop}%
\bibitem [{\citenamefont {{Rennela}}\ and\ \citenamefont {{Ollivier}}(2024)}]{biaspother2}%
  \BibitemOpen
  \bibfield  {author} {\bibinfo {author} {\bibfnamefont {M.}~\bibnamefont {{Rennela}}}\ and\ \bibinfo {author} {\bibfnamefont {H.}~\bibnamefont {{Ollivier}}},\ }\bibfield  {title} {\bibinfo {title} {{Low bit-flip rate probabilistic error cancellation}},\ }\href {https://doi.org/10.48550/arXiv.2411.06422} {\bibfield  {journal} {\bibinfo  {journal} {arXiv e-prints}\ ,\ \bibinfo {eid} {arXiv:2411.06422}} (\bibinfo {year} {2024})},\ \Eprint {https://arxiv.org/abs/2411.06422} {arXiv:2411.06422 [quant-ph]} \BibitemShut {NoStop}%
\bibitem [{\citenamefont {Johansson}\ \emph {et~al.}(2012)\citenamefont {Johansson}, \citenamefont {Nation},\ and\ \citenamefont {Nori}}]{qutip}%
  \BibitemOpen
  \bibfield  {author} {\bibinfo {author} {\bibfnamefont {J.}~\bibnamefont {Johansson}}, \bibinfo {author} {\bibfnamefont {P.}~\bibnamefont {Nation}},\ and\ \bibinfo {author} {\bibfnamefont {F.}~\bibnamefont {Nori}},\ }\bibfield  {title} {\bibinfo {title} {Qutip: An open-source python framework for the dynamics of open quantum systems},\ }\href {https://doi.org/https://doi.org/10.1016/j.cpc.2012.02.021} {\bibfield  {journal} {\bibinfo  {journal} {Computer Physics Communications}\ }\textbf {\bibinfo {volume} {183}},\ \bibinfo {pages} {1760} (\bibinfo {year} {2012})}\BibitemShut {NoStop}%
\bibitem [{\citenamefont {Flammia}\ and\ \citenamefont {Wallman}(2020)}]{flammiaHadTransf}%
  \BibitemOpen
  \bibfield  {author} {\bibinfo {author} {\bibfnamefont {S.~T.}\ \bibnamefont {Flammia}}\ and\ \bibinfo {author} {\bibfnamefont {J.~J.}\ \bibnamefont {Wallman}},\ }\bibfield  {title} {\bibinfo {title} {Efficient estimation of pauli channels},\ }\bibfield  {journal} {\bibinfo  {journal} {ACM Transactions on Quantum Computing}\ }\textbf {\bibinfo {volume} {1}},\ \href {https://doi.org/10.1145/3408039} {10.1145/3408039} (\bibinfo {year} {2020})\BibitemShut {NoStop}%
\bibitem [{\citenamefont {van~den Berg}\ \emph {et~al.}(2023)\citenamefont {van~den Berg}, \citenamefont {Minev}, \citenamefont {Kandala},\ and\ \citenamefont {Temme}}]{PEC}%
  \BibitemOpen
  \bibfield  {author} {\bibinfo {author} {\bibfnamefont {E.}~\bibnamefont {van~den Berg}}, \bibinfo {author} {\bibfnamefont {Z.~K.}\ \bibnamefont {Minev}}, \bibinfo {author} {\bibfnamefont {A.}~\bibnamefont {Kandala}},\ and\ \bibinfo {author} {\bibfnamefont {K.}~\bibnamefont {Temme}},\ }\bibfield  {title} {\bibinfo {title} {Probabilistic error cancellation with sparse pauli--lindblad models on noisy quantum processors},\ }\href {https://doi.org/10.1038/s41567-023-02042-2} {\bibfield  {journal} {\bibinfo  {journal} {Nature Physics}\ }\textbf {\bibinfo {volume} {19}},\ \bibinfo {pages} {1116} (\bibinfo {year} {2023})}\BibitemShut {NoStop}%
\bibitem [{\citenamefont {Burkard}\ \emph {et~al.}(2023)\citenamefont {Burkard}, \citenamefont {Ladd}, \citenamefont {Pan}, \citenamefont {Nichol},\ and\ \citenamefont {Petta}}]{RevModPhys.95.025003}%
  \BibitemOpen
  \bibfield  {author} {\bibinfo {author} {\bibfnamefont {G.}~\bibnamefont {Burkard}}, \bibinfo {author} {\bibfnamefont {T.~D.}\ \bibnamefont {Ladd}}, \bibinfo {author} {\bibfnamefont {A.}~\bibnamefont {Pan}}, \bibinfo {author} {\bibfnamefont {J.~M.}\ \bibnamefont {Nichol}},\ and\ \bibinfo {author} {\bibfnamefont {J.~R.}\ \bibnamefont {Petta}},\ }\bibfield  {title} {\bibinfo {title} {Semiconductor spin qubits},\ }\href {https://doi.org/10.1103/RevModPhys.95.025003} {\bibfield  {journal} {\bibinfo  {journal} {Rev. Mod. Phys.}\ }\textbf {\bibinfo {volume} {95}},\ \bibinfo {pages} {025003} (\bibinfo {year} {2023})}\BibitemShut {NoStop}%
\bibitem [{\citenamefont {Veldhorst}\ \emph {et~al.}(2015{\natexlab{a}})\citenamefont {Veldhorst}, \citenamefont {Yang}, \citenamefont {Hwang}, \citenamefont {Huang}, \citenamefont {Dehollain}, \citenamefont {Muhonen}, \citenamefont {Simmons}, \citenamefont {Laucht}, \citenamefont {Hudson}, \citenamefont {Itoh}, \citenamefont {Morello},\ and\ \citenamefont {Dzurak}}]{ESR}%
  \BibitemOpen
  \bibfield  {author} {\bibinfo {author} {\bibfnamefont {M.}~\bibnamefont {Veldhorst}}, \bibinfo {author} {\bibfnamefont {C.~H.}\ \bibnamefont {Yang}}, \bibinfo {author} {\bibfnamefont {J.~C.~C.}\ \bibnamefont {Hwang}}, \bibinfo {author} {\bibfnamefont {W.}~\bibnamefont {Huang}}, \bibinfo {author} {\bibfnamefont {J.~P.}\ \bibnamefont {Dehollain}}, \bibinfo {author} {\bibfnamefont {J.~T.}\ \bibnamefont {Muhonen}}, \bibinfo {author} {\bibfnamefont {S.}~\bibnamefont {Simmons}}, \bibinfo {author} {\bibfnamefont {A.}~\bibnamefont {Laucht}}, \bibinfo {author} {\bibfnamefont {F.~E.}\ \bibnamefont {Hudson}}, \bibinfo {author} {\bibfnamefont {K.~M.}\ \bibnamefont {Itoh}}, \bibinfo {author} {\bibfnamefont {A.}~\bibnamefont {Morello}},\ and\ \bibinfo {author} {\bibfnamefont {A.~S.}\ \bibnamefont {Dzurak}},\ }\bibfield  {title} {\bibinfo {title} {A two-qubit logic gate in silicon},\ }\href {https://doi.org/10.1038/nature15263} {\bibfield  {journal} {\bibinfo  {journal} {Nature}\ }\textbf {\bibinfo {volume} {526}},\
  \bibinfo {pages} {410–414} (\bibinfo {year} {2015}{\natexlab{a}})}\BibitemShut {NoStop}%
\bibitem [{\citenamefont {Watson}\ \emph {et~al.}(2018)\citenamefont {Watson}, \citenamefont {Philips}, \citenamefont {Kawakami}, \citenamefont {Ward}, \citenamefont {Scarlino}, \citenamefont {Veldhorst}, \citenamefont {Savage}, \citenamefont {Lagally}, \citenamefont {Friesen}, \citenamefont {Coppersmith}, \citenamefont {Eriksson},\ and\ \citenamefont {Vandersypen}}]{EDRS1}%
  \BibitemOpen
  \bibfield  {author} {\bibinfo {author} {\bibfnamefont {T.~F.}\ \bibnamefont {Watson}}, \bibinfo {author} {\bibfnamefont {S.~G.~J.}\ \bibnamefont {Philips}}, \bibinfo {author} {\bibfnamefont {E.}~\bibnamefont {Kawakami}}, \bibinfo {author} {\bibfnamefont {D.~R.}\ \bibnamefont {Ward}}, \bibinfo {author} {\bibfnamefont {P.}~\bibnamefont {Scarlino}}, \bibinfo {author} {\bibfnamefont {M.}~\bibnamefont {Veldhorst}}, \bibinfo {author} {\bibfnamefont {D.~E.}\ \bibnamefont {Savage}}, \bibinfo {author} {\bibfnamefont {M.~G.}\ \bibnamefont {Lagally}}, \bibinfo {author} {\bibfnamefont {M.}~\bibnamefont {Friesen}}, \bibinfo {author} {\bibfnamefont {S.~N.}\ \bibnamefont {Coppersmith}}, \bibinfo {author} {\bibfnamefont {M.~A.}\ \bibnamefont {Eriksson}},\ and\ \bibinfo {author} {\bibfnamefont {L.~M.~K.}\ \bibnamefont {Vandersypen}},\ }\bibfield  {title} {\bibinfo {title} {A programmable two-qubit quantum processor in silicon},\ }\href {https://doi.org/10.1038/nature25766} {\bibfield  {journal} {\bibinfo  {journal}
  {Nature}\ }\textbf {\bibinfo {volume} {555}},\ \bibinfo {pages} {633–637} (\bibinfo {year} {2018})}\BibitemShut {NoStop}%
\bibitem [{\citenamefont {Zajac}\ \emph {et~al.}(2018)\citenamefont {Zajac}, \citenamefont {Sigillito}, \citenamefont {Russ}, \citenamefont {Borjans}, \citenamefont {Taylor}, \citenamefont {Burkard},\ and\ \citenamefont {Petta}}]{EDSR2}%
  \BibitemOpen
  \bibfield  {author} {\bibinfo {author} {\bibfnamefont {D.~M.}\ \bibnamefont {Zajac}}, \bibinfo {author} {\bibfnamefont {A.~J.}\ \bibnamefont {Sigillito}}, \bibinfo {author} {\bibfnamefont {M.}~\bibnamefont {Russ}}, \bibinfo {author} {\bibfnamefont {F.}~\bibnamefont {Borjans}}, \bibinfo {author} {\bibfnamefont {J.~M.}\ \bibnamefont {Taylor}}, \bibinfo {author} {\bibfnamefont {G.}~\bibnamefont {Burkard}},\ and\ \bibinfo {author} {\bibfnamefont {J.~R.}\ \bibnamefont {Petta}},\ }\bibfield  {title} {\bibinfo {title} {Resonantly driven cnot gate for electron spins},\ }\href {https://doi.org/10.1126/science.aao5965} {\bibfield  {journal} {\bibinfo  {journal} {Science}\ }\textbf {\bibinfo {volume} {359}},\ \bibinfo {pages} {439} (\bibinfo {year} {2018})},\ \Eprint {https://arxiv.org/abs/https://www.science.org/doi/pdf/10.1126/science.aao5965} {https://www.science.org/doi/pdf/10.1126/science.aao5965} \BibitemShut {NoStop}%
\bibitem [{\citenamefont {Russ}\ \emph {et~al.}(2018)\citenamefont {Russ}, \citenamefont {Zajac}, \citenamefont {Sigillito}, \citenamefont {Borjans}, \citenamefont {Taylor}, \citenamefont {Petta},\ and\ \citenamefont {Burkard}}]{PhysRevB.97.085421}%
  \BibitemOpen
  \bibfield  {author} {\bibinfo {author} {\bibfnamefont {M.}~\bibnamefont {Russ}}, \bibinfo {author} {\bibfnamefont {D.~M.}\ \bibnamefont {Zajac}}, \bibinfo {author} {\bibfnamefont {A.~J.}\ \bibnamefont {Sigillito}}, \bibinfo {author} {\bibfnamefont {F.}~\bibnamefont {Borjans}}, \bibinfo {author} {\bibfnamefont {J.~M.}\ \bibnamefont {Taylor}}, \bibinfo {author} {\bibfnamefont {J.~R.}\ \bibnamefont {Petta}},\ and\ \bibinfo {author} {\bibfnamefont {G.}~\bibnamefont {Burkard}},\ }\bibfield  {title} {\bibinfo {title} {High-fidelity quantum gates in si/sige double quantum dots},\ }\href {https://doi.org/10.1103/PhysRevB.97.085421} {\bibfield  {journal} {\bibinfo  {journal} {Phys. Rev. B}\ }\textbf {\bibinfo {volume} {97}},\ \bibinfo {pages} {085421} (\bibinfo {year} {2018})}\BibitemShut {NoStop}%
\bibitem [{\citenamefont {Meunier}\ \emph {et~al.}(2011)\citenamefont {Meunier}, \citenamefont {Calado},\ and\ \citenamefont {Vandersypen}}]{PhysRevB.83.121403}%
  \BibitemOpen
  \bibfield  {author} {\bibinfo {author} {\bibfnamefont {T.}~\bibnamefont {Meunier}}, \bibinfo {author} {\bibfnamefont {V.~E.}\ \bibnamefont {Calado}},\ and\ \bibinfo {author} {\bibfnamefont {L.~M.~K.}\ \bibnamefont {Vandersypen}},\ }\bibfield  {title} {\bibinfo {title} {Efficient controlled-phase gate for single-spin qubits in quantum dots},\ }\href {https://doi.org/10.1103/PhysRevB.83.121403} {\bibfield  {journal} {\bibinfo  {journal} {Phys. Rev. B}\ }\textbf {\bibinfo {volume} {83}},\ \bibinfo {pages} {121403} (\bibinfo {year} {2011})}\BibitemShut {NoStop}%
\bibitem [{\citenamefont {Veldhorst}\ \emph {et~al.}(2015{\natexlab{b}})\citenamefont {Veldhorst}, \citenamefont {Yang}, \citenamefont {Hwang}, \citenamefont {Huang}, \citenamefont {Dehollain}, \citenamefont {Muhonen}, \citenamefont {Simmons}, \citenamefont {Laucht}, \citenamefont {Hudson}, \citenamefont {Itoh}, \citenamefont {Morello},\ and\ \citenamefont {Dzurak}}]{Veldhorst2015}%
  \BibitemOpen
  \bibfield  {author} {\bibinfo {author} {\bibfnamefont {M.}~\bibnamefont {Veldhorst}}, \bibinfo {author} {\bibfnamefont {C.~H.}\ \bibnamefont {Yang}}, \bibinfo {author} {\bibfnamefont {J.~C.~C.}\ \bibnamefont {Hwang}}, \bibinfo {author} {\bibfnamefont {W.}~\bibnamefont {Huang}}, \bibinfo {author} {\bibfnamefont {J.~P.}\ \bibnamefont {Dehollain}}, \bibinfo {author} {\bibfnamefont {J.~T.}\ \bibnamefont {Muhonen}}, \bibinfo {author} {\bibfnamefont {S.}~\bibnamefont {Simmons}}, \bibinfo {author} {\bibfnamefont {A.}~\bibnamefont {Laucht}}, \bibinfo {author} {\bibfnamefont {F.~E.}\ \bibnamefont {Hudson}}, \bibinfo {author} {\bibfnamefont {K.~M.}\ \bibnamefont {Itoh}}, \bibinfo {author} {\bibfnamefont {A.}~\bibnamefont {Morello}},\ and\ \bibinfo {author} {\bibfnamefont {A.~S.}\ \bibnamefont {Dzurak}},\ }\bibfield  {title} {\bibinfo {title} {A two-qubit logic gate in silicon},\ }\href {https://doi.org/10.1038/nature15263} {\bibfield  {journal} {\bibinfo  {journal} {Nature}\ }\textbf {\bibinfo {volume} {526}},\
  \bibinfo {pages} {410} (\bibinfo {year} {2015}{\natexlab{b}})}\BibitemShut {NoStop}%
\bibitem [{\citenamefont {Huang}\ \emph {et~al.}(2024)\citenamefont {Huang}, \citenamefont {Su}, \citenamefont {Lim}, \citenamefont {Feng}, \citenamefont {van Straaten}, \citenamefont {Severin}, \citenamefont {Gilbert}, \citenamefont {Dumoulin~Stuyck}, \citenamefont {Tanttu}, \citenamefont {Serrano}, \citenamefont {Cifuentes}, \citenamefont {Hansen}, \citenamefont {Seedhouse}, \citenamefont {Vahapoglu}, \citenamefont {Leon}, \citenamefont {Abrosimov}, \citenamefont {Pohl}, \citenamefont {Thewalt}, \citenamefont {Hudson}, \citenamefont {Escott}, \citenamefont {Ares}, \citenamefont {Bartlett}, \citenamefont {Morello}, \citenamefont {Saraiva}, \citenamefont {Laucht}, \citenamefont {Dzurak},\ and\ \citenamefont {Yang}}]{dzurakBPCZ}%
  \BibitemOpen
  \bibfield  {author} {\bibinfo {author} {\bibfnamefont {J.~Y.}\ \bibnamefont {Huang}}, \bibinfo {author} {\bibfnamefont {R.~Y.}\ \bibnamefont {Su}}, \bibinfo {author} {\bibfnamefont {W.~H.}\ \bibnamefont {Lim}}, \bibinfo {author} {\bibfnamefont {M.}~\bibnamefont {Feng}}, \bibinfo {author} {\bibfnamefont {B.}~\bibnamefont {van Straaten}}, \bibinfo {author} {\bibfnamefont {B.}~\bibnamefont {Severin}}, \bibinfo {author} {\bibfnamefont {W.}~\bibnamefont {Gilbert}}, \bibinfo {author} {\bibfnamefont {N.}~\bibnamefont {Dumoulin~Stuyck}}, \bibinfo {author} {\bibfnamefont {T.}~\bibnamefont {Tanttu}}, \bibinfo {author} {\bibfnamefont {S.}~\bibnamefont {Serrano}}, \bibinfo {author} {\bibfnamefont {J.~D.}\ \bibnamefont {Cifuentes}}, \bibinfo {author} {\bibfnamefont {I.}~\bibnamefont {Hansen}}, \bibinfo {author} {\bibfnamefont {A.~E.}\ \bibnamefont {Seedhouse}}, \bibinfo {author} {\bibfnamefont {E.}~\bibnamefont {Vahapoglu}}, \bibinfo {author} {\bibfnamefont {R.~C.~C.}\ \bibnamefont {Leon}}, \bibinfo {author}
  {\bibfnamefont {N.~V.}\ \bibnamefont {Abrosimov}}, \bibinfo {author} {\bibfnamefont {H.-J.}\ \bibnamefont {Pohl}}, \bibinfo {author} {\bibfnamefont {M.~L.~W.}\ \bibnamefont {Thewalt}}, \bibinfo {author} {\bibfnamefont {F.~E.}\ \bibnamefont {Hudson}}, \bibinfo {author} {\bibfnamefont {C.~C.}\ \bibnamefont {Escott}}, \bibinfo {author} {\bibfnamefont {N.}~\bibnamefont {Ares}}, \bibinfo {author} {\bibfnamefont {S.~D.}\ \bibnamefont {Bartlett}}, \bibinfo {author} {\bibfnamefont {A.}~\bibnamefont {Morello}}, \bibinfo {author} {\bibfnamefont {A.}~\bibnamefont {Saraiva}}, \bibinfo {author} {\bibfnamefont {A.}~\bibnamefont {Laucht}}, \bibinfo {author} {\bibfnamefont {A.~S.}\ \bibnamefont {Dzurak}},\ and\ \bibinfo {author} {\bibfnamefont {C.~H.}\ \bibnamefont {Yang}},\ }\bibfield  {title} {\bibinfo {title} {High-fidelity spin qubit operation and algorithmic initialization above 1 k},\ }\href {https://doi.org/10.1038/s41586-024-07160-2} {\bibfield  {journal} {\bibinfo  {journal} {Nature}\ }\textbf {\bibinfo {volume}
  {627}},\ \bibinfo {pages} {772} (\bibinfo {year} {2024})}\BibitemShut {NoStop}%
\bibitem [{\citenamefont {Jonathan}\ \emph {et~al.}(2000)\citenamefont {Jonathan}, \citenamefont {Plenio},\ and\ \citenamefont {Knight}}]{PhysRevA.62.042307}%
  \BibitemOpen
  \bibfield  {author} {\bibinfo {author} {\bibfnamefont {D.}~\bibnamefont {Jonathan}}, \bibinfo {author} {\bibfnamefont {M.~B.}\ \bibnamefont {Plenio}},\ and\ \bibinfo {author} {\bibfnamefont {P.~L.}\ \bibnamefont {Knight}},\ }\bibfield  {title} {\bibinfo {title} {Fast quantum gates for cold trapped ions},\ }\href {https://doi.org/10.1103/PhysRevA.62.042307} {\bibfield  {journal} {\bibinfo  {journal} {Phys. Rev. A}\ }\textbf {\bibinfo {volume} {62}},\ \bibinfo {pages} {042307} (\bibinfo {year} {2000})}\BibitemShut {NoStop}%
\bibitem [{\citenamefont {Manovitz}\ \emph {et~al.}(2022)\citenamefont {Manovitz}, \citenamefont {Shapira}, \citenamefont {Gazit}, \citenamefont {Akerman},\ and\ \citenamefont {Ozeri}}]{Manovitz2022}%
  \BibitemOpen
  \bibfield  {author} {\bibinfo {author} {\bibfnamefont {T.}~\bibnamefont {Manovitz}}, \bibinfo {author} {\bibfnamefont {Y.}~\bibnamefont {Shapira}}, \bibinfo {author} {\bibfnamefont {L.}~\bibnamefont {Gazit}}, \bibinfo {author} {\bibfnamefont {N.}~\bibnamefont {Akerman}},\ and\ \bibinfo {author} {\bibfnamefont {R.}~\bibnamefont {Ozeri}},\ }\bibfield  {title} {\bibinfo {title} {Trapped-ion quantum computer with robust entangling gates and quantum coherent feedback},\ }\bibfield  {journal} {\bibinfo  {journal} {PRX Quantum}\ }\textbf {\bibinfo {volume} {3}},\ \href {https://doi.org/10.1103/prxquantum.3.010347} {10.1103/prxquantum.3.010347} (\bibinfo {year} {2022})\BibitemShut {NoStop}%
\bibitem [{\citenamefont {Leibfried}\ \emph {et~al.}(2003)\citenamefont {Leibfried}, \citenamefont {DeMarco}, \citenamefont {Meyer}, \citenamefont {Lucas}, \citenamefont {Barrett}, \citenamefont {Britton}, \citenamefont {Itano}, \citenamefont {Jelenkovi{\'c}}, \citenamefont {Rosenband},\ and\ \citenamefont {Wineland}}]{Leibfried2003-bf}%
  \BibitemOpen
  \bibfield  {author} {\bibinfo {author} {\bibfnamefont {D.}~\bibnamefont {Leibfried}}, \bibinfo {author} {\bibfnamefont {B.}~\bibnamefont {DeMarco}}, \bibinfo {author} {\bibfnamefont {V.}~\bibnamefont {Meyer}}, \bibinfo {author} {\bibfnamefont {D.}~\bibnamefont {Lucas}}, \bibinfo {author} {\bibfnamefont {M.}~\bibnamefont {Barrett}}, \bibinfo {author} {\bibfnamefont {J.}~\bibnamefont {Britton}}, \bibinfo {author} {\bibfnamefont {W.~M.}\ \bibnamefont {Itano}}, \bibinfo {author} {\bibfnamefont {C.}~\bibnamefont {Jelenkovi{\'c}}, \bibfnamefont {Band~Langer}}, \bibinfo {author} {\bibfnamefont {T.}~\bibnamefont {Rosenband}},\ and\ \bibinfo {author} {\bibfnamefont {D.~J.}\ \bibnamefont {Wineland}},\ }\bibfield  {title} {\bibinfo {title} {Experimental demonstration of a robust, high-fidelity geometric two ion-qubit phase gate},\ }\href@noop {} {\bibfield  {journal} {\bibinfo  {journal} {Nature}\ }\textbf {\bibinfo {volume} {422}},\ \bibinfo {pages} {412} (\bibinfo {year} {2003})}\BibitemShut {NoStop}%
\bibitem [{\citenamefont {Milburn}\ \emph {et~al.}(2000)\citenamefont {Milburn}, \citenamefont {Schneider},\ and\ \citenamefont {James}}]{Milburn2000}%
  \BibitemOpen
  \bibfield  {author} {\bibinfo {author} {\bibfnamefont {G.}~\bibnamefont {Milburn}}, \bibinfo {author} {\bibfnamefont {S.}~\bibnamefont {Schneider}},\ and\ \bibinfo {author} {\bibfnamefont {D.}~\bibnamefont {James}},\ }\bibfield  {title} {\bibinfo {title} {Ion trap quantum computing with warm ions},\ }\href {https://doi.org/10.1002/1521-3978(200009)48:9/11<801::aid-prop801>3.0.co;2-1} {\bibfield  {journal} {\bibinfo  {journal} {Fortschritte der Physik}\ }\textbf {\bibinfo {volume} {48}},\ \bibinfo {pages} {801–810} (\bibinfo {year} {2000})}\BibitemShut {NoStop}%
\bibitem [{\citenamefont {Jaksch}\ \emph {et~al.}(2000)\citenamefont {Jaksch}, \citenamefont {Cirac}, \citenamefont {Zoller}, \citenamefont {Rolston}, \citenamefont {C\^ot\'e},\ and\ \citenamefont {Lukin}}]{PhysRevLett.85.2208}%
  \BibitemOpen
  \bibfield  {author} {\bibinfo {author} {\bibfnamefont {D.}~\bibnamefont {Jaksch}}, \bibinfo {author} {\bibfnamefont {J.~I.}\ \bibnamefont {Cirac}}, \bibinfo {author} {\bibfnamefont {P.}~\bibnamefont {Zoller}}, \bibinfo {author} {\bibfnamefont {S.~L.}\ \bibnamefont {Rolston}}, \bibinfo {author} {\bibfnamefont {R.}~\bibnamefont {C\^ot\'e}},\ and\ \bibinfo {author} {\bibfnamefont {M.~D.}\ \bibnamefont {Lukin}},\ }\bibfield  {title} {\bibinfo {title} {Fast quantum gates for neutral atoms},\ }\href {https://doi.org/10.1103/PhysRevLett.85.2208} {\bibfield  {journal} {\bibinfo  {journal} {Phys. Rev. Lett.}\ }\textbf {\bibinfo {volume} {85}},\ \bibinfo {pages} {2208} (\bibinfo {year} {2000})}\BibitemShut {NoStop}%
\bibitem [{\citenamefont {Lukin}\ \emph {et~al.}(2001)\citenamefont {Lukin}, \citenamefont {Fleischhauer}, \citenamefont {Cote}, \citenamefont {Duan}, \citenamefont {Jaksch}, \citenamefont {Cirac},\ and\ \citenamefont {Zoller}}]{PhysRevLett.87.037901}%
  \BibitemOpen
  \bibfield  {author} {\bibinfo {author} {\bibfnamefont {M.~D.}\ \bibnamefont {Lukin}}, \bibinfo {author} {\bibfnamefont {M.}~\bibnamefont {Fleischhauer}}, \bibinfo {author} {\bibfnamefont {R.}~\bibnamefont {Cote}}, \bibinfo {author} {\bibfnamefont {L.~M.}\ \bibnamefont {Duan}}, \bibinfo {author} {\bibfnamefont {D.}~\bibnamefont {Jaksch}}, \bibinfo {author} {\bibfnamefont {J.~I.}\ \bibnamefont {Cirac}},\ and\ \bibinfo {author} {\bibfnamefont {P.}~\bibnamefont {Zoller}},\ }\bibfield  {title} {\bibinfo {title} {Dipole blockade and quantum information processing in mesoscopic atomic ensembles},\ }\href {https://doi.org/10.1103/PhysRevLett.87.037901} {\bibfield  {journal} {\bibinfo  {journal} {Phys. Rev. Lett.}\ }\textbf {\bibinfo {volume} {87}},\ \bibinfo {pages} {037901} (\bibinfo {year} {2001})}\BibitemShut {NoStop}%
\bibitem [{\citenamefont {Saffman}\ \emph {et~al.}(2010)\citenamefont {Saffman}, \citenamefont {Walker},\ and\ \citenamefont {M\o{}lmer}}]{RevModPhys.82.2313}%
  \BibitemOpen
  \bibfield  {author} {\bibinfo {author} {\bibfnamefont {M.}~\bibnamefont {Saffman}}, \bibinfo {author} {\bibfnamefont {T.~G.}\ \bibnamefont {Walker}},\ and\ \bibinfo {author} {\bibfnamefont {K.}~\bibnamefont {M\o{}lmer}},\ }\bibfield  {title} {\bibinfo {title} {Quantum information with rydberg atoms},\ }\href {https://doi.org/10.1103/RevModPhys.82.2313} {\bibfield  {journal} {\bibinfo  {journal} {Rev. Mod. Phys.}\ }\textbf {\bibinfo {volume} {82}},\ \bibinfo {pages} {2313} (\bibinfo {year} {2010})}\BibitemShut {NoStop}%
\bibitem [{\citenamefont {Childress}\ and\ \citenamefont {Hanson}(2013)}]{Childress2013}%
  \BibitemOpen
  \bibfield  {author} {\bibinfo {author} {\bibfnamefont {L.}~\bibnamefont {Childress}}\ and\ \bibinfo {author} {\bibfnamefont {R.}~\bibnamefont {Hanson}},\ }\bibfield  {title} {\bibinfo {title} {Diamond nv centers for quantum computing and quantum networks},\ }\href {https://doi.org/10.1557/mrs.2013.20} {\bibfield  {journal} {\bibinfo  {journal} {MRS Bulletin}\ }\textbf {\bibinfo {volume} {38}},\ \bibinfo {pages} {134–138} (\bibinfo {year} {2013})}\BibitemShut {NoStop}%
\bibitem [{\citenamefont {Dolde}\ \emph {et~al.}(2013)\citenamefont {Dolde}, \citenamefont {Jakobi}, \citenamefont {Naydenov}, \citenamefont {Zhao}, \citenamefont {Pezzagna}, \citenamefont {Trautmann}, \citenamefont {Meijer}, \citenamefont {Neumann}, \citenamefont {Jelezko},\ and\ \citenamefont {Wrachtrup}}]{Dolde2013}%
  \BibitemOpen
  \bibfield  {author} {\bibinfo {author} {\bibfnamefont {F.}~\bibnamefont {Dolde}}, \bibinfo {author} {\bibfnamefont {I.}~\bibnamefont {Jakobi}}, \bibinfo {author} {\bibfnamefont {B.}~\bibnamefont {Naydenov}}, \bibinfo {author} {\bibfnamefont {N.}~\bibnamefont {Zhao}}, \bibinfo {author} {\bibfnamefont {S.}~\bibnamefont {Pezzagna}}, \bibinfo {author} {\bibfnamefont {C.}~\bibnamefont {Trautmann}}, \bibinfo {author} {\bibfnamefont {J.}~\bibnamefont {Meijer}}, \bibinfo {author} {\bibfnamefont {P.}~\bibnamefont {Neumann}}, \bibinfo {author} {\bibfnamefont {F.}~\bibnamefont {Jelezko}},\ and\ \bibinfo {author} {\bibfnamefont {J.}~\bibnamefont {Wrachtrup}},\ }\bibfield  {title} {\bibinfo {title} {Room-temperature entanglement between single defect spins in diamond},\ }\href {https://doi.org/10.1038/nphys2545} {\bibfield  {journal} {\bibinfo  {journal} {Nature Physics}\ }\textbf {\bibinfo {volume} {9}},\ \bibinfo {pages} {139–143} (\bibinfo {year} {2013})}\BibitemShut {NoStop}%
\bibitem [{\citenamefont {Finsterhoelzl}\ \emph {et~al.}(2024)\citenamefont {Finsterhoelzl}, \citenamefont {Hannes},\ and\ \citenamefont {Burkard}}]{NV-hyper}%
  \BibitemOpen
  \bibfield  {author} {\bibinfo {author} {\bibfnamefont {R.}~\bibnamefont {Finsterhoelzl}}, \bibinfo {author} {\bibfnamefont {W.-R.}\ \bibnamefont {Hannes}},\ and\ \bibinfo {author} {\bibfnamefont {G.}~\bibnamefont {Burkard}},\ }\href {https://arxiv.org/abs/2403.11553} {\bibinfo {title} {High-fidelity entangling gates for electron and nuclear spin qubits in diamond}} (\bibinfo {year} {2024}),\ \Eprint {https://arxiv.org/abs/2403.11553} {arXiv:2403.11553 [quant-ph]} \BibitemShut {NoStop}%
\bibitem [{\citenamefont {Everitt}\ \emph {et~al.}(2014)\citenamefont {Everitt}, \citenamefont {Devitt}, \citenamefont {Munro},\ and\ \citenamefont {Nemoto}}]{PhysRevA.89.052317}%
  \BibitemOpen
  \bibfield  {author} {\bibinfo {author} {\bibfnamefont {M.~S.}\ \bibnamefont {Everitt}}, \bibinfo {author} {\bibfnamefont {S.}~\bibnamefont {Devitt}}, \bibinfo {author} {\bibfnamefont {W.~J.}\ \bibnamefont {Munro}},\ and\ \bibinfo {author} {\bibfnamefont {K.}~\bibnamefont {Nemoto}},\ }\bibfield  {title} {\bibinfo {title} {High-fidelity gate operations with the coupled nuclear and electron spins of a nitrogen-vacancy center in diamond},\ }\href {https://doi.org/10.1103/PhysRevA.89.052317} {\bibfield  {journal} {\bibinfo  {journal} {Phys. Rev. A}\ }\textbf {\bibinfo {volume} {89}},\ \bibinfo {pages} {052317} (\bibinfo {year} {2014})}\BibitemShut {NoStop}%
\bibitem [{\citenamefont {Shim}\ \emph {et~al.}(2013)\citenamefont {Shim}, \citenamefont {Niemeyer}, \citenamefont {Zhang},\ and\ \citenamefont {Suter}}]{PhysRevA.87.012301}%
  \BibitemOpen
  \bibfield  {author} {\bibinfo {author} {\bibfnamefont {J.~H.}\ \bibnamefont {Shim}}, \bibinfo {author} {\bibfnamefont {I.}~\bibnamefont {Niemeyer}}, \bibinfo {author} {\bibfnamefont {J.}~\bibnamefont {Zhang}},\ and\ \bibinfo {author} {\bibfnamefont {D.}~\bibnamefont {Suter}},\ }\bibfield  {title} {\bibinfo {title} {Room-temperature high-speed nuclear-spin quantum memory in diamond},\ }\href {https://doi.org/10.1103/PhysRevA.87.012301} {\bibfield  {journal} {\bibinfo  {journal} {Phys. Rev. A}\ }\textbf {\bibinfo {volume} {87}},\ \bibinfo {pages} {012301} (\bibinfo {year} {2013})}\BibitemShut {NoStop}%
\bibitem [{\citenamefont {Liang}\ \emph {et~al.}(2016)\citenamefont {Liang}, \citenamefont {Yue}, \citenamefont {Lv}, \citenamefont {Du}, \citenamefont {Huang}, \citenamefont {Yan},\ and\ \citenamefont {Zhu}}]{Liang_2016}%
  \BibitemOpen
  \bibfield  {author} {\bibinfo {author} {\bibfnamefont {Z.-T.}\ \bibnamefont {Liang}}, \bibinfo {author} {\bibfnamefont {X.}~\bibnamefont {Yue}}, \bibinfo {author} {\bibfnamefont {Q.}~\bibnamefont {Lv}}, \bibinfo {author} {\bibfnamefont {Y.-X.}\ \bibnamefont {Du}}, \bibinfo {author} {\bibfnamefont {W.}~\bibnamefont {Huang}}, \bibinfo {author} {\bibfnamefont {H.}~\bibnamefont {Yan}},\ and\ \bibinfo {author} {\bibfnamefont {S.-L.}\ \bibnamefont {Zhu}},\ }\bibfield  {title} {\bibinfo {title} {Proposal for implementing universal superadiabatic geometric quantum gates in nitrogen-vacancy centers},\ }\bibfield  {journal} {\bibinfo  {journal} {Physical Review A}\ }\textbf {\bibinfo {volume} {93}},\ \href {https://doi.org/10.1103/physreva.93.040305} {10.1103/physreva.93.040305} (\bibinfo {year} {2016})\BibitemShut {NoStop}%
\bibitem [{\citenamefont {Hays}\ \emph {et~al.}(2025)\citenamefont {Hays}, \citenamefont {Kim},\ and\ \citenamefont {Oliver}}]{oliver25}%
  \BibitemOpen
  \bibfield  {author} {\bibinfo {author} {\bibfnamefont {M.}~\bibnamefont {Hays}}, \bibinfo {author} {\bibfnamefont {J.}~\bibnamefont {Kim}},\ and\ \bibinfo {author} {\bibfnamefont {W.~D.}\ \bibnamefont {Oliver}},\ }\bibfield  {title} {\bibinfo {title} {{Non-degenerate noise-resilient superconducting qubit}},\ }\href@noop {} {\bibfield  {journal} {\bibinfo  {journal} {arXiv}\ } (\bibinfo {year} {2025})},\ \Eprint {https://arxiv.org/abs/2502.15459} {arXiv:2502.15459 [quant-ph]} \BibitemShut {NoStop}%
\bibitem [{\citenamefont {Rigetti}\ and\ \citenamefont {Devoret}(2010)}]{cr1}%
  \BibitemOpen
  \bibfield  {author} {\bibinfo {author} {\bibfnamefont {C.}~\bibnamefont {Rigetti}}\ and\ \bibinfo {author} {\bibfnamefont {M.}~\bibnamefont {Devoret}},\ }\bibfield  {title} {\bibinfo {title} {Fully microwave-tunable universal gates in superconducting qubits with linear couplings and fixed transition frequencies},\ }\href {https://doi.org/10.1103/PhysRevB.81.134507} {\bibfield  {journal} {\bibinfo  {journal} {Phys. Rev. B}\ }\textbf {\bibinfo {volume} {81}},\ \bibinfo {pages} {134507} (\bibinfo {year} {2010})}\BibitemShut {NoStop}%
\bibitem [{\citenamefont {Chow}\ \emph {et~al.}(2011)\citenamefont {Chow}, \citenamefont {C\'orcoles}, \citenamefont {Gambetta}, \citenamefont {Rigetti}, \citenamefont {Johnson}, \citenamefont {Smolin}, \citenamefont {Rozen}, \citenamefont {Keefe}, \citenamefont {Rothwell}, \citenamefont {Ketchen},\ and\ \citenamefont {Steffen}}]{cr2}%
  \BibitemOpen
  \bibfield  {author} {\bibinfo {author} {\bibfnamefont {J.~M.}\ \bibnamefont {Chow}}, \bibinfo {author} {\bibfnamefont {A.~D.}\ \bibnamefont {C\'orcoles}}, \bibinfo {author} {\bibfnamefont {J.~M.}\ \bibnamefont {Gambetta}}, \bibinfo {author} {\bibfnamefont {C.}~\bibnamefont {Rigetti}}, \bibinfo {author} {\bibfnamefont {B.~R.}\ \bibnamefont {Johnson}}, \bibinfo {author} {\bibfnamefont {J.~A.}\ \bibnamefont {Smolin}}, \bibinfo {author} {\bibfnamefont {J.~R.}\ \bibnamefont {Rozen}}, \bibinfo {author} {\bibfnamefont {G.~A.}\ \bibnamefont {Keefe}}, \bibinfo {author} {\bibfnamefont {M.~B.}\ \bibnamefont {Rothwell}}, \bibinfo {author} {\bibfnamefont {M.~B.}\ \bibnamefont {Ketchen}},\ and\ \bibinfo {author} {\bibfnamefont {M.}~\bibnamefont {Steffen}},\ }\bibfield  {title} {\bibinfo {title} {Simple all-microwave entangling gate for fixed-frequency superconducting qubits},\ }\href {https://doi.org/10.1103/PhysRevLett.107.080502} {\bibfield  {journal} {\bibinfo  {journal} {Phys. Rev. Lett.}\ }\textbf {\bibinfo {volume}
  {107}},\ \bibinfo {pages} {080502} (\bibinfo {year} {2011})}\BibitemShut {NoStop}%
\bibitem [{\citenamefont {Yan}\ \emph {et~al.}(2018)\citenamefont {Yan}, \citenamefont {Krantz}, \citenamefont {Sung}, \citenamefont {Kjaergaard}, \citenamefont {Campbell}, \citenamefont {Orlando}, \citenamefont {Gustavsson},\ and\ \citenamefont {Oliver}}]{stc1}%
  \BibitemOpen
  \bibfield  {author} {\bibinfo {author} {\bibfnamefont {F.}~\bibnamefont {Yan}}, \bibinfo {author} {\bibfnamefont {P.}~\bibnamefont {Krantz}}, \bibinfo {author} {\bibfnamefont {Y.}~\bibnamefont {Sung}}, \bibinfo {author} {\bibfnamefont {M.}~\bibnamefont {Kjaergaard}}, \bibinfo {author} {\bibfnamefont {D.~L.}\ \bibnamefont {Campbell}}, \bibinfo {author} {\bibfnamefont {T.~P.}\ \bibnamefont {Orlando}}, \bibinfo {author} {\bibfnamefont {S.}~\bibnamefont {Gustavsson}},\ and\ \bibinfo {author} {\bibfnamefont {W.~D.}\ \bibnamefont {Oliver}},\ }\bibfield  {title} {\bibinfo {title} {Tunable coupling scheme for implementing high-fidelity two-qubit gates},\ }\href {https://doi.org/10.1103/PhysRevApplied.10.054062} {\bibfield  {journal} {\bibinfo  {journal} {Phys. Rev. Appl.}\ }\textbf {\bibinfo {volume} {10}},\ \bibinfo {pages} {054062} (\bibinfo {year} {2018})}\BibitemShut {NoStop}%
\bibitem [{\citenamefont {Li}\ \emph {et~al.}(2024)\citenamefont {Li}, \citenamefont {Kubo}, \citenamefont {Ho}, \citenamefont {Yan}, \citenamefont {Nakamura},\ and\ \citenamefont {Goto}}]{stc2}%
  \BibitemOpen
  \bibfield  {author} {\bibinfo {author} {\bibfnamefont {R.}~\bibnamefont {Li}}, \bibinfo {author} {\bibfnamefont {K.}~\bibnamefont {Kubo}}, \bibinfo {author} {\bibfnamefont {Y.}~\bibnamefont {Ho}}, \bibinfo {author} {\bibfnamefont {Z.}~\bibnamefont {Yan}}, \bibinfo {author} {\bibfnamefont {Y.}~\bibnamefont {Nakamura}},\ and\ \bibinfo {author} {\bibfnamefont {H.}~\bibnamefont {Goto}},\ }\bibfield  {title} {\bibinfo {title} {Realization of high-fidelity cz gate based on a double-transmon coupler},\ }\href {https://doi.org/10.1103/PhysRevX.14.041050} {\bibfield  {journal} {\bibinfo  {journal} {Phys. Rev. X}\ }\textbf {\bibinfo {volume} {14}},\ \bibinfo {pages} {041050} (\bibinfo {year} {2024})}\BibitemShut {NoStop}%
\bibitem [{\citenamefont {Dua}\ \emph {et~al.}(2024)\citenamefont {Dua}, \citenamefont {Kubica}, \citenamefont {Jiang}, \citenamefont {Flammia},\ and\ \citenamefont {Gullans}}]{cliffordDeformation}%
  \BibitemOpen
  \bibfield  {author} {\bibinfo {author} {\bibfnamefont {A.}~\bibnamefont {Dua}}, \bibinfo {author} {\bibfnamefont {A.}~\bibnamefont {Kubica}}, \bibinfo {author} {\bibfnamefont {L.}~\bibnamefont {Jiang}}, \bibinfo {author} {\bibfnamefont {S.~T.}\ \bibnamefont {Flammia}},\ and\ \bibinfo {author} {\bibfnamefont {M.~J.}\ \bibnamefont {Gullans}},\ }\bibfield  {title} {\bibinfo {title} {Clifford-deformed surface codes},\ }\href {https://doi.org/10.1103/PRXQuantum.5.010347} {\bibfield  {journal} {\bibinfo  {journal} {PRX Quantum}\ }\textbf {\bibinfo {volume} {5}},\ \bibinfo {pages} {010347} (\bibinfo {year} {2024})}\BibitemShut {NoStop}%
\bibitem [{\citenamefont {Acharya}\ \emph {et~al.}(2023)\citenamefont {Acharya}, \citenamefont {Aleiner}, \citenamefont {Allen}, \citenamefont {Andersen}, \citenamefont {Ansmann}, \citenamefont {Arute}, \citenamefont {Arya}, \citenamefont {Asfaw}, \citenamefont {Atalaya}, \citenamefont {Babbush}, \citenamefont {Bacon} \emph {et~al.}}]{googleSurf}%
  \BibitemOpen
  \bibfield  {author} {\bibinfo {author} {\bibfnamefont {R.}~\bibnamefont {Acharya}}, \bibinfo {author} {\bibfnamefont {I.}~\bibnamefont {Aleiner}}, \bibinfo {author} {\bibfnamefont {R.}~\bibnamefont {Allen}}, \bibinfo {author} {\bibfnamefont {T.~I.}\ \bibnamefont {Andersen}}, \bibinfo {author} {\bibfnamefont {M.}~\bibnamefont {Ansmann}}, \bibinfo {author} {\bibfnamefont {F.}~\bibnamefont {Arute}}, \bibinfo {author} {\bibfnamefont {K.}~\bibnamefont {Arya}}, \bibinfo {author} {\bibfnamefont {A.}~\bibnamefont {Asfaw}}, \bibinfo {author} {\bibfnamefont {J.}~\bibnamefont {Atalaya}}, \bibinfo {author} {\bibfnamefont {R.}~\bibnamefont {Babbush}}, \bibinfo {author} {\bibfnamefont {D.}~\bibnamefont {Bacon}}, \emph {et~al.},\ }\bibfield  {title} {\bibinfo {title} {Suppressing quantum errors by scaling a surface code logical qubit},\ }\href {https://doi.org/10.1038/s41586-022-05434-1} {\bibfield  {journal} {\bibinfo  {journal} {Nature}\ }\textbf {\bibinfo {volume} {614}},\ \bibinfo {pages} {676} (\bibinfo {year}
  {2023})}\BibitemShut {NoStop}%
\bibitem [{\citenamefont {{Gong}}\ \emph {et~al.}(2024)\citenamefont {{Gong}}, \citenamefont {{Cammerer}},\ and\ \citenamefont {{Renes}}}]{bpgdg}%
  \BibitemOpen
  \bibfield  {author} {\bibinfo {author} {\bibfnamefont {A.}~\bibnamefont {{Gong}}}, \bibinfo {author} {\bibfnamefont {S.}~\bibnamefont {{Cammerer}}},\ and\ \bibinfo {author} {\bibfnamefont {J.~M.}\ \bibnamefont {{Renes}}},\ }\bibfield  {title} {\bibinfo {title} {{Toward Low-latency Iterative Decoding of QLDPC Codes Under Circuit-Level Noise}},\ }\href {https://doi.org/10.48550/arXiv.2403.18901} {\bibfield  {journal} {\bibinfo  {journal} {arXiv e-prints}\ ,\ \bibinfo {eid} {arXiv:2403.18901}} (\bibinfo {year} {2024})},\ \bibinfo {note} {\url{https://arxiv.org/abs/2403.18901}},\ \Eprint {https://arxiv.org/abs/2403.18901} {arXiv:2403.18901 [quant-ph]} \BibitemShut {NoStop}%
\bibitem [{\citenamefont {Siegel}\ \emph {et~al.}(2024)\citenamefont {Siegel}, \citenamefont {Strikis},\ and\ \citenamefont {Fogarty}}]{strikisSC}%
  \BibitemOpen
  \bibfield  {author} {\bibinfo {author} {\bibfnamefont {A.}~\bibnamefont {Siegel}}, \bibinfo {author} {\bibfnamefont {A.}~\bibnamefont {Strikis}},\ and\ \bibinfo {author} {\bibfnamefont {M.}~\bibnamefont {Fogarty}},\ }\bibfield  {title} {\bibinfo {title} {Towards early fault tolerance on a $2\ifmmode\times\else\texttimes\fi{}n$ array of qubits equipped with shuttling},\ }\href {https://doi.org/10.1103/PRXQuantum.5.040328} {\bibfield  {journal} {\bibinfo  {journal} {PRX Quantum}\ }\textbf {\bibinfo {volume} {5}},\ \bibinfo {pages} {040328} (\bibinfo {year} {2024})}\BibitemShut {NoStop}%
\bibitem [{\citenamefont {{deMarti iOlius}}\ and\ \citenamefont {{Etxezarreta Martinez}}(2024)}]{closedbranch}%
  \BibitemOpen
  \bibfield  {author} {\bibinfo {author} {\bibfnamefont {A.}~\bibnamefont {{deMarti iOlius}}}\ and\ \bibinfo {author} {\bibfnamefont {J.}~\bibnamefont {{Etxezarreta Martinez}}},\ }\bibfield  {title} {\bibinfo {title} {{The closed-branch decoder for quantum LDPC codes}},\ }\href {https://doi.org/10.48550/arXiv.2402.01532} {\bibfield  {journal} {\bibinfo  {journal} {arXiv e-prints}\ ,\ \bibinfo {eid} {arXiv:2402.01532}} (\bibinfo {year} {2024})},\ \Eprint {https://arxiv.org/abs/2402.01532} {arXiv:2402.01532 [quant-ph]} \BibitemShut {NoStop}%
\bibitem [{\citenamefont {{Hillmann}}\ \emph {et~al.}(2024)\citenamefont {{Hillmann}}, \citenamefont {{Berent}}, \citenamefont {{Quintavalle}}, \citenamefont {{Eisert}}, \citenamefont {{Wille}},\ and\ \citenamefont {{Roffe}}}]{LSD}%
  \BibitemOpen
  \bibfield  {author} {\bibinfo {author} {\bibfnamefont {T.}~\bibnamefont {{Hillmann}}}, \bibinfo {author} {\bibfnamefont {L.}~\bibnamefont {{Berent}}}, \bibinfo {author} {\bibfnamefont {A.~O.}\ \bibnamefont {{Quintavalle}}}, \bibinfo {author} {\bibfnamefont {J.}~\bibnamefont {{Eisert}}}, \bibinfo {author} {\bibfnamefont {R.}~\bibnamefont {{Wille}}},\ and\ \bibinfo {author} {\bibfnamefont {J.}~\bibnamefont {{Roffe}}},\ }\bibfield  {title} {\bibinfo {title} {{Localized statistics decoding: A parallel decoding algorithm for quantum low-density parity-check codes}},\ }\href {https://doi.org/10.48550/arXiv.2406.18655} {\bibfield  {journal} {\bibinfo  {journal} {arXiv e-prints}\ ,\ \bibinfo {eid} {arXiv:2406.18655}} (\bibinfo {year} {2024})},\ \Eprint {https://arxiv.org/abs/2406.18655} {arXiv:2406.18655 [quant-ph]} \BibitemShut {NoStop}%
\bibitem [{\citenamefont {Higgott}()}]{pymatching}%
  \BibitemOpen
  \bibfield  {author} {\bibinfo {author} {\bibfnamefont {O.}~\bibnamefont {Higgott}},\ }\href@noop {} {\bibinfo {title} {Pymatching}},\ \bibinfo {howpublished} {\url{https://github.com/oscarhiggott/PyMatching}},\ \bibinfo {note} {accessed: 2024-07-26}\BibitemShut {NoStop}%
\bibitem [{Note4()}]{Note4}%
  \BibitemOpen
  \bibinfo {note} {We only discuss Hadamard gates for the noise model since those are generally the ones appearing in QEC syndrome extraction circuits. In a more generic case, single qubit gates related to rotations involving $X$ or $Y$ basis will be followed by depolarizing noise, while rotations strictly over the $Z$ axis can be done in a bias-preserving manner.}\BibitemShut {Stop}%
\bibitem [{Note5()}]{Note5}%
  \BibitemOpen
  \bibinfo {note} {Note that the standard depolarizing case does not occur at $\eta =1/2$ since at such value, the two-qubit gate errors will occur with probability $p/9$ for the biased errors and with $p/18$ for the rest. Thus, the standard depolarizing case has to be defined on its own.}\BibitemShut {Stop}%
\bibitem [{\citenamefont {{Tan}}\ \emph {et~al.}(2024)\citenamefont {{Tan}}, \citenamefont {{Pattison}}, \citenamefont {{McEwen}},\ and\ \citenamefont {{Preskill}}}]{teraBurst}%
  \BibitemOpen
  \bibfield  {author} {\bibinfo {author} {\bibfnamefont {S.~J.~S.}\ \bibnamefont {{Tan}}}, \bibinfo {author} {\bibfnamefont {C.~A.}\ \bibnamefont {{Pattison}}}, \bibinfo {author} {\bibfnamefont {M.}~\bibnamefont {{McEwen}}},\ and\ \bibinfo {author} {\bibfnamefont {J.}~\bibnamefont {{Preskill}}},\ }\bibfield  {title} {\bibinfo {title} {{Resilience of the surface code to error bursts}},\ }\href {https://doi.org/10.48550/arXiv.2406.18897} {\bibfield  {journal} {\bibinfo  {journal} {arXiv e-prints}\ ,\ \bibinfo {eid} {arXiv:2406.18897}} (\bibinfo {year} {2024})},\ \Eprint {https://arxiv.org/abs/2406.18897} {arXiv:2406.18897 [quant-ph]} \BibitemShut {NoStop}%
\bibitem [{\citenamefont {Gidney}(2021)}]{stim}%
  \BibitemOpen
  \bibfield  {author} {\bibinfo {author} {\bibfnamefont {C.}~\bibnamefont {Gidney}},\ }\bibfield  {title} {\bibinfo {title} {Stim: a fast stabilizer circuit simulator},\ }\href {https://doi.org/10.22331/q-2021-07-06-497} {\bibfield  {journal} {\bibinfo  {journal} {{Quantum}}\ }\textbf {\bibinfo {volume} {5}},\ \bibinfo {pages} {497} (\bibinfo {year} {2021})}\BibitemShut {NoStop}%
\bibitem [{\citenamefont {Horsman}\ \emph {et~al.}(2012)\citenamefont {Horsman}, \citenamefont {Fowler}, \citenamefont {Devitt},\ and\ \citenamefont {Meter}}]{surgery}%
  \BibitemOpen
  \bibfield  {author} {\bibinfo {author} {\bibfnamefont {D.}~\bibnamefont {Horsman}}, \bibinfo {author} {\bibfnamefont {A.~G.}\ \bibnamefont {Fowler}}, \bibinfo {author} {\bibfnamefont {S.}~\bibnamefont {Devitt}},\ and\ \bibinfo {author} {\bibfnamefont {R.~V.}\ \bibnamefont {Meter}},\ }\bibfield  {title} {\bibinfo {title} {Surface code quantum computing by lattice surgery},\ }\href {https://doi.org/10.1088/1367-2630/14/12/123011} {\bibfield  {journal} {\bibinfo  {journal} {New Journal of Physics}\ }\textbf {\bibinfo {volume} {14}},\ \bibinfo {pages} {123011} (\bibinfo {year} {2012})}\BibitemShut {NoStop}%
\end{thebibliography}%

\end{document}



\title{Supplementary Material for Leveraging biased noise for more efficient quantum error correction at the circuit-level with two-level system qubits}

\author{Josu {Etxezarreta Martinez}}
\email{jetxezarreta@unav.es}
\affiliation{Department of Basic Sciences, Tecnun - University of Navarra, 20018 San Sebastian, Spain.}
\affiliation{Cavendish Laboratory, Department of Physics, University of Cambridge, Cambridge CB3 0HE, UK.}
\author{Paul Schnabl}
\affiliation{Department of Basic Sciences, Tecnun - University of Navarra, 20018 San Sebastian, Spain.}
\affiliation{Institute for Theoretical Physics, University of Innsbruck, A-6020 Innsbruck, Austria.}
\author{Javier {Oliva del Moral}}
\affiliation{Department of Basic Sciences, Tecnun - University of Navarra, 20018 San Sebastian, Spain.}
\affiliation{Donostia International Physics Center, 20018 San Sebastian, Spain.}
\author{Reza Dastbasteh}
\affiliation{Department of Basic Sciences, Tecnun - University of Navarra, 20018 San Sebastian, Spain.}
\author{Pedro M. Crespo}
\affiliation{Department of Basic Sciences, Tecnun - University of Navarra, 20018 San Sebastian, Spain.}
\author{Ruben M. Otxoa}
\affiliation{Hitachi Cambridge Laboratory, J. J. Thomson Avenue, Cambridge, CB3 0HE, United Kingdom.}

\maketitle

\section*{Figures for footprint estimations of Section III B}
\begin{figure}[!ht]
\centering
\includegraphics[width=12cm]{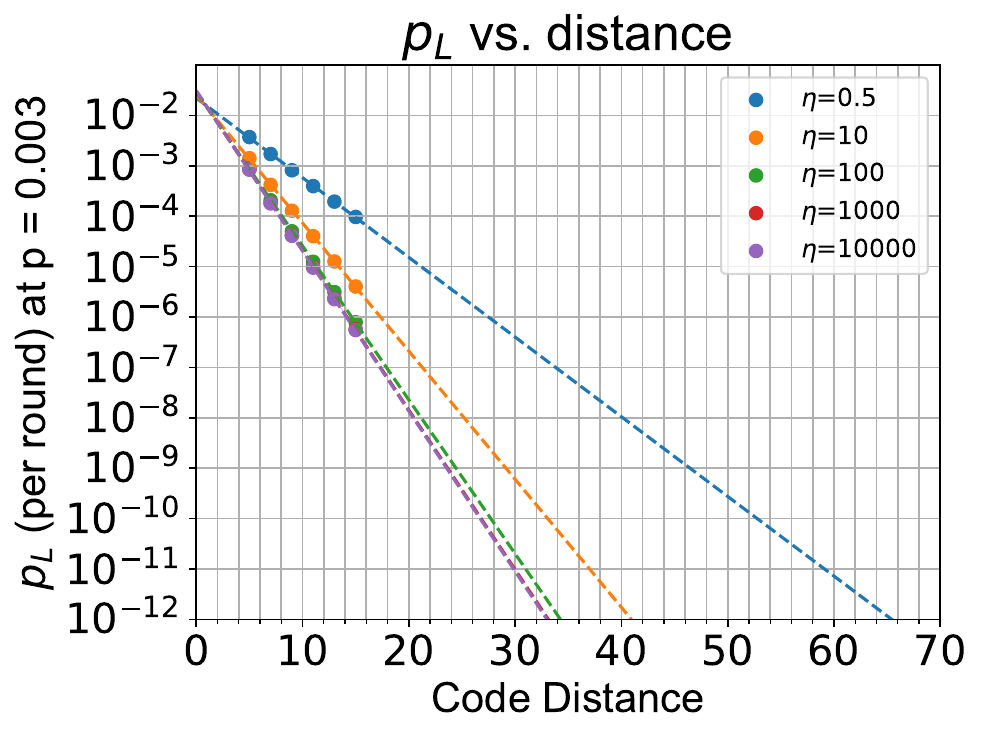}
\caption{Logical error rates per round versus rotated XZZX surface code distances with bias-preserving CZ gates and CNOTs with the residual biases discussed in Section II. The plot is done for physical error rate $p=0.003$.}
\label{fig:sub1}
\end{figure}

\begin{figure}[!ht]
\centering
\includegraphics[width=12cm]{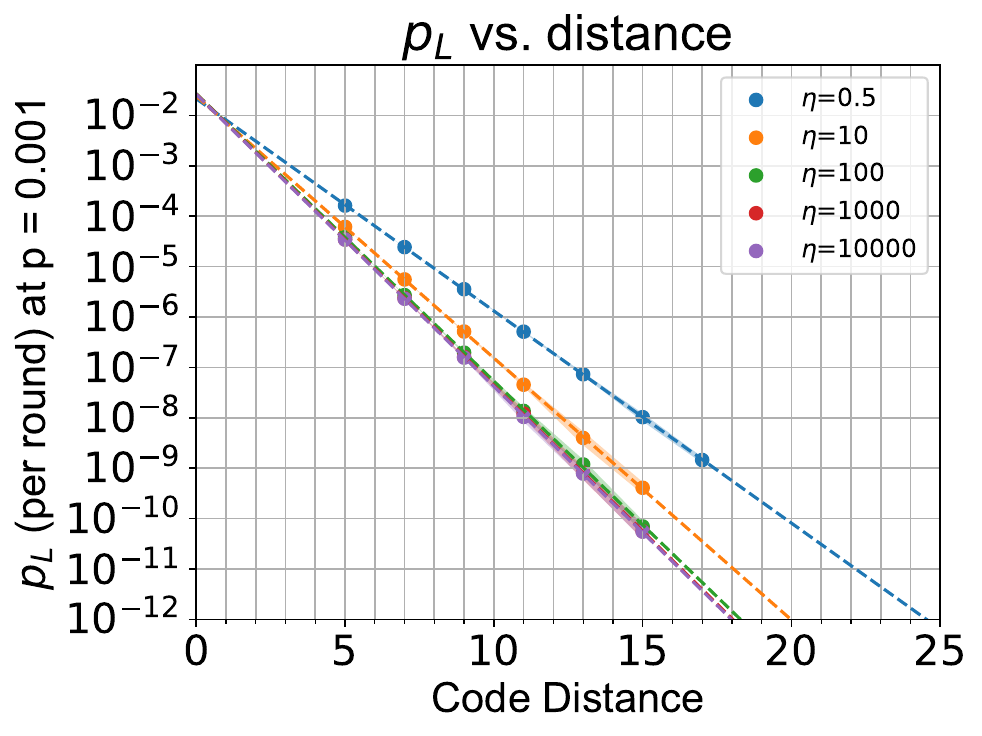}
\caption{Logical error rates per round versus rotated XZZX surface code distances with bias-preserving CZ gates and CNOTs with the residual biases discussed in Section II. The plot is done for physical error rate $p=0.001$.}
\label{fig:sub2}
\end{figure}

Figures \ref{fig:sub1} and \ref{fig:sub2} present the numerically obtained logical error probabilities per round as a function of the distance as well as the projections done to estimate the physical qubit footprints presented in Section III B in the main text. Details on the numerical methods used can be found in the Methods section in the main text. The logical error rates in these figures are for a Horizontal (H) memory experiment. Both memories behave almost similarly deep below threshold, with the H memory showing a slightly worse logical error probability, $p_L$ and, thus, it is the limiting one.

\section*{Compiling the XZZX code with CZ gates}

In many qubit platforms, CNOT gates are not usually enabled in a direct manner by means of pulses such as the one described in Section II. The CNOTs are compiled by means of a CZ gate sandwiched by Hadamard gates on the target qubit. This is the case, for example, for silicon spin qubits or neutral atoms, among others. Thus, it is relevant to compile the extraction circuit by means of those gates. Here, we numerically study how this compilation does allow leveraging system bias. For our numerical evaluation we consider the HBD circuit-level noise model proposed in this work.

\subsubsection{Threshold}
\begin{figure}[!h]
\centering
\includegraphics[width=\columnwidth]{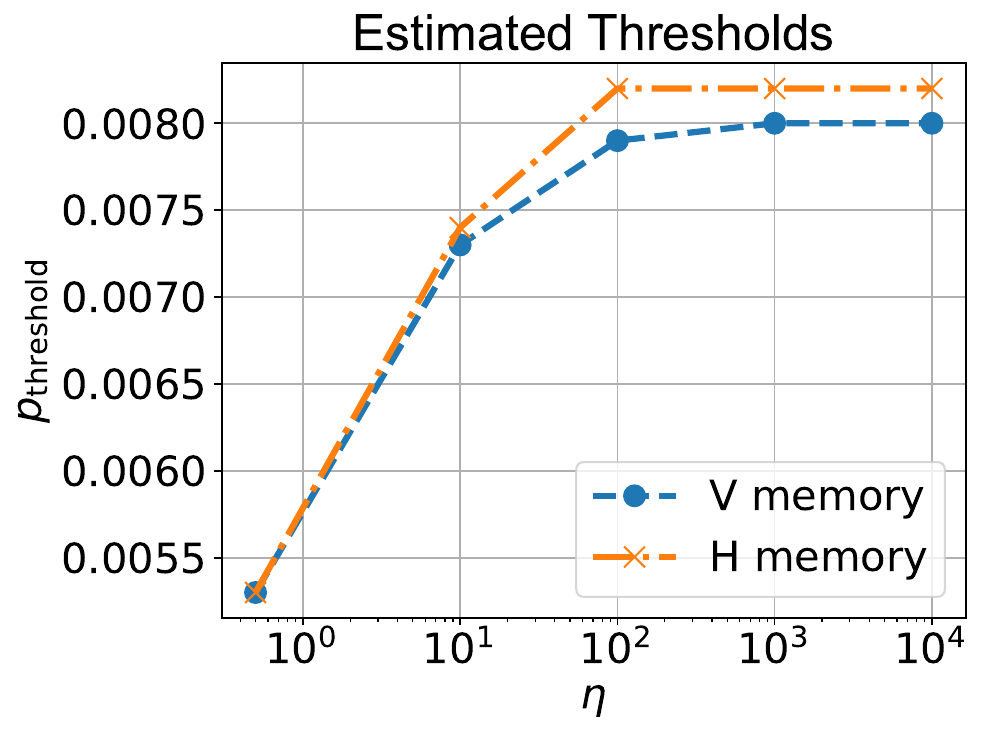}
\caption{Threshold of the rotated XZZX code over the hybrid biased-depolarizing (HBD) model with a syndrome extraction circuit with only CZ as entangling gates. The threshold improved from around $0.53\%$ for the standard depolarizing (SD) noise model up to $0.8\%$. This is a $51\%$ improvement. Furthermore, a saturation of the threshold can still be observed for high bias values.}
\label{fig:ThreshonlyCZ}
\end{figure}

In Figure \ref{fig:ThreshonlyCZ} we show the threshold of the XZZX surface code as a function of the system bias for the extraction circuit compilation using exclusively CZ gates as entangling gates. The threshold here improves from a physical error rate around $0.53\%$ up to a $0.8\%$ for system biases above $\eta = 100$. This is a $50\%$ improvement in threshold. Note that, here, the threshold values both for the standard depolarizing and biased cases are lower than for the previous sections. This occurs due to the extra error locations introduced by the layers of Hadamard gates on the data qubits required for $X$ stabiliser extractions. Since the hybrid bias depolarizing (HBD) model we consider assumes all error locations to fail with the same error rate, the errors introduced by said Hadamard gates have a considerable impact in code performance. Note, however, that in realistic hardware, single qubit gates present error rates around an order of magnitude lower than entangling gates. Thus, the performance will be better when considering such non-identically distributed noise scenarios \cite{toninid,willow,teraHoneycomb,rotvsunrot,spinhex}.

\begin{table}[h!]
 \centering
\caption{Improvement of the threshold as a function of the bias for the HBD circuit-level noise model and CZ compilation.}
\label{tab:CZthreshGains}
\begin{tabular}{ |c|c|c| }
 \hline
 $\eta$ & $p_{threshold}$ & Improvement ($\approx$) \\ 
 \hline\hline
$1/2$ (SD) & $0.53\%$ & - \\ 
 \hline
  $10$ & $0.73\%$ & $37.7\%$ \\ 
 \hline
  $100$ & $0.79\%$ & $49\%$ \\ 
 \hline
  $1000$ & $0.8\%$ & $51\%$ \\ 
 \hline
  $10000$ & $0.8\%$ & $51\%$ \\ 
 \hline
\end{tabular}
\end{table}

\begin{table*}[t]
 \centering
\caption{Qubit footprints and relative decrease in percentage with respect to the standard depolarizing (SD) circuit-level noise model for each bias value considered at $p=0.003$ with the compilation only using CZ gates. The footprints are computed by selecting the first odd distance value below the target logical error probability in Supplementary Figure \ref{fig:sub3} and using $d^2+(d^2-1)$ as the number of qubits (data and check qubits) required for such distance, where d is the code distance. }
\label{tab:foot0_003_CZ}
\begin{tabular}{|c|c|c|c|c|c|c| }
 \hline
 $\eta$ & Megaquop footprint & Decrease ($\approx$) & Gigaquop footprint & Decrease ($\approx$) & Teraquop footprint & Decrease ($\approx$) \\ 
 \hline\hline
$1/2$ (SD) & $3697$ & - &  $10657$ & - & $21217$ & -\\ 
 \hline
  $10$ & $881$ & $76\%$ & $3041$ & $72\%$ & $6049$ & $72\%$\\ 
 \hline
  $100$ & $881$ & $76\%$ & $2449$ & $77\%$ & $4417$ & $79\%$\\ 
 \hline
  $1000$ & $881$ & $76\%$ & $2449$ & $77\%$ & $4417$ & $79\%$\\ 
 \hline
  $10000$ & $881$ & $76\%$ & $2449$ &  $77\%$ & $4417$ & $79\%$\\ 
 \hline
\end{tabular}
\end{table*}

\begin{table*}[t]
 \centering
\caption{Qubit footprints and relative decrease in percentage with respect to the symmetric SD circuit-level noise model for each bias value considered at $p=0.001$ with the compilation only using CZ gates. The footprints are computed by selecting the first odd distance value below the target logical error probability in Supplementary Figure \ref{fig:sub4} and using $d^2+(d^2-1)$ as the number of qubits (data and check qubits) required for such distance. }
\label{tab:foot0_001_CZ}
\begin{tabular}{|c|c|c|c|c|c|c| }
 \hline
 $\eta$ & Megaquop footprint & Decrease ($\approx$) & Gigaquop footprint & Decrease ($\approx$) & Teraquop footprint & Decrease ($\approx$) \\ 
 \hline\hline
$1/2$ (SD) & $337$ & - &  $881$ & - & $1681$ & -\\ 
 \hline
  $10$ & $241$ & $29\%$ & $577$ & $35\%$ & $1057$ & $37\%$\\ 
 \hline
  $100$ & $161$ & $52\%$ & $449$ & $49\%$ & $881$ & $48\%$\\ 
 \hline
  $1000$ & $161$ & $52\%$ & $449$ & $49\%$ & $881$ & $48\%$\\ 
 \hline
  $10000$ & $161$ & $52\%$ & $449$ &  $49\%$ & $881$ & $48\%$\\ 
 \hline
\end{tabular}
\end{table*}
\subsubsection{Footprints}
Tables \ref{tab:foot0_003_CZ} and \ref{tab:foot0_001_CZ} present the qubit footprints required to reach the three quantum computational regimes discussed before, for physical error rates of $p=0.003$ and $p=0.001$, respectively. The logical error rates per round as a function of the code distance obtained by numerical simulations are presented in Supplementary Figures \ref{fig:sub3} and \ref{fig:sub4}. Similar to section IIIB, we observe that the required footprints reduce in a significant manner, ranging from $34\%$ up to $79\%$. As discussed before, these reductions in qubit footprints come with no other cost than constructing CZ gates that preserve the bias of the system. Importantly, note that the footprints obtained in this case using only CZ entangling gates result in higher qubit numbers than in Section IIIB. This arises from the extra error locations introduced by the layers of Hadamard gates on the data qubits required for $X$ stabiliser extractions, as explained for the code thresholds. 

\begin{figure}[!ht]
\centering
\includegraphics[width=12cm]{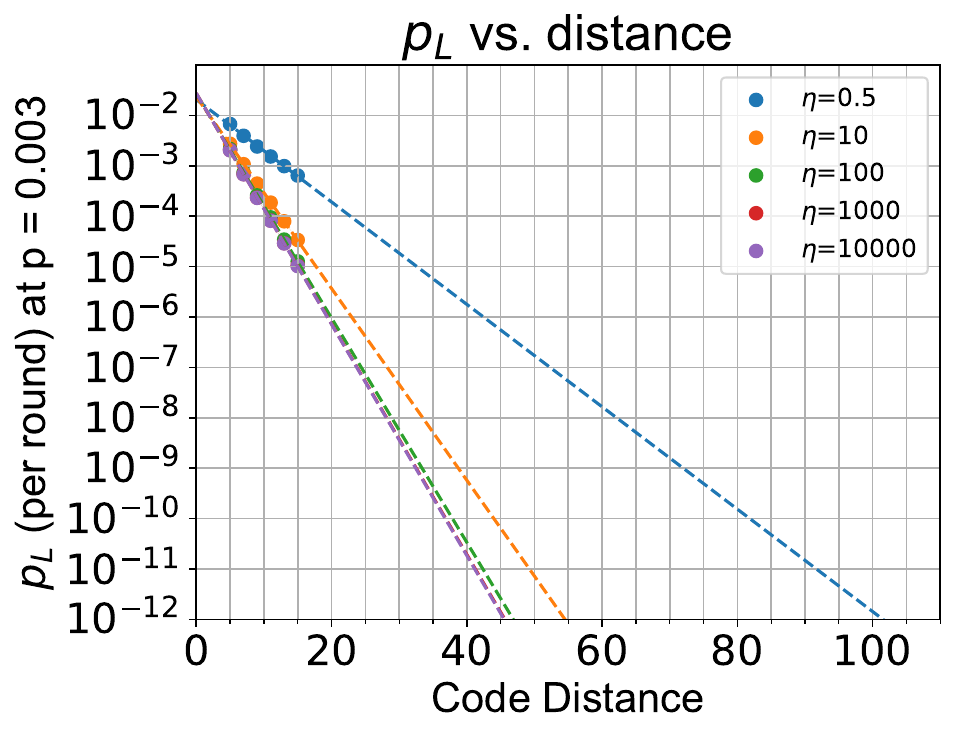}
\caption{Logical error rates per round versus rotated XZZX surface code distances with the extraction circuit using only bias-preserving CZ gates as entangling gates discussed in Section II. The plot is done for physical error rate $p=0.003$.}
\label{fig:sub3}
\end{figure}

\begin{figure}[!ht]
\centering
\includegraphics[width=12cm]{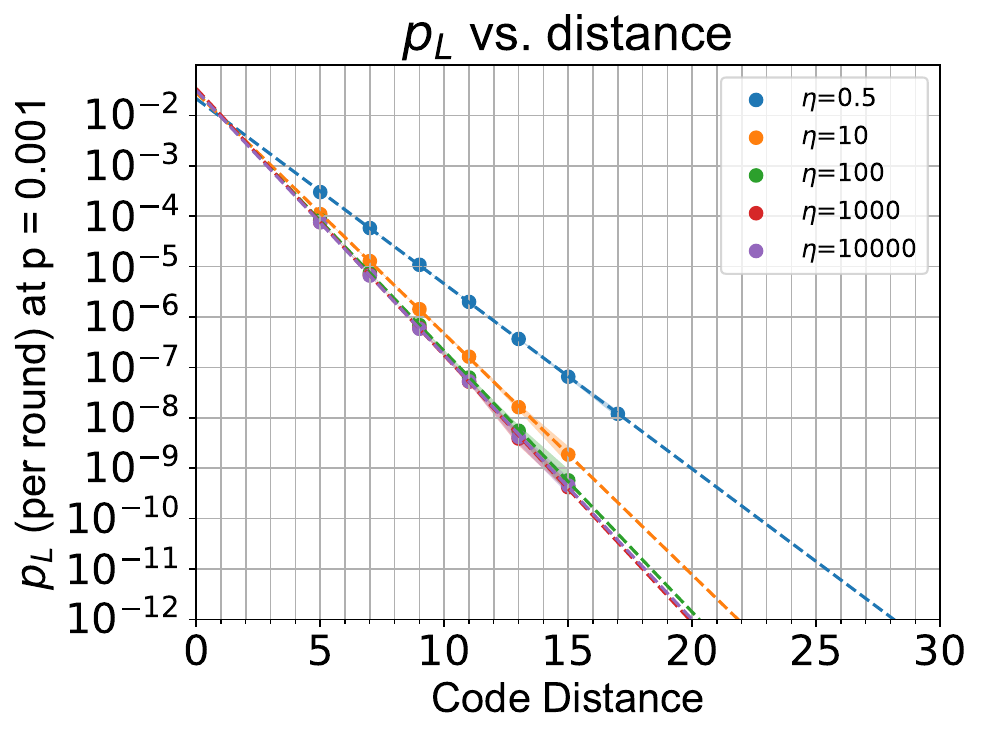}
\caption{Logical error rates per round versus rotated XZZX surface code distances with the extraction circuit using only bias-preserving CZ gates as entangling gates discussed in Section II. The plot is done for physical error rate $p=0.001$.}
\label{fig:sub4}
\end{figure}